\documentclass[aps,prb,reprint]{revtex4-1}
\usepackage{epsfig}
\usepackage{subfigure}
\usepackage{dcolumn}
\usepackage{graphicx,graphics}
\usepackage{amsmath}
\usepackage{amssymb}
\usepackage{bm}
\usepackage{color}
\usepackage[utf8]{inputenc}
\usepackage{pifont}
\usepackage[colorlinks,linkcolor={blue},citecolor={blue},urlcolor={blue}]{hyperref}
\usepackage{mathtools}
\usepackage{booktabs}
\newcommand{\Top}{{\rm T}}
\newcommand{\Bot}{{\rm B}}

\newcommand{\bk}{{\bm k}}
\newcommand{\bq}{{\bm q}}

\newcommand{\zh}{\hat{z}}
\newcommand{\nh}{\hat{n}}

\newcommand{\bkpar}{{\bm k}_{ \scriptscriptstyle \parallel}}
\newcommand{\bqpar}{{\bm q}_{ \scriptscriptstyle \parallel}}
\newcommand{\bppar}{{\bm p}_{ \scriptscriptstyle \parallel}}
\newcommand{\kpar}{ k_{ \scriptscriptstyle \parallel}}
\newcommand{\sbkpar}{{ \bm k}_{{\tiny{\text{$ \scriptscriptstyle \parallel$}}}}}
\newcommand{\qpar}{q_{ \scriptscriptstyle \parallel}}
\newcommand{\ppar}{p_{ \scriptscriptstyle \parallel}}
\newcommand{\bsigma}{{\bm \sigma}}
\newcommand{\bqz}{{\bm q}_{z}}
\newcommand{\rS}{{\rm S}}
\newcommand{\hn}{\hat{n}}
\begin{document}

\title{Interlayer RKKY Coupling in Bulk Rashba Semiconductors \\
under Topological Phase Transition}

\author{Mahmoud M. Asmar}
\affiliation{Department of Physics and Astronomy, Center for Materials for Information Technology, The University of Alabama, Tuscaloosa, AL 35487, USA }
\author{Wang-Kong Tse }
\affiliation{Department of Physics and Astronomy, Center for Materials for Information Technology, The University of Alabama, Tuscaloosa, AL 35487, USA }
\date\today

\begin{abstract}
The bulk Rashba semiconductors BiTeX (X=I, Cl and Br) with intrinsically enhanced Rashba spin-orbit coupling provide a new platform for
investigation of spintronic and magnetic phenomena in materials.
We theoretically investigate the interlayer exchange interaction
between two ferromagnets deposited on opposite surfaces of a bulk Rashba
semiconductor BiTeI in its trivial and topological insulator phases.
In the trivial phase BiTeI, we find that for ferromagnets with a
magnetization orthogonal to the interface, the exchange coupling is
reminiscent of that of a conventional three-dimensional
metal. Remarkably, ferromagnets with a magnetization parallel to the
interface display a magnetic exchange qualitatively different from
that of conventional three-dimensional metal due to the spin-orbit
coupling. In this case,
the interlayer exchange interaction acquires two periods of oscillations
and decays as the inverse of the thickness of the BiTeI layer. For
topological BiTeI, the magnetic exchange interaction becomes mediated
only by the helical surface states and acts between the one-dimensional
spin chains at the edges of the sample.
The surface state-mediated interlayer exchange interaction allows for the coupling of ferromagnets with non-collinear magnetization and displays a decay power different from that of trivial BiTeI, allowing the detection of the topological phase transition in this material. Our work provides insights into the magnetic properties of these newly discovered materials and their possible functionalization.
\end{abstract}

\maketitle

\section{Introduction} \label{intro}
The Rashba spin-orbit coupling (SOC) in materials arises from
broken spatial inversion symmetry. It has been extensively
studied in
two-dimensional (2D) electron systems such as heterointerfaces of
semiconductors, thin films and surfaces of heavy
metals~\cite{Rashba1,Rashba2,Rashba3,Rashba4,Rashba5,Rashba6}.
Due to broken inversion symmetry in these systems, electrons experience a
perpendicular electric field $\bm{E}$, which generates a spin-orbit
coupling $\bm{E}\cdot(\bsigma\times\bk)$
that depends on the electrons' momentum $\bk$ and spin $\bsigma$. This Rashba effect in a 2D electron gas splits the spin-degenerate parabolic bands into dispersions with opposite spin polarizations~\cite{Rashba}.

Three-dimensional (3D) materials with \textit{intrinsically} broken
inversion symmetry can also exhibit Rashba SOC. Recent
theoretical and experimental studies have revealed a giant Rashba splitting
in bismuth tellurohalides BiTeX (X=I,Cl and Br) due to the large
internal electric field between the constituent layers of these
materials~\cite{BiTeI1,BiTeI2,BiTeI3,BiTeI4,BiTeI5,BiTeI6}. These
materials became known as bulk Rashba semiconductors. The Rashba
semiconductor BiTeI, in particular, has been shown to undergo a
pressure-controlled topological phase
transition~\cite{BiTeITop,BiTeITop1,BiTeITopExp,BiTeITopExp3,BiTeITopExp2}. As
this material is subjected to an increasing hydrostatic pressure it
transitions from a non-topological phase to a strong topological
insulator phase, at approximately 3 GPa~\cite{BiTeITop,BiTeITop1,BiTeITopExp,BiTeITopExp3}, and these two phases are separated by an intermediate Weyl phase~\cite{BiTeITopTheory,BiTeITopTheory1,BiTeITopTheory2}. Rashba semiconductors have generated much interest as a new material platform for spintronics and controlled topological phenomena~\cite{Spintronics1,Spintronics2,Rev1,Rev2,Bernevig}.

Heterostructures composed of magnetic and non-magnetic materials are
an important platform that allow controlled information transfer
between spins~\cite{layered1,layered2}. The key ingredient to this
transfer is the effective interaction between the magnetic moments
mediated by the conduction electrons of the non-magnetic host material. This effective interaction is known as the
Ruderman-Kittel-Kasuya-Yosida (RKKY) or the indirect carrier-mediated exchange
interaction~\cite{RKKY1,RKKY2,RKKY3}. The theory of RKKY interaction
was initially formulated to address the problem of the interaction between magnetic
impurities in the bulk of a metal. Because it is carrier-mediated,
RKKY interaction depends on the dimensionality of the host metal and the nature of its low-energy
fermionic
excitations~\cite{Graphenerkky,MOS2RKKY,MOS2RKKY2,DiracSemimetaRKKY,DiracSemimetaRKKY2}.
In 3D conventional metals characterized by a single spin-degenerate parabolic band,
the strength of this interaction undergoes oscillations as a function of
the impurities' separation $r$ at a period $\lambda_{F}/2$ given by
the Fermi wavelength $\lambda_{F}$, while the envelope of the oscillations
decays as $r^{-3}$~\cite{RKKY1,RKKY2,RKKY3}. The same physical
mechanism occurs in a ferromagnet-normal metal-ferromagnet (FM/NM/FM) trilayer
structure, with each of the ferromagnetic layers forming a 2D
collection of spins at the interface and the metal spacer mediating
the indirect exchange interaction. The RKKY theory was generalized in the seminal work by Bruno and
Chappert ~\cite{LayersRKKY1} to describe the oscillatory
interlayer exchange coupling between the ferromagnetic layers. For a conventional metallic
spacer such as Au or Cu, it predicts multiple
oscillation periods of the coupling decaying as the inverse square of
the spacer thickness, \textit{i.e.}
$z^{-2}$~\cite{LayersRKKY2}. A review of interlayer exchange coupling
in magnetic multilayers can be found in Ref.~\cite{Stiles2005}.

The interplay between magnetic and spin-orbit effects provides the
basis for a number of wide-ranging phenomena, such as topological
phases of matter, magnetic domain walls, Majorana bound states and
magnetic
skyrmions~\cite{Rev1,Rev2,Bernevig,Majoranawire,domainwall,skyrmions}.
Hence, the strong and intrinsically generated Rashba SOC in bismuth
tellurohalides and their pressure-controlled topological phases promise to provide important insights on the heretofore unexplored interlayer exchange coupling mediated by these materials.

In this paper, we consider the problem of interlayer exchange coupling
mediated by conduction electrons in a bulk Rashba semiconductor, in its trivial and topological phases,
between ferromagnetic layers.
The main ingredient of our theory is the $q_z$-dependent static spin
susceptibility, which we
have obtained analytically.
The closed-form
result of the spin susceptbility facilitates identification of the
Fermi surface singularities (Kohn
anomalies) and enables us to capture the salient long-range dependence of
the interlayer exchange coupling, including the periods of its oscillations and the spatial decay of its envelope.
In the non-topological phase, our theory predicts that the interlayer coupling is strongly dependent on the
magnetization directions of the ferromagnets. For ferromagnets
with magnetization orthogonal to the interface, we find that
SOC effects are not prominent and the interlayer exchange coupling
behaves as in conventional 3D
metals, decaying with the spacer thickness $z$ as $z^{-2}$.
In contrast, SOC effects are found to play an important role when
the ferromagnets' magnetization directions are parallel to
interface; the dominant contribution of the interlayer exchange
coupling is proportional to the Rashba SOC parameter $\alpha$ and decays as
$z^{-1}$.  In the topological phase, where the bulk of these materials becomes insulating and the conduction only happens through the surface electrons, the magnetic exchange in the system becomes limited to the magnetic chains at the sample's edges and is mediated by the 2D helical surface states.  Unlike the non-topological phase, the interlayer exchange interaction in the topological phase not only couples collinear spins but also non-collinear spins via the Dzyaloshinskii-Moriya (DM) interaction, decaying with the thickness as $z^{-3/2}$. The qualitative differences exhibited by the arrangement of the ferromagnets' magnetization directions highlight the role of the
spacer's Rashba SOC and band topology in the magnetic trilayer geometry.

The remainder of this paper is organized as follows. In Sec.~\ref{RSC}
we introduce the low-energy effective model for the Rashba
semiconductor BiTeI and describe its Fermi surface and the associated
spin textures. We then
develop the formalism for the interlayer exchange coupling between two ferromagnets sandwiching the BiTeI layer in
Sec~\ref{sec2}. In Sec.~\ref{sec3}, we first employ this formalism to
study the interlayer exchange coupling mediated by a 3D electron
gas with an anisotropic Fermi surface but without SOC. In Sec.~\ref{S45}
we turn our attention to the
case with BiTeI as the spacer and compare critically the obtained results including Rashba SOC with those
obtained without SOC from Sec.~\ref{sec3}.
The case of two ferromagnets with magnetizations orthogonal to
the interface is studied in Sec.~\ref{sec4} and the case of two
ferromagnets with magnetizations parallel to the interface in
Sec.~\ref{sec5}. In Secs.~\ref{sec6} and \ref{sec7} we study the
interlayer exchange interaction in the topological phase of BiTeI. In
Sec.~\ref{sec6} the interlayer exchange coupling formalism is adapted to the
helical surface states of topological BiTeI, the spin susceptibility
of the helical surface states is found and a generic form of the
interlayer exchange interaction is derived. In Sec.~\ref{sec7} the dependence of
the interlayer exchange coupling on the thickness of topological BiTeI is derived.
Finally, Secs.~\ref{discuss} and \ref{conc}  provide a discussion of possible experimental realizations and our concluding remarks.

\section{Bulk Rashba Semiconductor, B\lowercase{i}T\lowercase{e}I} \label{RSC}
\begin{figure}[t]
  \centering
  \includegraphics[width=\columnwidth]{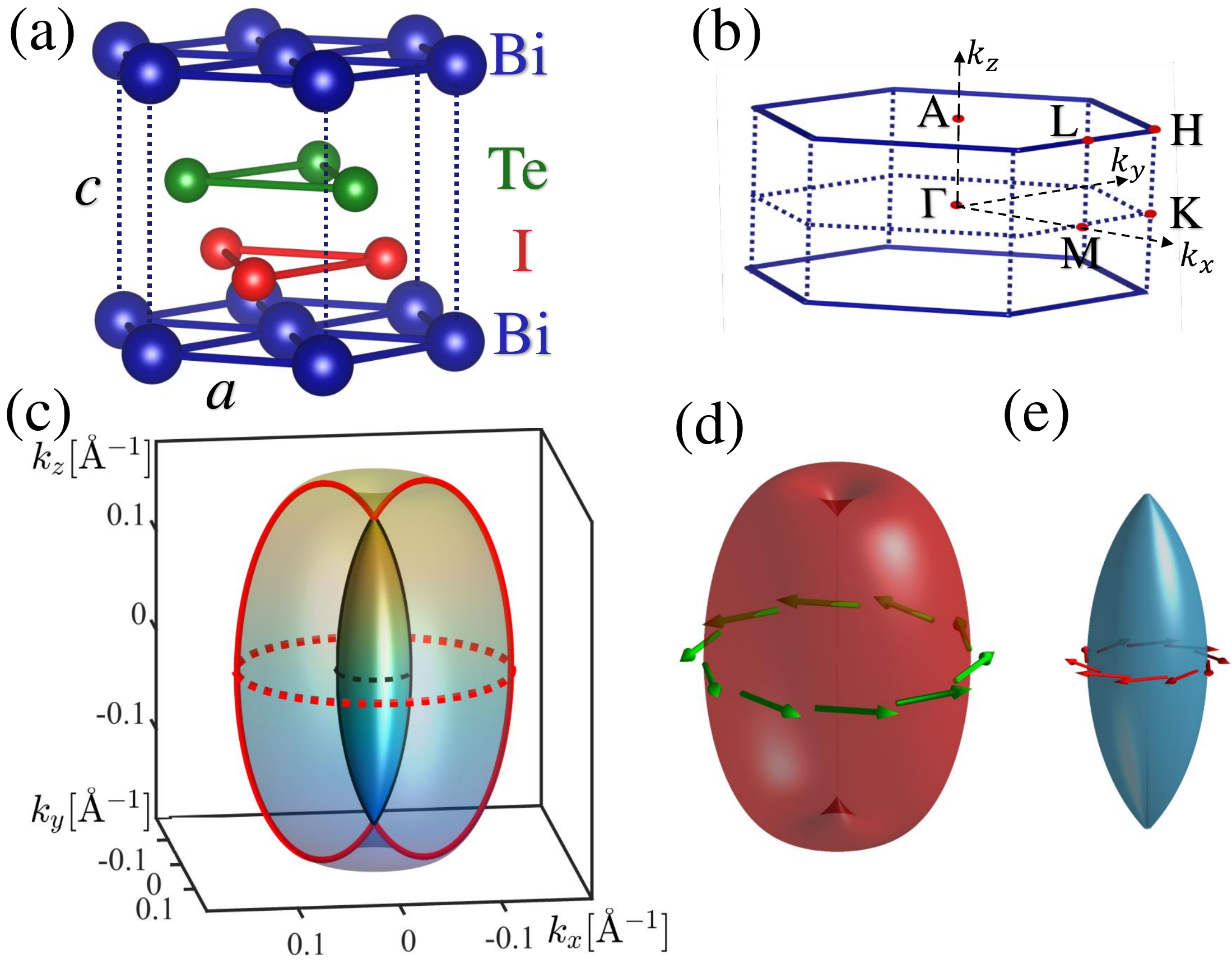}
  \caption{(a) BiTeI crystal structure. (b) Brillouin zone of BiTeI. (c) Fermi surface of BiTeI centred at the A point for $E_{F}>0$. (d) Outer portion of the Fermi surface with negative helicity electrons. (e) Inner portion of the Fermi surface with positive helicity electrons. }\label{fig1}
\end{figure}
Bismuth tellurohalides BiTeX (X=I,Cl and Br) generally have a layered lattice
structure.
In particular, the crystal of BiTeI is composed of
a series of asymmetrically stacked triangular layers of Bi, Te and I
atoms which leads to an intrinsically broken inversion symmetry along its
stacking, $c$, axis. Fig.~\ref{fig1}(a) shows the crystal structure of BiTeI and its
corresponding Brillouin zone (BZ) is shown in Fig.~\ref{fig1}(b). The asymmetric stacking of the Bi,
Te and I layers causes a net polarity along the stacking direction~\cite{BiTeI1,BiTeI2,BiTeI3}.
Due to the net polarity along the $c$-axis, electrons in the $a$-$b$
plane experience an SOC. The symmetry properties of this SOC are
constrained by the space symmetries of the $a$-$b$ plane. Owing to the threefold symmetry of this plane in BiTeI, the intrinsic 2D SOC in this material inherits this symmetry.
At low energies, this symmetry restricts the form of the SOC to
$\alpha(\bsigma\times \bppar)_{\zh}$, where  $\bsigma=(\sigma_{x},\sigma_{y},\sigma_{z})$ is the vector of Pauli
matrix representing spins,
$\bppar=(p_{x},p_{y},0)$ is the in-plane momentum and $\alpha$ is the
Rashba SOC strength. The low-energy electrons of this material are
centered in the vicinity of the BZ's hexagonal face center known as the
A-point [A$=(0,0,\pm\pi/c)$]~\cite{BiTeI1,BiTeI2,BiTeI3,BiTeI4,BiTeI5,BiTeI6,Transport2BiTeI,OpticalBiTeI,Transport2BiTeI}, and are described by the effective Hamiltonian
\begin{equation}\label{effective1}
H_{{\rm BiTeI}}=Ap_{z}^{2}+B\bppar^{2}+\alpha(\bsigma\times \bppar)_{\zh}\;.
\end{equation}
The material parameters $A\approx 8.04$ eV$\rm\AA^{2}$, $B\approx
40.21$ eV$\rm\AA^{2}$,~\cite{BiTeI1,BiTeI2,Transport2BiTeI,OpticalBiTeI,Transport2BiTeI}
$\alpha\approx 3.85$ eV$\rm\AA$, $c= 6.854\;\rm\AA$ and
$a=b=4.34\;\rm\AA$ have been reported in the literature from
photoemission, transport, optical experiments and density functional
theory
studies~\cite{BiTeI1,BiTeI2,BiTeI3,BiTeI4,BiTeI5,BiTeI6,Transport2BiTeI,OpticalBiTeI,Transport2BiTeI}.
Eq.~\eqref{effective1} is valid up to a cutoff energy $E_c =
0.2$eV, beyond which the dispersion  acquires trigonal warping effects and two additional bands. Since the low-energy Hamiltonian of BiTeI commutes with the helicity
operator, $\hat{h}=({\bsigma}\times \bppar)_{\zh}/\ppar$, the
helicity eigenstates diagonalize $H_{{\rm BiTeI}}$ and  are given by
 \begin{equation}\label{eigenstates}
\vert \bm{k},\mu\rangle =\frac{1}{\sqrt{2}}\left(
                        \begin{array}{c}
                          i \\
                          \mu e^{i\phi_{k}} \\
                        \end{array}
                      \right)e^{i(\sbkpar\cdot{\bm r}+k_{z}z)}\;,
\end{equation}
where $\bk=(k_{x},k_{y},k_{z})$, $\phi_{k}=\tan^{-1}(k_{y}/k_{x})$ is the azimuthal angle of $\bkpar=(k_{x},k_{y},0)$
and $\mu=\pm$ is the helicity of the eigenstate. The corresponding
energy eigenvalues  are
\begin{equation}\label{muvalues}
E_{\bk,\mu}=Ak^{2}_{z}+B\kpar^{2}+\mu\alpha \kpar\;,
\end{equation}
where $\kpar=\sqrt{k_x^{2}+k^{2}_{y}}$. Stoichiometric BiTeI is an
n-doped semiconductor where the Fermi energy is located above the
Dirac node that results from the Rashba
SOC~\cite{BiTeI1,BiTeI2,TransportBiTeI,Transport2BiTeI}.
As we explain in Fig.~\ref{fig1}, the Fermi surface for $E_{F}>0$
consists of two segments characterized by states with opposite
helicities. The in-plane momenta that span these
two parts of the Fermi surface individually can be obtained by solving
for $\kpar$ from the dispersion $Ak^{2}_{z}+B\kpar^{2}+\mu\alpha \kpar=E_{F}$ for a
particular value of $k_z$.
The radii $k_{\mu}$ for a given $k_z $ plane for the positive (negative)
helicity segment of the Fermi surface is determined as
\begin{subequations}\label{cases}
\begin{eqnarray}
  k_{\mu}= k_{F}-\mu\frac{\alpha}{2B},\; & \mbox{for } k_{z}\le k_{D} \\
  k_{-}=\pm k_{F}+\frac{\alpha}{2B},\; & \mbox{for } |k_{z}|\ge k_{D}\;.
\end{eqnarray}
\end{subequations}
where
$\pm k_{D}$, with $k_{D}=\sqrt{E_{F}/A}$, specify the locations
of the two Dirac points along the $k_z$ axis and $k_{F} = \sqrt{E_{F}+{\alpha^{2}}/{4B}-Ak^{2}_{z}}/\sqrt{B}$.
Additionally, the upper bound on
$k_{z}$ follows from the condition that  $k_{-}$ and hence $k_F$ must
be real,
leading to
 \begin{equation}\label{kzhight}
|k_{z}|\le\sqrt{k^{2}_{D}+\frac{\alpha^{2}_{R}}{4AB}}\equiv k_{m}\;.
 \end{equation}
 The helicity-resolved segments of the Fermi surface are shown in
 Fig.~\ref{fig1}(c)-(e). Outside the Dirac nodes
 ($k_{D}<|k_{z}|<k_{m}$), only the negative helicity states
 exist. Between the Dirac nodes however ($|k_{z}|<k_{D}$), states with
 oppositive helicities coexist in the inner and outer sections of the
 Fermi surface. The spin textures associated with the inner ($\mu=+$)
 and outer ($\mu=-$) portions of the Fermi surface are given by
 \begin{equation}\label{spntexturesf}
 \langle\sigma_{x}\rangle=\mu\sin\phi_{k}\;\; {\rm and}\;\;  \langle\sigma_{y}\rangle=-\mu\cos\phi_{k}\;.
 \end{equation}
 As displayed in Fig.~\ref{fig1}(d)-(e), the two helical branches of Fermi
 surface are characterized by opposite sense of rotations of the
 electron spins.

\section{Formalism of Interlayer exchange Interaction: Trivial Phase}~\label{sec2}
 \begin{figure}[t]
  \centering
  \includegraphics[width=\columnwidth]{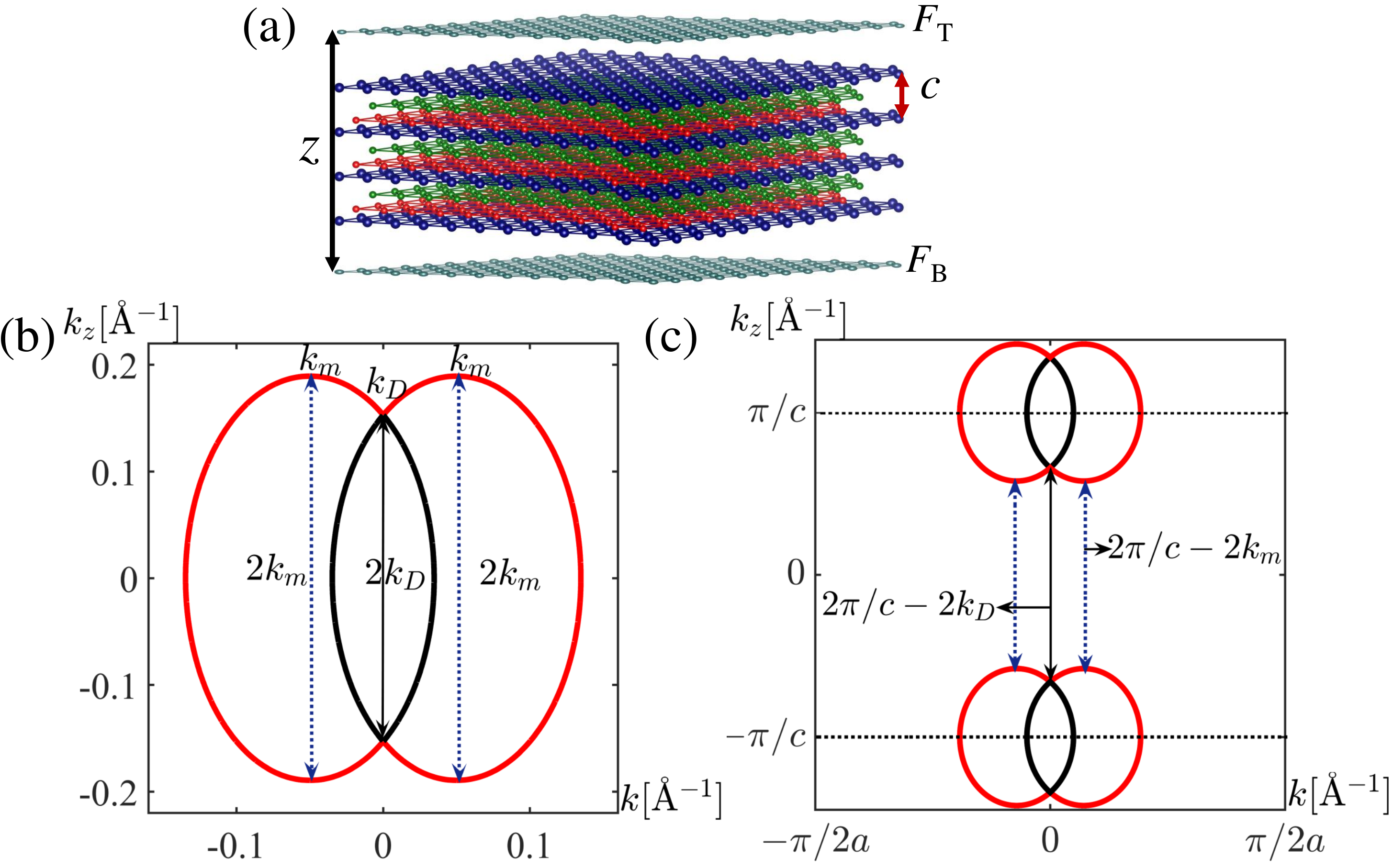}
  \caption{ (a) Atomic configuration of the
    Ferromagnet-BiTeI-Ferromagnet multilayered system. The top
    (bottom) ferromagnetic layer is indicated by $F_{\Top}$
    ($F_{\Bot}$), and the distance between $F_{\Top}$ and $F_{\Bot}$
    is $z=(N+1)c$, where $c$ is the thickness of a BiTeI single layer and $N$ is an integer. (b) BiTeI Fermi surface projection centered at the A point. The arrows indicate the critical spanning vectors determined by the Fermi function and the spin textures. (c) BiTeI Fermi surface projection defined in its periodic BZ. The critical spanning vectors resulting from the connection between the extrema of Fermi surfaces at $\pm\pi/c$ are equivalent to those in (b) and result in the same period of oscillation.}\label{fig2}
\end{figure}
We employ the RKKY formalism for interlayer exchange coupling in
Refs.~\cite{LayersRKKY1,LayersRKKY2}. Our system consists of two
ferromagnets ($F_{\Top}$ and $F_{\Bot}$) sandwiching BiTeI along its
stacking direction, as shown in Fig~\ref{fig2}(a).
The distance between the ferromagnets is $z=(N+1)c$, where $c$ thickness of a BiTeI unit cell and $N$ is an integer, see Fig.~\ref{fig1}(a).
The ferromagnetic layer adjacent to BiTeI is assumed to consist of classical spins $\bm S_{i}$
located at the atomic position $\bm R_{i}$ of the spacer material BiTeI.
The spins $\bm S_{i}$ of the ferromagnetic layer are coupled to the electron spins of the BiTeI via a contact potential at the interface,
$\mathcal{V}_{i}=J_{0}\delta(\bm r-\bm R_{i})\mathcal{S}\cdot\bm
S_{i}$,
where $J_{0}$ is the amplitude of the potential and $\mathcal{S}$ is the electron spin operator of the BiTeI
spacer. Within these considerations,
the interlayer coupling
can be expressed as follows
\begin{eqnarray}\label{internonsim}
I(z)=-\sum_{a,b=x,y,z}\frac{J^{2}_{0}S^{\Top}_{a}S^{\Bot}_{b}c}{2(2\pi)^{3}V_{0}}\int_{-{\pi}/{c}}^{{\pi}/{c}}dq_{z}e^{iq_z z}\nonumber\\ \times \int_{{\rm2DBZ}}d^{2}{\bqpar}\chi_{ab}(\bqpar,q_{z}) \sum_{\bm R\in  F_{{\rm T}}}e^{i\bqpar\cdot \bm R}\;,
\end{eqnarray}
where $V_{0}$ is the volume of the unit-cell,
$S^{(\Top,\Bot)}_{x,y,z}$ are the spin projections of the
top (T) and bottom (B) ferromagnetic layers, and $\chi_{ab}(\bqpar,q_{z})$ is the $ab$
component of the static spin susceptibility tensor. The
planar dimensions of the ferromagnetic layers satisfy periodic
boundary conditions
since they are assumed to be large compared to the interlayer
distance. The last sum in Eq.~\eqref{internonsim} is then nonzero only for $\bqpar=0$.
Recalling that the area of the projected 2D BZ for BiTeI is $(2\pi)^{3}c/(2\pi
V_{0})$ [where $(2\pi)^3/V_{0}$ is the volume of the 3D BZ],
the interlayer exchange coupling can be written as
\begin{eqnarray}\label{internonsim1}
I_{ab}(z)=-\frac{1}{2}\left(\frac{J_{0}}{V_{0}}\right)^2\frac{S^{\Top}_{a}S^{\Bot}_{b}c^{2}}{2\pi }\int_{-\pi/c}^{\pi/c}{dq_{z}}\nonumber\\ \times e^{iq_{z}z}\chi_{ab}({\bqpar}=0,q_{z}).
\end{eqnarray}
The components of spin susceptibility, $\chi_{ab}$, for BiTeI
consist of intraband and interband contributions. For simplicity, in the
sections running up to Sec.~\ref{sec5}
we omit the $\bqpar=0$ argument in $\chi_{ab}$ and denote
$\chi_{ab}(q_{z}) \equiv \chi_{ab}(\bqpar=0,q_{z})$. Since the
bands of BiTeI are characterized by their helicity, the spin
susceptibility can be written as
$\chi_{ab}(q_{z})=\sum_{\mu,\nu=\pm}\chi^{\mu\nu}_{ab}(q_{z})$ with
\begin{eqnarray}\label{chispin}
\chi^{\mu\nu}_{ab}(q_{z}) &=& \frac{-\mu^{2}_{B}}{(2\pi)^{3}}\int_{-\pi/c}^{\pi/c}dk_{z} \int_{{\rm2DBZ}}{ d^{2}\bkpar }\nonumber\\&&\times\frac{f(E_{\bk,\mu})-f(E_{\bk+q_{z},\nu})}{E_{\bk,\mu}-E_{\bk+\bqz,\nu}+i\delta}\mathcal{F}^{\mu\nu}_{ab}(\bk,\bk+\bqz),\nonumber\\
\end{eqnarray}
where $\mu,\nu=\pm$, $f(E_{\bk,\mu})$ is the Fermi function,
$\mu_{B}$ is the Bohr magneton, $\delta$ is a positive infinitesimal,
and $\mathcal{F}^{\mu\nu}_{ab}(\bk,\bk+\bq)=\langle
\mu,\bk |\sigma_{a}|\bk+\bq, \nu\rangle \langle
\nu,\bk+\bq|\sigma_{b}|\bk, \mu\rangle$ is a form factor (see
Appendix~\ref{appa}). In this work, we consider low temperatures
$k_{\mathrm{B}}T \ll E_F$ and take $T = 0$ in Eq.~(\ref{chispin}).

The oscillatory nature of interlayer RKKY interaction is due to the spatial
oscillations of the induced spin density by the localized moments of
the ferromagnetic layers and shares the same physical origin as the
Kohn anomaly~\cite{kohnanomaly}.
It stems from the sharpness of the Fermi surface
at zero temperature
and is measured by the critical spanning vectors
(also called calipers). A critical spanning vector is one that
connects a pair of extremum points of the Fermi surface along
$k_z$. These critical spanning vectors determine the
periods of oscillations as a function of the spacer thickness. Because
the thinnest possible spacer is one with a single unit cell, the
smallest observable period is twice the unit cell thickness
$2c$ corresponding to a critical spanning vector of $\pi/c$. If the
Fermi surface yields a critical spanning vector that is larger than $\pi/c$, a
period that is longer than that given directly by the critical
spanning vector will be observed instead because the latter cannot be
sampled. This effect is known as aliasing and was observed in the
interlayer exchange coupling mediated by noble metals, \textit{e.g.},
Fe/Cu, Fe/Cr and Co/Cu~\cite{exp1,exp2}.

In BiTeI, the largest critical spanning vector defined by its Fermi
surface is $2k_{m}$ [see Eq.~\eqref{kzhight}]
corresponding to the period $\pi/k_{m}$, as shown in
Fig.~\ref{fig2}(b).
The effect of aliasing would manifest in the case
when the period resulting from the largest spanning vector is shorter than $2c$.
Within our low-energy model for BiTeI, the Fermi energy is bounded
from above by the cutoff energy $E_c = 0.2$ eV, and the corresponding
$k_{m}$ is small such that $\pi/k_{m}$  exceeds $2c$.
Therefore, aliasing does not occur for the critical spanning vectors $2k_m,
2k_D$ calipering the Fermi surface from the \textit{inside}.
As shown in Fig.~\ref{fig2}(c), the other two critical spanning vectors $2\pi/c-2k_m,
2\pi/c-2k_D$ calipering the Fermi surface from the \textit{outside}
are equivalent to $2k_m,2k_D$, respectively.
Hence, the limits of integration in Eq.~\eqref{chispin} are only
determined by the boundaries of the two helical Fermi surfaces through
the Fermi functions. To make this explicit,
it is useful to define a function
$g^{\mu\nu}(k_{z})$ that captures the $k_{z}$ dependence of the
helicity content of the bands, with $g^{\mu\nu}(k_{z})=\Theta(k_{D}-k_{z})\Theta(k_{D}+k_{z})$
for $\mu=-\nu$ and $\mu=\nu=+$ and $g^{\mu\nu}=1$ for $\mu=\nu=-$,
where $\Theta$ is the Heaviside step function. We can then write
Eq.~\eqref{chispin} as an integral over the entire momentum space
constrained by $g^{\mu\nu}(k_{z})$,
\begin{eqnarray}\label{chispin2}
\chi^{\mu\nu}_{ab}(q_{z}) &=& \frac{-\mu^{2}_{B}}{(2\pi)^{3}}\int dk_{z}\,g^{\mu\nu}(k_{z})\int{ d^{2}\bkpar }\nonumber\\ &&\times\frac{f(E_{\bk,\mu})-f(E_{\bk+\bqz,\nu})}{E_{\bk,\mu}-E_{\bk+\bqz,\nu}+i\delta}\mathcal{F}^{\mu\nu}_{ab}(\bk,\bk+\bqz),\nonumber \\
\end{eqnarray}
The form factor $\mathcal{F}^{\mu\nu}_{ab}$ is independent of the
momentum along $z$ and is given by
\begin{eqnarray}
 && \mathcal{F}^{\mu\nu}_{zz} =\frac{1-\mu\nu}{2},\;\mathcal{F}^{\mu\nu}_{xx} =\frac{1-\mu\nu\cos(2\phi_{k})}{2} \;, \nonumber \\
  &&\mathcal{F}^{\mu\nu}_{yy} =\frac{1+\mu\nu\cos(2\phi_{k})}{2} ,\;\mathcal{F}^{\mu\nu}_{xy} =\mathcal{F}^{\mu\nu}_{yx}=-\frac{\sin(2\phi_{k})}{2}\;,\nonumber\\
  &&\mathcal{F}^{\mu\nu}_{zx} =-\mathcal{F}^{\mu\nu}_{xz}=\frac{\mu-\nu}{2i}\cos(\phi_{k}),\nonumber\\
    &&\mathcal{F}^{\mu\nu}_{zy} =-\mathcal{F}^{\mu\nu}_{yz}=\frac{\mu-\nu}{2i}\sin(\phi_{k})\;.\label{formfactorsxyz}
\end{eqnarray}
Upon angular integration, Eq.~\eqref{chispin2} gives
\begin{widetext}
\begin{equation}\label{chispin22}
\chi_{ab}(q_{z})=C\begin{dcases}
                    2\int_{-k_{D} }^{k_{D} }dk_{z}\sum_{\mu}\mathcal{P}\int_{0}^{k_{\mu}}\left(\frac{1}{E_{\bk+\bqz,-\mu,}-E_{\bk,\mu}}-\frac{1}{E_{\bk,\mu}-E_{\bk-\bqz,-\mu}}\right)\kpar d\kpar, & \mbox{if } a=b=z \\
                       \frac{\chi_{zz}(q_{z})}{2C}+\int_{-k_{m}}^{k_{m}}dk_{z}\mathcal{P}\int_{0}^{k_{-}}\left(\frac{1}{E_{\bk+\bqz,-}-E_{\bk,-}}-\frac{1}{E_{\bk,-}-E_{\bk-\bqz,-}}\right)\kpar d\kpar\\ +\int_{-k_{D}}^{k_{D}}dk_{z}\mathcal{P}\int_{0}^{k_{+}}\left(\frac{1}{E_{\bk+\bqz,+}-E_{\bk,+}}-\frac{1}{E_{\bk,+}-E_{\bk-\bqz,+}}\right)\kpar d\kpar , & \mbox{if } a=b=(x,y) \\
                       0, & \mbox{if } a\ne b,
                     \end{dcases}
\end{equation}
\end{widetext}
where $C=\pi \mu^{2}_{B}/(2\pi)^{3}$, $k_{\mu}$ is given in Eq.~\eqref{cases}, $k_{D}=\sqrt{E_{F}/A}$ and $k_{m}$ is given in Eq.~\eqref{kzhight} (see Appendix~\ref{app}).

The Rashba SOC in BiTeI is 2D in nature and only couples to in-plane
momentum. The spin susceptibility tensor is therefore anisotropic with
diagonal components $\chi_{xx}(q_{z}) = \chi_{yy}(q_{z}) \ne
\chi_{zz}(q_{z})$ and  vanishing off-diagonal components $\chi_{ab}(q_{z})$ ($a \ne b$).
For ferromagnets with spins normal to the interface, the spin
susceptibility only has contributions from interband transitions, limiting these contributions to the part of the Fermi surface
that hosts both helicities. When the ferromagnets have spins parallel
to interface, the spin susceptibility has contributions from  both
intraband and interband processes and all regions of the Fermi surface
become relevant. RKKY mechanism does not contribute to interlayer magnetic exchange when the spins of the ferromagnets
are orthogonal to each other due to the vanishing off-diagonal
components of $\chi_{ab}(q_{z})$.
Consequently, the interlayer exchange coupling between the ferromagnetic layers is given by
\begin{eqnarray}\label{internonsim2}
I(z)=-\sum_{a=x,y,z}\frac{S^{\Top}_{a}S^{\Bot}_{a}}{2}\left(\frac{J_{0}}{V_{0}}\right)^2\frac{c^{2}}{2\pi }\int_{-\infty}^{\infty}{dq_{z}}e^{iq_{z}z}\chi_{aa}(q_{z}).\nonumber\\
\end{eqnarray}
We note that the integration over $q_{z}$ in the above can be extended to $\pm
\infty$, since all critical spanning vectors are much smaller than
$\pi/c$ within the range of Fermi energy $E_{F}<0.2$ eV considered in
the low-energy effective theory for BiTeI.

Having laid out the formalism for the interlayer exchange coupling
and evaluated the spin susceptibility, we first consider the simpler case
without SOC in order to establish a reference scenario to which the
SOC effects from a BiTeI spacer (Sec.~\ref{S45}) will be compared.

\section{Special Case: Spin-Degenerate Metal}\label{sec3}
\begin{figure}[t]
    \begin{center}
            \subfigure{
            \includegraphics[width=0.95\columnwidth]{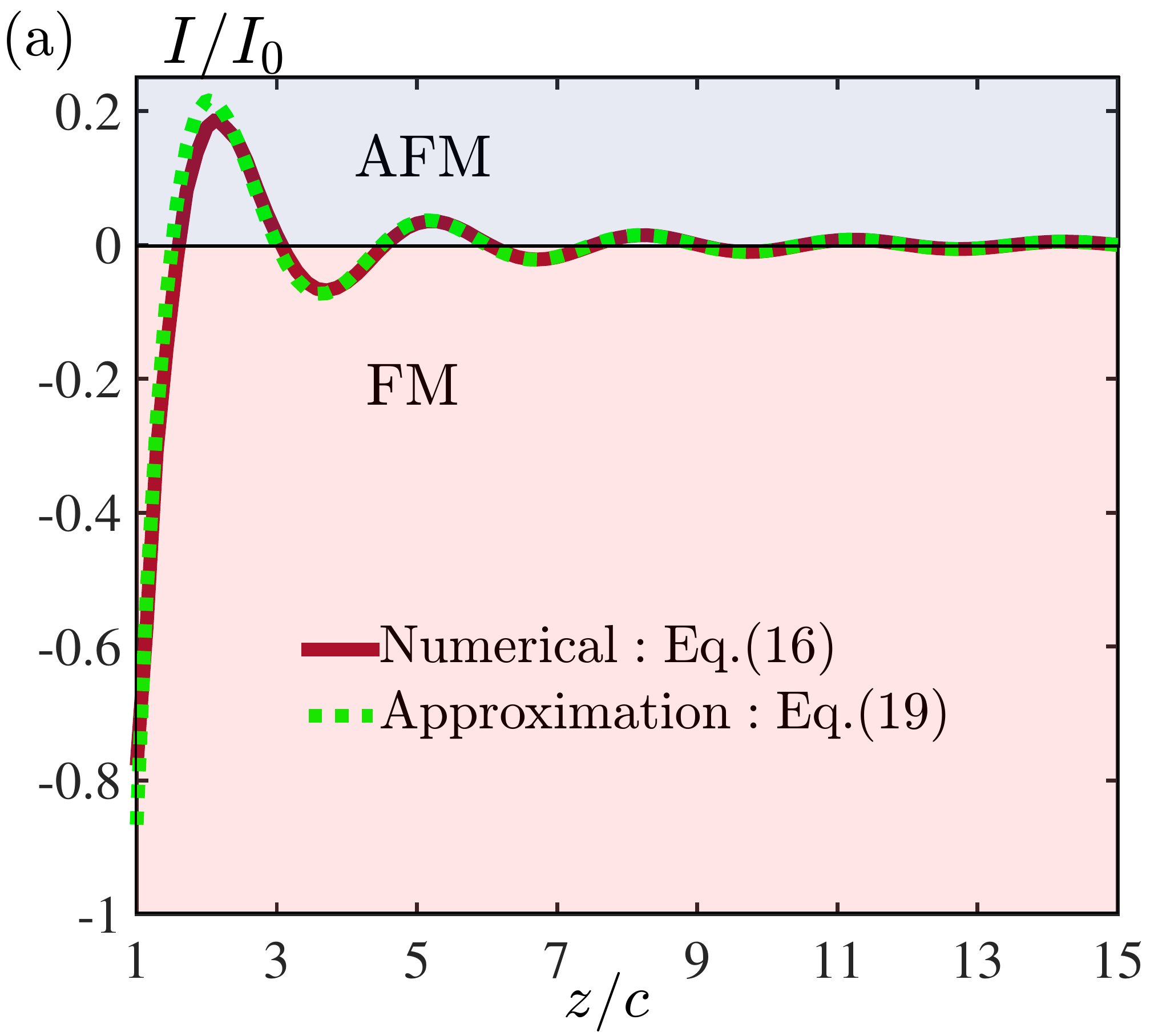}}
        \subfigure{
            \includegraphics[width=0.95\columnwidth]{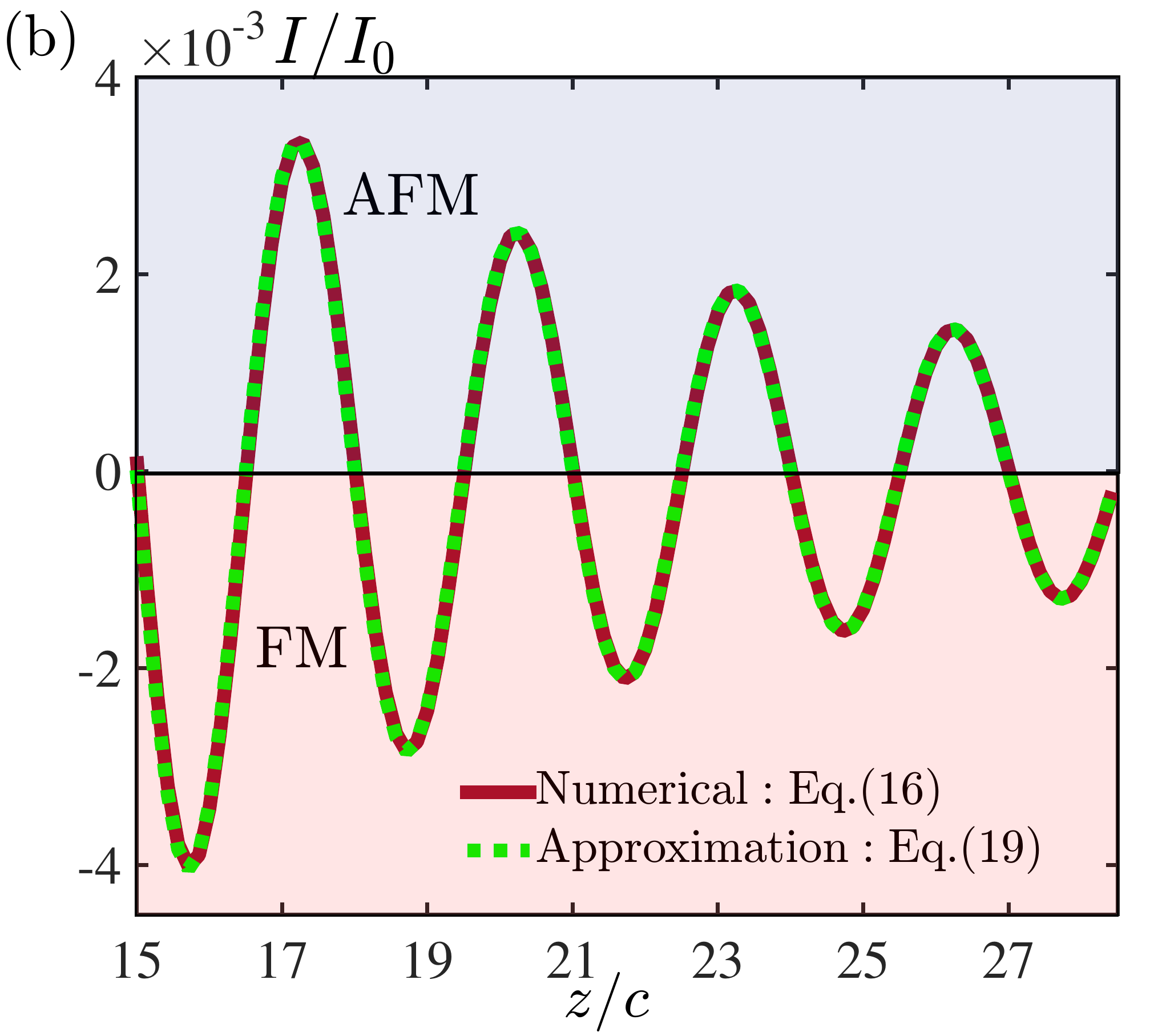}}
                \end{center}
    \caption{Exchange coupling between two ferromagnets with parallel
      spins mediated by a spin-degenerate spacer metal as a function of
      the ferromagnets' separation $z$. (a) Shows the $z$ dependence
      of $I(z)$ for thin spacers. (b) Shows the behaviour of $I(z)$ as
      function of $z$ for thick spacers. In both (a) and (b) $I(z)$
      oscillates with a period $\pi/k_{D}$ and this coupling
      transitions from ferromagnetic (FM) to anti-ferromagnetic (AFM)
      while decaying as $z^{-2}$. The Fermi energy for both (a) and (b) is $E_{F}=0.18$ eV.}
\label{fig3}
\end{figure}
\begin{figure}[t]
    \begin{center}
            \subfigure{
            \includegraphics[width=0.95\columnwidth]{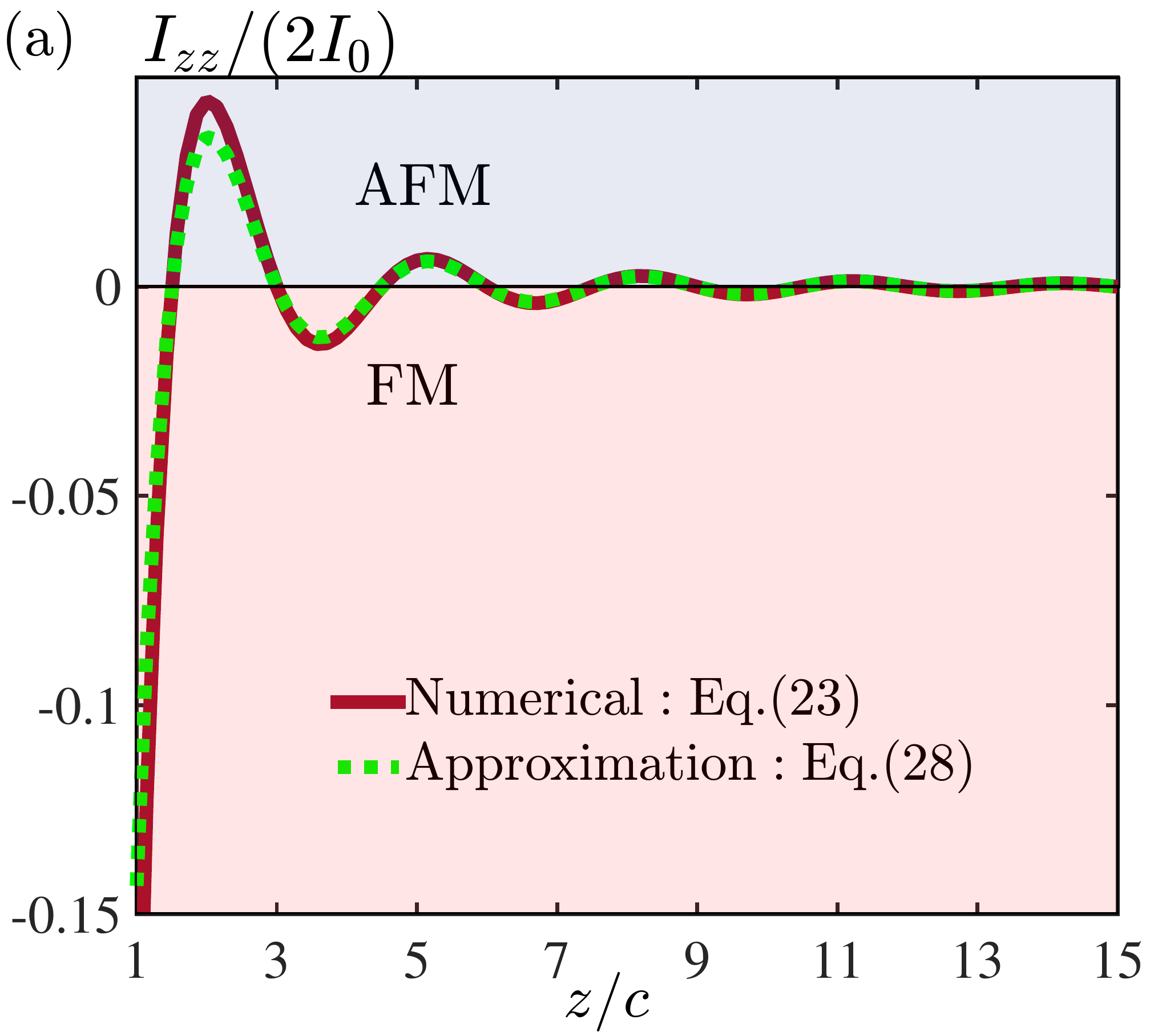}}
        \subfigure{
            \includegraphics[width=0.95\columnwidth ]{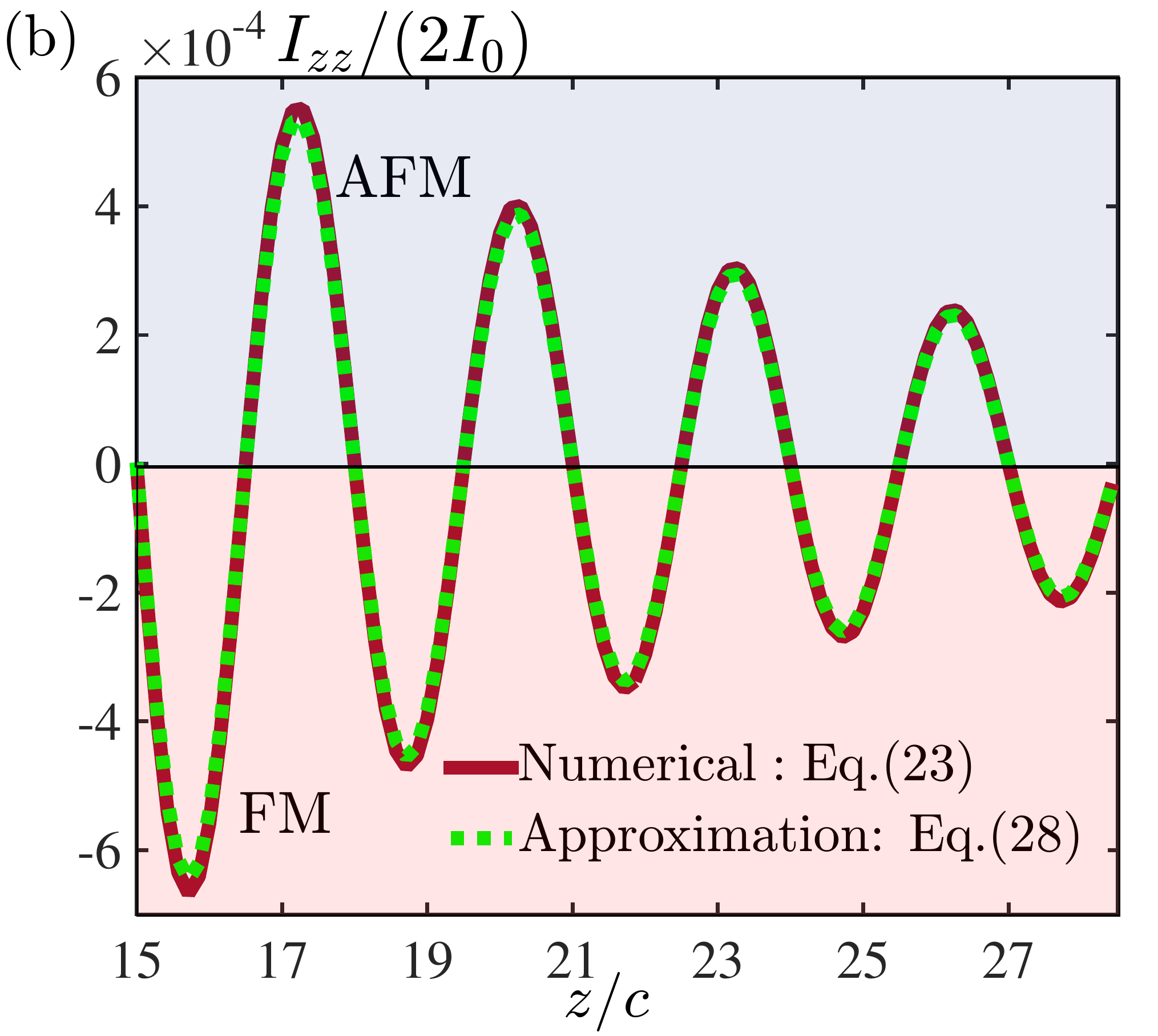}}
                \end{center}
    \caption{Thickness dependence of the BiTeI-mediated interlayer
      magnetic exchange between two $z$-polarized ferromagnets (Fig.~\ref{fig2}) for $E_{F}=0.18$ eV. (a) Shows the
      interlayer exchange coupling for relatively thin samples. (b) Shows the
      long-range behaviour of the interlayer exchange coupling, \textit{i.e.}
      relatively thick samples. In both (a) and (b) $I_{zz}(z)$
      oscillates with a period $\pi/k_{D}$ and this coupling
      transitions from ferromagnetic (FM) to anti-ferromagnetic (AFM) while decaying as $z^{-2}$. [Panels (a) and (b) also represent the interband contributions to $I_{xx}(z)$ in Sec.~\ref{sec5}].}
\label{fig4}
\end{figure}
\begin{figure}[t]
            \includegraphics[width=0.95\columnwidth ]{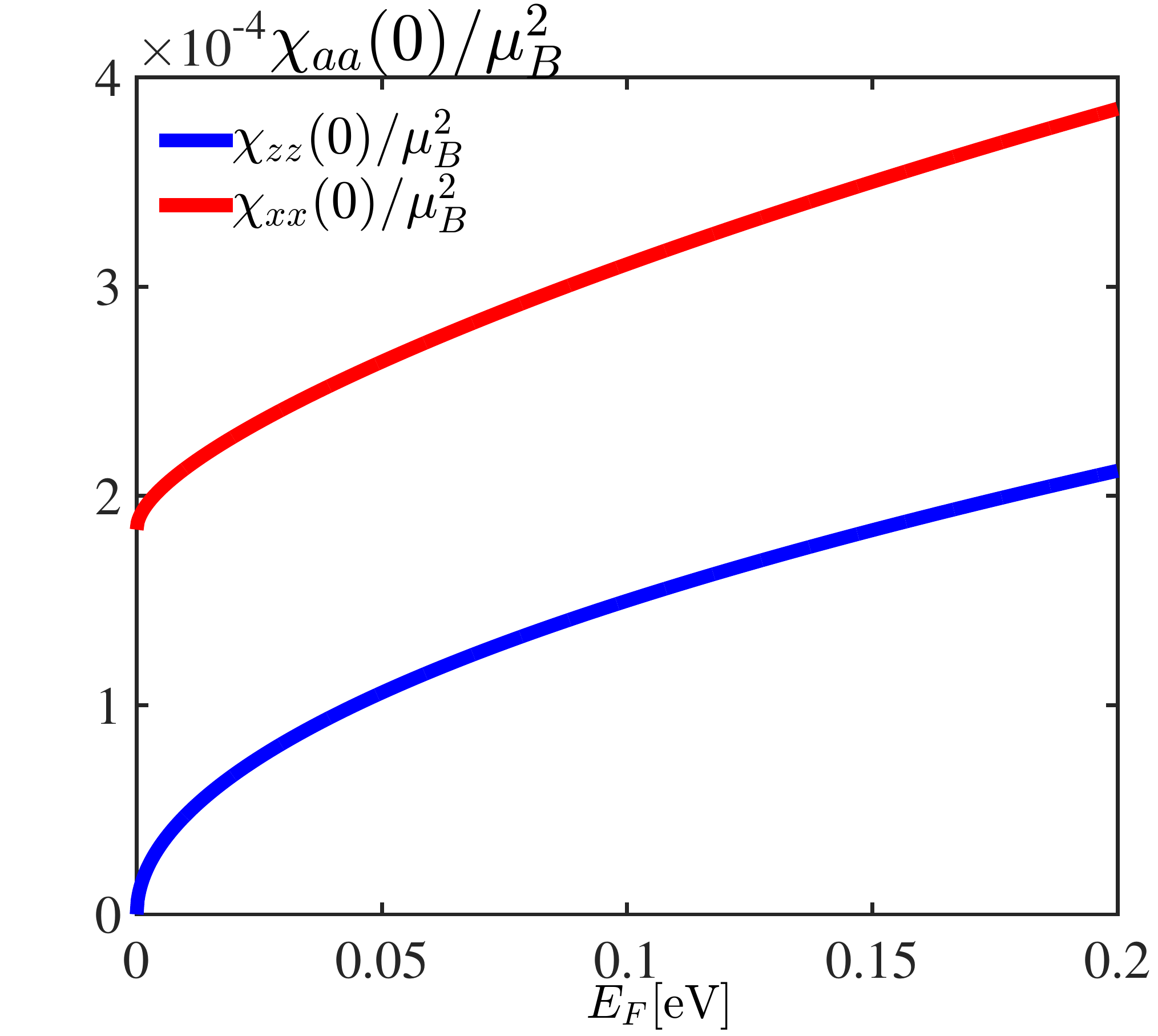}
    \caption{Long-wavelength behaviour of the spin susceptibility components $\chi_{zz}(0)$ and $\chi_{xx}(0)$ ($\chi_{xx}=\chi_{yy}$) as a function of the Fermi energy. One can notice that $\chi_{xx}(0)>\chi_{zz}(0)$, and that $\chi_{xx}(0)$ is non-vanishing for $E_{F}=0$. }
\label{pauli}
\end{figure}
\begin{figure}[th!]
    \begin{center}
            \subfigure{
            \includegraphics[width=0.95\columnwidth ]{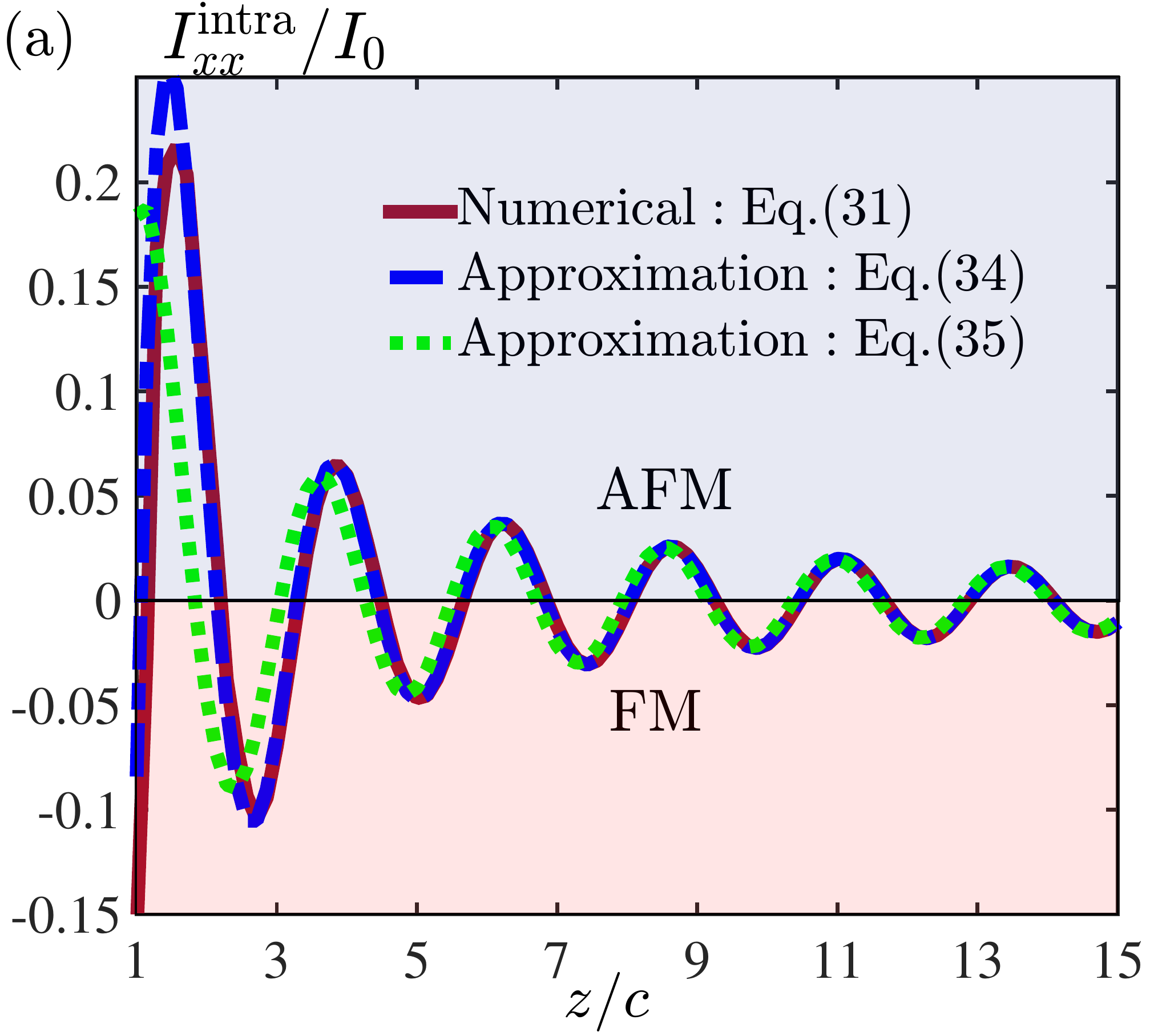}}
        \subfigure{
            \includegraphics[width=0.95\columnwidth]{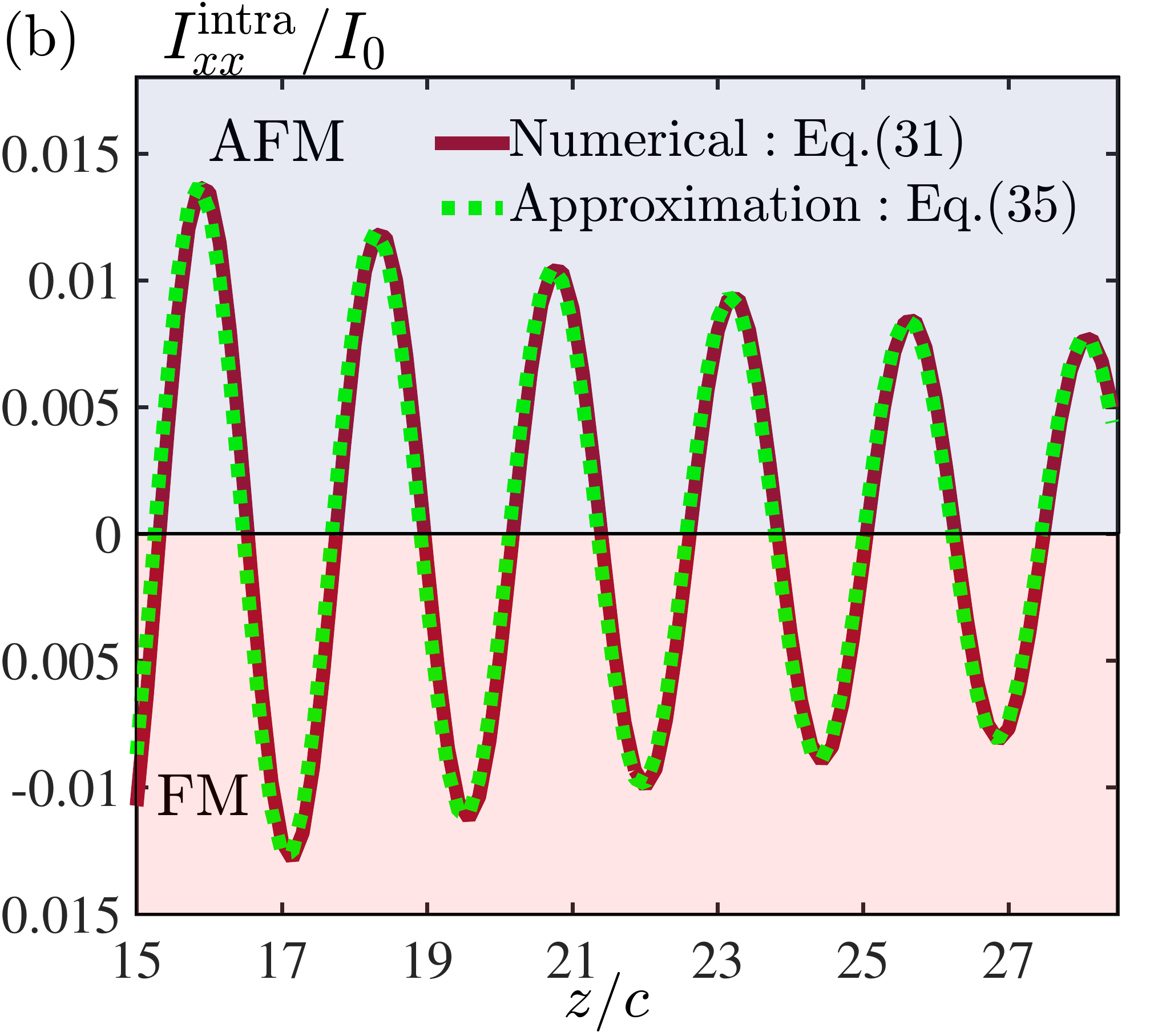}}
                \end{center}
    \caption{Intraband contributions to the interlayer exchange coupling
      $I_{xx}(z)$ for $E_{F}=0.18$ eV. Panel (a) shows the dependence
      for thin BiTeI films. In this case these contributions oscillate
      with a period $\pi/k_m$, the coupling transitions from FM to
      AFM, and both powers $z^{-2}$ and $z^{-1}$ are necessary to describe its decay.
        (b) For relatively thick BiTeI samples, the intraband contributions are dominated by a single decay power of $z^{-1}$, while oscillating with the same period as in (a).}
\label{fig5}
\end{figure}
In this section we
consider a 3D spin-degenerate metal with an anisotropic energy dispersion
described by the Hamiltonian in Eq.~\eqref{effective1} with
$\alpha=0$. The components of the spin
susceptibility tensor become equal
and Eq.~\eqref{chispin22} becomes
\begin{equation}\label{freechi}
\chi(q_{z})= \frac{2\mu^{2}_{B}}{(2\pi)^{3}}\int dk_{z}\mathcal{P}\int d^{2}\bkpar\frac{f(E_{\bk})-f(E_{\bk+\bqz})}{E_{\bk+\bqz}-E_{\bk}},
\end{equation}
where the factor of $2$ results from spin degeneracy.
The integration over $\kpar$ and $k_{z}$ leads to
 \begin{equation}\label{freechi3}
 \chi(q_{z})= \frac{\mu^{2}_{B}}{(2\pi)^{2}B}\begin{dcases}
 k_{D}-\frac{q^{2}_{z}-4k^{2}_{D}}{4q_{z}}\ln\left|\frac{q_{z}+2k_{D}}{q_{z}-2k_{D}}\right|& \\\;\; \qquad\qquad\qquad\qquad\mbox{for } q_{z}\ne0\\
 2k_{D}\;\;\;\qquad\qquad\qquad \mbox{for } q_{z}=0,
 \end{dcases}
 \end{equation}
with $k_{D}=\sqrt{E_{F}/A}$.
The expression above yields a critical spanning vector $|q_{z}|=2k_{D}$ at which
 $\partial\chi(q_{z})/\partial q_{z}$ has a logarithmic
singularity. This singularity indicates the presence of a Kohn
 anomaly which gives rise to spatial oscillations of induced spin
 densities and consequently the RKKY mediated interlayer
 exchange~\cite{kohnanomaly}.
In the limit of $q_{z}=0$ and a spherical Fermi surface with
$A=B=\hbar^{2}/2m$ and $E_{F}=\hbar^{2}k_{F}^{2}/2m$,
Eq.~\eqref{freechi3} recovers the Pauli susceptibility of a
non-interacting 3D Fermi gas $\chi(0)=
3n\mu^{2}_{B}/(2E_{F})\equiv\mu^{2}_{B}\mathcal{D}(E_{F})$, where $n=k^{3}_{F}/(3\pi^2)$ is the number of electrons per unit volume and $\mathcal{D}(E_{F})$ is the density of states~\cite{RKKY1}.

 We now evaluate the interlayer exchange
 coupling Eq.~\eqref{internonsim2} between the ferromagnetic
 layers.  Since the system is spin-degenerate and $\chi(q_{z})$ is an
 even function of $q_{z}$, we can write Eq.~\eqref{internonsim2}
for ferromagnetic layers with parallel spins as
\begin{equation}\label{Cuplingz2}
I(z)=-\frac{1}{2}\left(\frac{J_{0}}{V^{2}_{0}}\right)^2S^{\Top}_{a}S^{\Bot}_{a}\frac{c^{2}}{2\pi}\int_{0}^{\infty}{dq_{z}}\cos(q_{z}z)\chi(q_{z})\;.
\end{equation}
The dominant contribution to Eq.~\eqref{Cuplingz2} can be extracted
analytically by expanding the integrand near the Kohn anomaly $q_z \in
[2k_{D}-\epsilon,2k_{D}+\epsilon]$  (where $\epsilon \ll 2k_{D}$)
\begin{eqnarray}\label{khon1}
I(z)\approx-&&\frac{S^{\Top}_{a}S^{\Bot}_{a}}{(2\pi)^{3}B}\left(\frac{\mu_{B}J_{0}c}{V_{0}}\right)^{2}\int_{2k_{D}-\epsilon}^{2k_{D}+\epsilon}dq_{z}\cos(q_{z}z)\nonumber\\&&\times\left[k_{D}+\frac{(q_{z}-2k_{D})}{2}\ln|q_{z}-2k_{D}|\right]\;.
\end{eqnarray}
Changing variable $q{'}=q_{z}-2k_{D}$ and integrating by parts twice
lead to
\begin{eqnarray}\label{khon3}
I(z)\approx-\frac{S^{\Top}_{a}S^{\Bot}_{a}}{2(2\pi)^{3}B}&&\left(\frac{\mu_{B}J_{0}}{V_{0}}\right)^{2}\left(\frac{c}{z}\right)^{2}\sin(2k_{D}z)\nonumber\\&&\times\int_{-\epsilon}^{\epsilon}dq{'}\left[\frac{\sin(q{'}z)}{q{'}}\right]\;.
\end{eqnarray}
Since the main contribution of the integral above comes
from the vicinity of $q'=0$, one can extend the limits of this integral to
$\pm\infty$, arriving at
\begin{equation}\label{khon4}
\frac{I}{I_{0}}\approx-\left(\frac{c}{z}\right)^{2}\sin(2k_{D}z)\;,
\end{equation}
where
\begin{equation}\label{I0}
I_{0}=\frac{S^{\Top}_{a}S^{\Bot}_{a}}{16\pi^{2}B}\left(\frac{\mu_{B}J_{0}}{V_{0}}\right)^{2}\;.
\end{equation}
The analytic result Eq.~\eqref{khon4} indicates that the interlayer
exchange interaction has a single period of
oscillation determined by the critical spanning vector $2k_{D}$,
given by $\pi/k_{D}$, and the envelope of the oscillations decay as
$z^{-2}$.
In Figs.~\ref{fig3}(a) and (b), we plot and compare the interlayer exchange
coupling obtained from direct numerical evaluation
of Eq.~\eqref{Cuplingz2} and from Eq.~\eqref{khon4}. The excellent
agreement between the two results confirms that the dominant contribution to the
interlayer exchange interaction indeed originates from the critical spanning
vector given by the Kohn anomaly of the system, as captured by our
approximate analytic result Eq.~\eqref{khon4}.
In the isotropic limit  of a spherical Fermi surface with $A=B=\hbar^{2}/2m$ and
$E_{F}=\hbar^{2}k_{F}^{2}/2m$, one recovers the well-known result of
interlayer exchange interaction obtained by Yafet ~\cite{Yafet},
 \begin{equation}\label{khon5}
I(z)\approx-\frac{S^{\Top}_{a}S^{\Bot}_{a}m}{8\pi^{2}\hbar^{2}}\left(\frac{\mu_{B}J_{0}}{V_{0}}\right)^{2}\left(\frac{c}{z}\right)^{2}\sin(2k_{F}z)\;.
\end{equation}

In this section we have quantified the interlayer exchange coupling between two
ferromagnetic layers mediated by a spacer without SOC. In the next two
sections, we restore SOC effects and consider BiTeI as the spacer. We
will study the interlayer exchange coupling in
FM/B\lowercase{i}T\lowercase{e}I/FM for the two cases when the
ferromagnets' spins are aligned perpendicular to the plane and
parallel to the plane.

\section{Interlayer Exchange Interaction Mediated by
  B\lowercase{i}T\lowercase{e}I} \label{S45}
\subsection{Out-of-Plane Magnetization}\label{sec4}
When the spins of the ferromagnetic layers shown in
Fig.~\ref{fig2}(a) are in the $z$-direction, the exchange interaction
between $F_{\Top}$ and $F_{\Bot}$ is mediated by BiTeI electrons that are
spin polarized out of the plane and therefore depends on the
$\chi_{zz}(q_{z})$ component of the spin susceptibility.
Since the spin textures of the electronic states in BiTeI are helical
and have no out-of-plane components, only interband processes
contribute to $\chi_{zz}(q_{z})$
and its form factor $\mathcal{F}^{\mu\nu}_{zz} =(1-\mu\nu)/2$ vanishes
for $\mu=\nu$. Since interband transitions require a change in
helicity, the region of the Fermi surface that contributes to
$\chi_{zz}(q_{z})$ is limited to $|k_{z}|<k_{D}$,
leading to the form of $\chi_{zz}(q_{z})$ in Eq.~\eqref{chispin22}. Integrating $\chi_{zz}(q_{z})$ over $\kpar$ and $k_{z}$ leads to
\begin{widetext}
\begin{eqnarray}\label{chizzfinal}
  &&\frac{\chi_{zz}(q_{z})}{2C} = \frac{2k_{D}}{B}+\frac{Aq_{z}}{8} \left[\left(\frac{4k^{2}_{D}+q^{2}_{z}}{\alpha^{2}}-\frac{q^{2}_{z}-4k^{2}_{D}}{ABq^{2}_{z}+\alpha^{2}}\right) \ln\left|\frac{2k_{D}+q_{z}}{2k_{D}-q_{z}}\right|+\left(\frac{4k^{2}_{D}+q^{2}_{z}}{\alpha^{2}}+\frac{q^{2}_{z}-4k^{2}_{D}}{ABq^{2}_{z}+\alpha^{2}}\right)\right.\nonumber\\ &&\left.\times \ln\left|\frac{q_{z}(2k_{D}-q_{z})-2\alpha^{2}/(AB)}{q_{z}(2k_{D}+q_{z})+2\alpha^{2}/(AB)}\right|\right]+\frac{Ak_{D}q^{2}_{z}}{2\alpha^{2}}\ln\left|\frac{q_{z}^{2}(q^{2}_{z}-4k^{2}_{D})}{[q_{z}(2k_{D}+q_{z})+2\alpha^{2}/(AB)][q_{z}(2k_{D}-q_{z})-2\alpha^{2}/(AB)]}\right|\;,
\end{eqnarray}
\end{widetext}
where $C=\mu^{2}_{B}\pi/(2\pi)^{3}$. The long-wavelength limit of the
spin susceptibility due to $z$-polarized ferromagnets is
$\chi_{zz}(0)=4Ck_{D}/B$. This component of the spin susceptibility is
known as the van Vleck susceptibility, originating from virtual
interband transitions and in this case is identical to the case
without SOC [Eq.~\eqref{freechi3}]~\cite{vanvleck1,vanvleck2,vanvleck3,spinresp,vvnew}.
Since the spin susceptibility $\chi_{zz}(q_{z})$ is an even function of $q_z$,
we can express the interlayer exchange coupling as
\begin{equation}\label{CuplingRashZZ}
I_{zz}(z)=-\left(\frac{J_{0}}{V_{0}}\right)^2S^{\Top}_{z}S^{\Bot}_{z}\frac{c^{2}}{2\pi}\int_{0}^{\infty}{dq_{z}\cos(q_{z}z)\chi_{zz}(q_{z})}\;.
\end{equation}

The spin susceptibility in Eq.~\eqref{chizzfinal} reveals the presence
of three critical spanning vectors at $|q_{z}|=2k_{D}$, and
$|q_{z}|=k_{D}\pm\sqrt{k^{2}_{D}-2\alpha^{2}/(AB)}=k^{\pm}_{n}$. When
$|q_{z}|$ takes the value of a critical spanning vector, $\partial
\chi_{zz}(q_{z})/\partial{q_{z}}$ has a logarithmic divergence,
\textit{i.e.} Kohn anomaly~\cite{kohnanomaly}. The contributions of
the Kohn anomalies to interlayer exchange interaction depend on the Fermi energy
of the system. The anomalies associated with the vectors $k^{\pm}_{n}$
contribute to the interlayer exchange interaction only when $E_{F}>2\alpha^{2}/B$,
while $2k_{D}$ contributes for any value of the Fermi energy. We
numerically evaluate the integral in Eq.~\eqref{CuplingRashZZ}, and
analytically determine the dominant behavior of $I_{zz}(z)$ by considering the
contributions $I^{(1)}_{zz}(z)$ and $I^{(2,3)}_{zz}(z)$ of the
Kohn anomalies associated with $2k_{D}$ and $k^{\pm}_{n}$, respectively. The latter is done by integration over a small interval enclosing these anomalies.

In the vicinity of  $q_{z}\approx 2k_{D}$, the interlayer exchange interaction is
\begin{eqnarray}\label{khonR1}
I^{(1)}_{zz}(z)\approx-\frac{S^{\Top}_{z}S^{\Bot}_{z}}{(2\pi)^{3}}\left(\frac{\mu_{B}J_{0}c}{V_{0}}\right)^{2}\int_{2k_{D}-\epsilon}^{2k_{D}+\epsilon}dq_{z}\bigg[\mathcal{C}+\nonumber\\ \left.\frac{Ak^{2}_{D}(q_z-2k_{D})}{4ABk^{2}_{D}+\alpha^{2}}\ln|q_z-2k_{D}|\right]\cos(q_{z}z)\;.\nonumber\\
\end{eqnarray}
Here $\epsilon \ll 2k_{D}$, and $\mathcal{C}$ is a constant. Following
similar steps that lead to Eq.~\eqref{khon4},
we obtain the contribution of $2k_{D}$ to the
interlayer exchange interaction
\begin{equation}\label{khonR4}
I^{(1)}_{zz}(z)\approx-2I_{0}\left(\frac{Ak^{2}_{D}}{4ABk^{2}_{D}+\alpha^{2}}\right)\left(\frac{c}{z}\right)^{2}\sin(2k_{D}z)\;,
\end{equation}
where $I_{0}$ is given in Eq.~\eqref{I0}. In the vicinity of
$k^{\pm}_{n}$ the interlayer exchange coupling $I^{(2,3)}_{zz}(z)$, where the
superscript $2$ $(3)$ denotes the contribution of $k^{+}_{n}$
($k^{-}_{n}$), takes the form
\begin{eqnarray}\label{khonR12}
I^{(2,3)}_{zz}(z)\approx-\frac{S^{\Top}_{z}S^{\Bot}_{z}}{(2\pi)^{3}}\left(\frac{\mu_{B}J_{0}c}{V_{0}}\right)^{2}\int_{k^{\pm}_{n}-\epsilon}^{k^{\pm}_{n}+\epsilon}dq_{z}\bigg[\mathcal{C}_{\pm} +\nonumber\\ \left.\frac{k^{\pm}_{n}-k_{D}}{B(2k_{D}+k^{\pm}_{n})}\ln|q_{z}-k^{\pm}_{n}|\right]\cos(q_{z}z)\;,\nonumber\\
\end{eqnarray}
where $\mathcal{C}_{\pm}$ is a constant.
Following a similar procedure
as in the case of $k_{D}$ we obtain the contributions of $k^{\pm}_{n}$ to
the interlayer exchange interaction
 \begin{equation}\label{khon22}
I^{(2,3)}_{zz}(z)\approx-2I_{0}\left[\frac{k^{\pm}_{n}-k_{D}}{B(2k_{D}+k^{\pm}_{n})}\right]\left(\frac{c}{z}\right)^{2}\sin(k^{\pm}_{n}z)\;.
\end{equation}
Hence, the total exchange interaction
$I_{zz}(z)=\sum_{i=1}^{3}I^{(i)}_{zz}(z)$ is given by
\begin{eqnarray}
 &&I_{zz}(z)\approx-2I_{0}\left[\left(\frac{Ak^{2}_{D}}{4ABk^{2}_{D}+\alpha^{2}}\right)\left(\frac{c}{z}\right)^{2}\sin(2k_{D}z)+\right. \nonumber \\ \label{interbandzz}
&&\left.\Theta\left(E_{F}-\frac{2\alpha^{2}}{B}\right)\sum_{j=\pm} \frac{k^{j}_{n}-k_{D}}{B(2k_{D}+k^{j}_{n})}\left(\frac{c}{z}\right)^{2}\sin(k^{j}_{n}z)\right]\;.\nonumber\\
\end{eqnarray}

Within our low-energy theory for BiTeI, the maximum Fermi energy given
by the energy cutoff of $0.2$ eV is smaller than the value of
$2\alpha^{2}/B =0.74$ eV,
and thus the second term in Eq.~\eqref{interbandzz} does not contribute.
Hence, as shown in Figs.~\ref{fig4}(a) and (b), the BiTeI-mediated exchange
between two $z$-polarized magnetic layers only has a single period of oscillation determined by $\pi/k_{D}$. In addition to the oscillatory behavior of the interlayer exchange coupling we notice that $I_{zz}(z)$ decays as $z^{-2}$. This behavior is reminiscent of conventional 3D metallic spacers. Moreover, in the limit of $\alpha=0$, $k^{+}_{n}=2k_{D}$ and $k^{-}_{n}=0$, we recover the interlayer exchange coupling in the absence of SOC in Eq.~\eqref{khon4}.

In this section we have seen that the interlayer exchange coupling
between two $z$-polarized ferromagnets mediated by BiTeI displays a
behavior similar to that of a conventional 3D metal and
that the SOC coupling only renormalizes the amplitude of this coupling
by a factor $2Ak^{2}_{D}/(4ABk^{2}_{D}+\alpha^{2})$. In
the next section, we will show that the SOC has a dramatic effect on the
interlayer exchange coupling between two ferromagnets with spin
polarization parallel to the interface with BITeI.

\subsection{In-Plane Magnetization}\label{sec5}
In order to study the interlayer exchange coupling between two ferromagnets
with spins parallel to the interface with BiTeI
[Fig.~\ref{fig2}(a)], we consider both
ferromagnets [$F_{\Top}$ and $F_{\Bot}$ in Fig.~\ref{fig2}(a)] to have spins
aligned in the $x$-direction without loss of generality. The interaction between the
ferromagnets $F_{\Top}$ and $F_{\Bot}$ depends on the element $\chi_{xx}(q_{z})$
of the spin susceptibility tensor in
Eq.~\eqref{chispin22}, which contains contributions from both interband and intraband transitions. Hence, unlike the case of the $z$-polarized ferromagnets, the spin susceptibility in this case has contributions from all regions of the Fermi surface as shown in Eq.~\eqref{chispin22}.

The interband contribution of $\chi_{xx}(q_{z})$ is given by
$\chi_{zz}(q_{z})/2$, and its contributions to the interlayer exchange coupling
is $I_{zz}(z)/2$, where $I_{zz}(z)$ is given in
Eq.~\eqref{interbandzz}. On the other hand, the intraband component of
the spin susceptibility, $\chi_{xx}(q_{z})$ in Eq.~\eqref{chispin22},
can be written as
\begin{eqnarray}
  \frac{\chi^{\rm intra}_{xx}(q_{z})}{C}= \int_{-k_{D}}^{k_{D}}dk_{z}\sum_{\mu=\pm}\mathcal{P}\int_{0}^{k_{\mu}}\left[\frac{2}{A(q^{2}_{z}-4k_{z}^2)}\right]kdk \nonumber \\
+2\int_{k_{D}}^{k_{m}}dk_{z}\sum_{\mu=\pm}\mathcal{P}\int_{0}^{-\mu k_{\mu}}\left[\frac{2}{A(q^{2}_{z}-4k_{z}^2)}\right]kdk\;,\nonumber\\\label{integral}
\end{eqnarray}
where $k_{D}=\sqrt{E_{F}/A}$, and $k_{\mu}$, $k_{m}$ are given in
Eq.~\eqref{cases} and Eq.~\eqref{kzhight}, respectively.
Integrating Eq.~\eqref{integral} over ${\kpar}$ and
$k_{z}$ gives
\begin{widetext}
\begin{equation}\label{chispinxx}
\chi^{{\rm intra}}_{xx}(q_{z})=\frac{C}{B}\begin{dcases}
                    k_{m}-\frac{(q^{2}_{z}-4k^{2}_{m})}{4q_{z}}\ln\left|\frac{q_{z}+2k_{m}}{q_{z}-2k_{m}}\right|+\frac{\alpha^{2}}{4ABq_{z}}\ln\left|\frac{q_{z}+2k_{m}}{q_{z}-2k_{m}}\right|\;, & \mbox{if } q_{z} \neq 0\\
                       2k_{m}+\frac{\alpha^{2}}{4ABk_{m}}, & \mbox{if } q_{z}=0.
                     \end{dcases}
\end{equation}
\end{widetext}
Before proceeding to the interlayer exchange interaction between the
$x$-polarized ferromagnets, we find the long-wavelength limit of the
spin susceptibility $\chi_{xx}(0)=\chi_{zz}(0)/2+\chi^{\rm
  intra}_{xx}(0)=C[2k_{m}+2k_{D}+\alpha^{2}/(4ABk_{m})]/B$, where
$C=\mu^{2}_{B}\pi/(2\pi)^{3}$. The latter is the sum of the van Vleck
and Pauli susceptibilities~\cite{spinresp,vanvleck1,vanvleck2,vanvleck3}. Moreover, we notice that
$\chi_{xx}(0)>\chi_{zz}(0)$ for $E_{F}\ge 0$ (see Fig.~\ref{pauli})
and therefore electron spins in BiTeI are more easily polarized in
response to a uniform in-plane magnetic field than an out-of-plane field.

The intraband contribution to  the
interlayer exchange interaction is
\begin{eqnarray}\label{CuplingRashZZ}
I^{{\rm intra}}_{xx}(z)=-\left(\frac{J_{0}}{V_{0}}\right)^2S^{\Top}_{x}S^{\Bot}_{x}\frac{c^{2}}{2\pi}\int_{0}^{\infty}dq_{z}\nonumber\\ \times \cos(q_{z}z)\chi^{{\rm intra}}_{xx}(q_{z})\;,
\end{eqnarray}
and the total interlayer exchange interaction is given by
\begin{equation}\label{totalIxx}
I_{xx}(z)=I^{{\rm intra}}_{xx}(z)+\left.\frac{I_{zz}(z)}{2}\right\vert_{S^{(\Top,\Bot)}_{z}\rightarrow S^{(\Top,\Bot)}_{x}}\;,
\end{equation}
where $I_{zz}(z)$ is given in Eq.~\eqref{interbandzz}.

The intraband component of $\chi_{xx}(q_{z})$ Eq.~\eqref{chispinxx}
reveals the presence of a critical spanning vector that leads to a
Kohn anomaly at $|q_{z}|=2k_{m}$. The contribution of this anomaly to
the exchange coupling is found by integrating around a small
interval containing it, such that for $|q_{z}|\approx 2k_{m}$,
 \begin{eqnarray}\label{kohnxx2}
 I^{{\rm intra}}_{xx}(z)\approx-&&\frac{S^{\Top}_{x}S^{\Bot}_{x}}{16\pi^{3}B}\left(\frac{\mu_{B}J_{0}c}{V_{0}}\right)^{2}\int_{2k_{m}-\epsilon}^{2k_{m}+\epsilon}{dq_{z}}\bigg[k_{m}+\nonumber\\ &&\left.\frac{(q_{z}-2k_{m})}{2}\ln|q_{z}-2k_{m}|-\right.\nonumber\\ &&\left.\frac{\alpha^{2}}{4ABk_{m}}\ln|q_{z}-2k_{m}|\right]\cos(q_{z}z)\;.
\end{eqnarray}
Following similar steps as before,
 \begin{eqnarray}\label{kohnxx6}
 I^{{\rm intra}}_{xx}(z)\approx -\frac{I_{0}}{2}\left[\left(\frac{c}{z}\right)^{2}\sin(2k_{m}z)+\right.\nonumber\\ \left.\frac{\alpha^{2}c}{4ABk_{m}}\left(\frac{c}{z}\right)\cos(2k_{m}z)\right]\;.
\end{eqnarray}

The numerical and analytical evaluations of intraband component of
$I(z)$ are shown in Figs.~\ref{fig5}(a) and (b). We find that this part of
the interlayer exchange interaction oscillates with a period $\pi/k_{m}$. For thin films of BiTeI [Fig.~\ref{fig5}(a)], both decay powers [$z^{-1}$ and
$z^{-2}$ in Eq.~\eqref{kohnxx6}] are essential for $I^{{\rm intra}}_{xx}(z)$. However, for relatively thick samples of BiTeI [Figs.~\ref{fig5}(a) and (b)], \textit{i.e.} $z\approx 5c$, the interlayer exchange interaction is solely determined by the spin-orbit dependent part of $ I^{{\rm intra}}_{xx}(z)$, \textit{i.e},
 \begin{equation}\label{kohnxx7}
 I^{{\rm intra}}_{xx}(z)\approx -\frac{I_{0}}{2}\left[\frac{\alpha^{2}c}{4ABk_{m}}\left(\frac{c}{z}\right)\cos(2k_{m}z)\right]\;,
\end{equation}
where the decay power law $z^{-1}$ is sufficient to describe $I^{{\rm intra}}_{xx}(z)$, as shown in Fig.~\ref{fig5}.

Combining the interband and intraband contributions as in
Eq.~\eqref{totalIxx},
the total exchange between these ferromagnets is given by
\begin{widetext}
\begin{eqnarray}\label{ixxfinal}
 I_{xx}(z)\approx -\frac{I_{0}}{2}\left\{\left(\frac{c}{z}\right)^{2}\left[\sin(2k_{m}z)+\frac{2A Bk^{2}_{D}\sin(2k_{D}z)}{4ABk^{2}_{D}+\alpha^{2}}+\Theta\left(E_{F}-\frac{2\alpha^{2}}{B}\right)\sum_{j=\pm}\frac{2(k^{j}_{n}-k_{D})\sin(k^{j}_{n}z)}{(2k_{D}+k^{j}_{n})}\right]\right.&&\nonumber\\
  +\left.\left(\frac{c}{z}\right)\frac{\alpha^{2}c}{4ABk_{m}}\cos(2k_{m}z)\right\}\;.
\end{eqnarray}
\end{widetext}
Here we notice that the two different periods of oscillation result from the interband and intraband contributions to the magnetic exchange interaction. In the limit $\alpha=0$, $k^{+}_{n}=2k_{c}$ and $k^{-}_{n}=0$, one
recovers the interlayer exchange coupling in the absence of SOC in Eq.~\eqref{khon4}.

The interlayer exchange coupling $I_{xx}(z)$ displays four periods of
oscillation for $E_{F}>2\alpha^{2}/B$. However, for the low-energy
electrons in BiTeI the Fermi energy $E_{F}<2\alpha^{2}/B$ restricts
the periods to oscillation to $\pi/k_{D}$ and $\pi/k_{m}$.

The presence of a SOC in BiTeI makes the interlayer exchange coupling $I_{xx}(z)$ mediated by this material,  Eq.~\eqref{ixxfinal}, qualitatively different from that of a conventional 3D metal, Eq.~\eqref{khon4}. These differences are reflected in the experimentally relevant characteristics of this exchange, \textit{i.e.} its oscillation periods and its dependence on the spacer thickness. Whereas the interlayer exchange interaction mediated by a conventional 3D metal  has a single oscillatory period and decays as $z^{-2}$, the interlayer exchange coupling $I_{xx}(z)$ mediated by BiTeI has two periods of oscillation and  more strikingly it decays as the inverse the spacer thickness $z^{-1}$.

\begin{figure}[t!]
    \begin{center}
            \subfigure{
            \includegraphics[width=0.95\columnwidth]{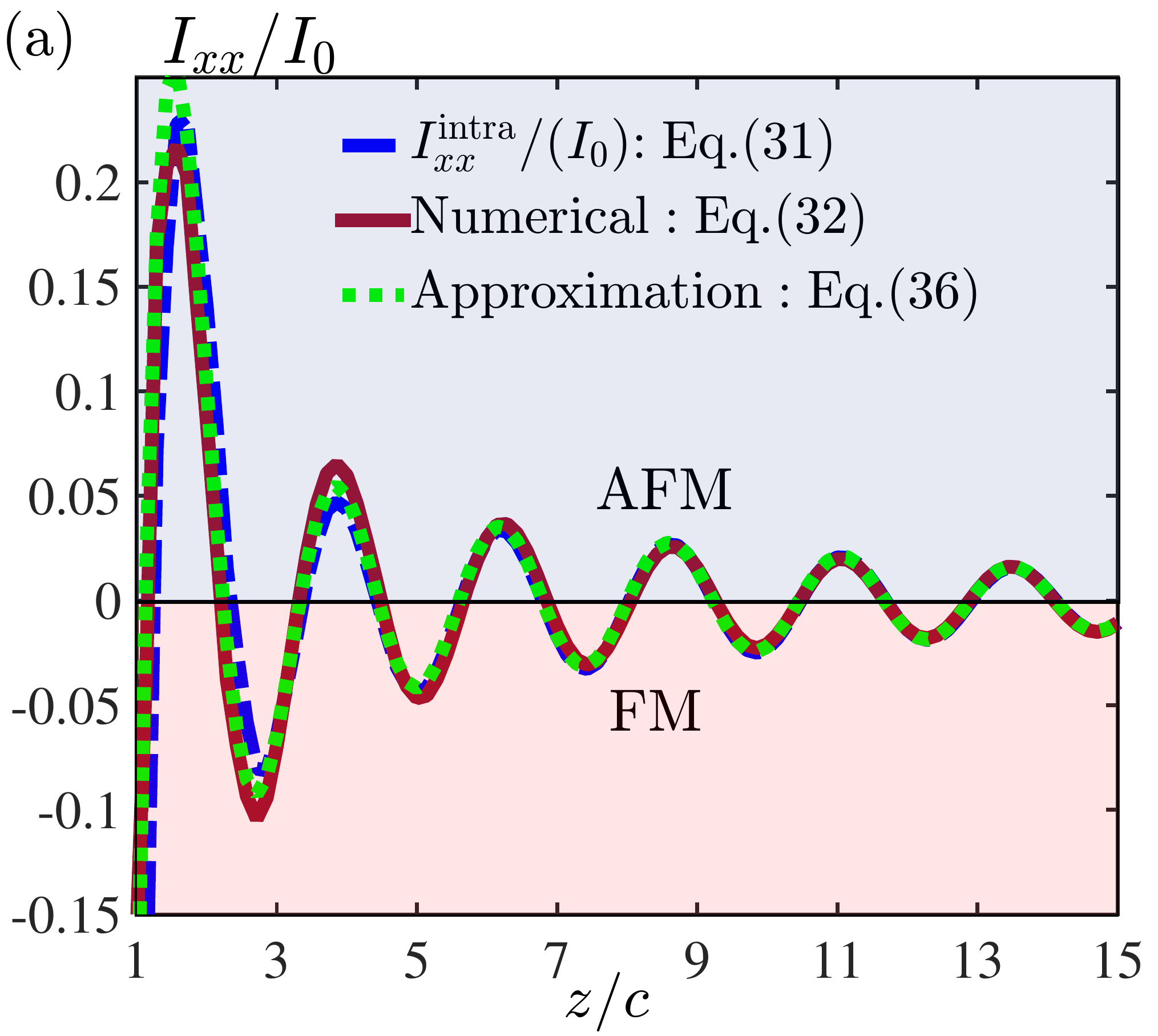}}
        \subfigure{
            \includegraphics[width=0.95\columnwidth]{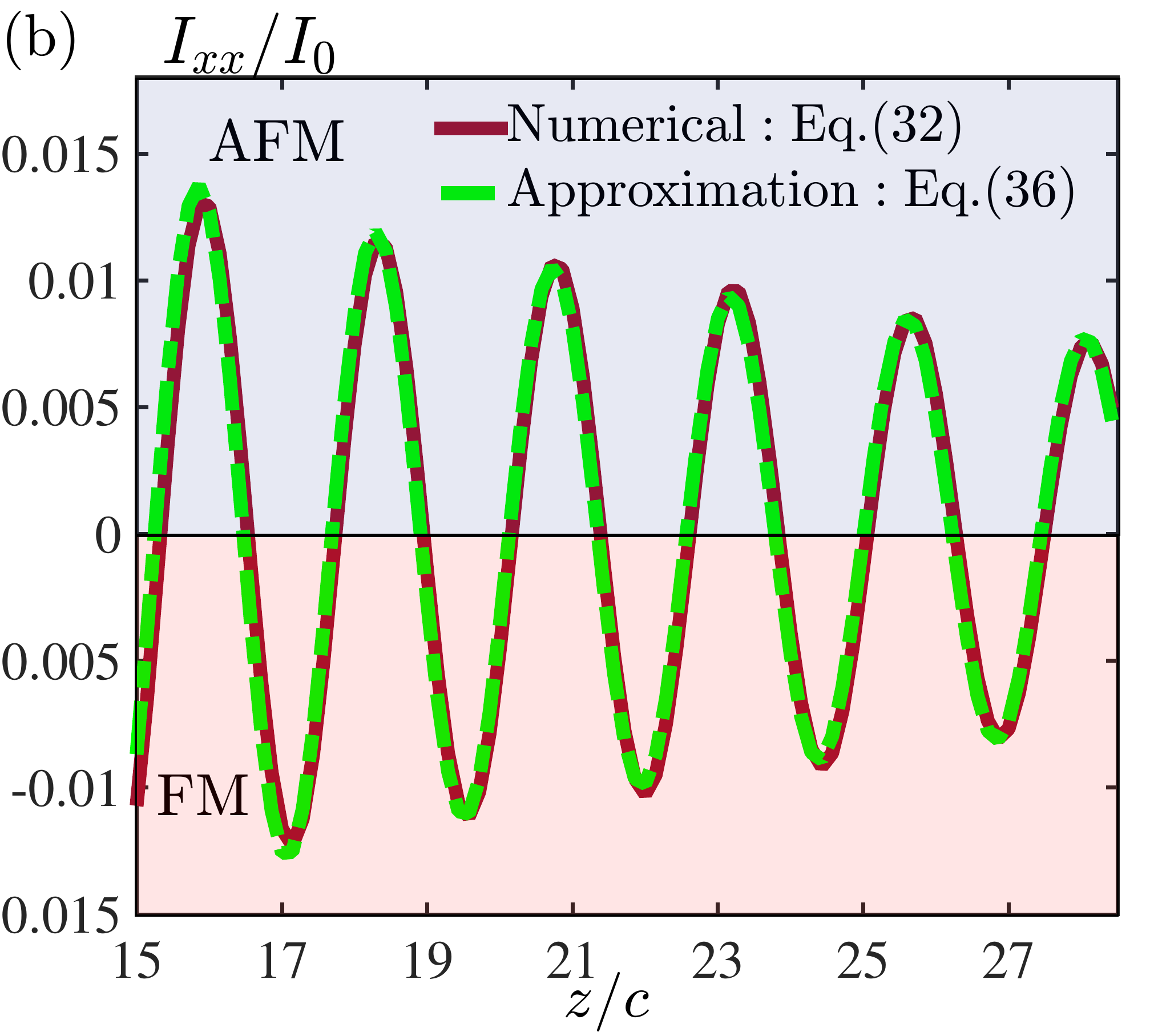}}
                \end{center}
    \caption{Thickness dependence of the interlayer exchange coupling between
      two ferromagnets with spins parallel to the interface
      (Fig.~\ref{fig2}) with $E_{F}=0.18$ eV. Panel (a) shows this
      dependence of $I_{xx}(z)$ for relatively thin samples of
      BiTeI. In this case $I_{xx}(z)$ oscillates between FM and AFM
      couplings with two periods, $\pi/k_{m}$ and
      $\pi/k_{D}$. Additionally, the  intraband and interband
      contributions to the coupling can be discerned in this regime
      since the coupling decays as $z^{-1}$ and $z^{-2}$,
      respectively.
       (b) For relatively thick films of BiTeI, the coupling is dominated by the intraband contributions. It oscillates with a period $\pi/k_{m}$ and decays as $z^{-1}$.}
\label{fig6}
\end{figure}

A close look at $I_{xx}(z)$, Fig.~\ref{fig6}(a), shows that the intraband
contribution of $I_{xx}(z)$ is dominant due to
its peculiar dependence on $z$, Eq.~\eqref{kohnxx7}. Hence, the interband
contribution to the interlayer exchange coupling can only be discerned for small BiTeI thickness,
$z<5c$.
Moreover, for relatively large thicknesses of BiTeI, $z>5c$ [Figs.~\ref{fig6}(a) and (b)],
the interlayer exchange coupling is uniquely determined  by the term
proportional to the SOC coupling that decays as $z^{-1}$ in
Eq.~\eqref{kohnxx7}.
Consequently, unlike the case in a conventional 3D
metal, the exchange coupling displays an
intriguing dependence on the thickness of the BiTeI decaying as
$z^{-1}$, which is
reminiscent of the coupling between two magnetic chains mediated by a
2D conventional metal~\cite{kittelbook,note}. We attribute this unusual dependence to the 2D nature of the
Rashba SOC coupling in BiTeI.

\section{Formalism of Interlayer Exchange Interaction: Topological Phase}\label{sec6}

BiTeI has been theoretically predicted to undergo a topological phase
transition under moderate hydrostatic
pressure~\cite{BiTeITop,BiTeITop1,BiTeITopExp,BiTeITopExp3}. Subjecting
this material to an increasing pressure leads to its transition from
its trivial phase to a strong topological insulator phase at approximately
$3$ GPa~\cite{BiTeITop,BiTeITop1,BiTeITopExp,BiTeITopExp3}. These two
topologically distinct phases are separated by Weyl semi-metallic
phase~\cite{BiTeITopTheory,BiTeITopTheory1,BiTeITopTheory2}. Experimentally,
transport measurements suggest that the topological phase transition
happens at the theoretically expected value~\cite{BiTeITopExp2},
$3$ GPa, while optical experiments suggest that this topological phase
transition happens at $9$ GPa~\cite{BiTeITop1,BiTeITopExp3}. In the
following two sections, we analyze the interlayer exchange interaction between two
ferromagnets mediated by BiTeI in its topological insulator phase and
show that it exhibits important differences compared to that in the trivial
phase.

\begin{figure}[t]
  \centering
  \includegraphics[width=\columnwidth]{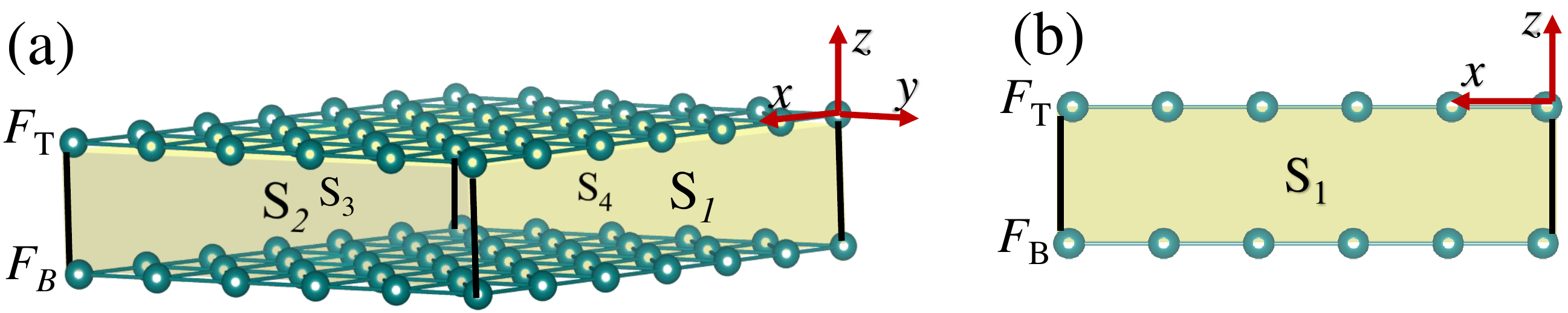}
  \caption{Schematic representation of (a) magnetic layers sandwiching
    topological BiTeI. The surface states at ${\rm S_{i=1,2,3,4}}$  mediate
    the exchange between the magnetic chains at the edges of the
    sample as shown in (b). The coordinate axes indicate the global
    coordinate frame of the system.}\label{fig8}
\end{figure}

In the previous sections we have shown that the interlayer exchange
interaction mediated by BiTeI in its trivial phase is dependent on the
spin orientation of the magnetic layers [Fig~\ref{fig2}(a)] and it
only couples collinear spins. The transition of BiTeI from
its trivial to its topological insulator phase crucially changes the nature of
the magnetic exchange in the system.
In its topological phase, BiTeI becomes insulating in the bulk and conducting
only through its surfaces. The exchange interaction between
magnetic impurities mediated by topological insulator surface states was previously
studied
~\cite{sufacesusc1,sufacesusc2,rkkysurf1,rkkysurf2,rkkysurf3, Ref21,Ref22,Ref23,ref11}.
In our case, the exchange interaction couples not only two magnetic impurities but
two one-dimensional chains of spins at the top
and bottom edges of the sample, mediated by the helical electrons residing on the side
surfaces. Fig.~\ref{fig8}(a) shows our setup with the side surfaces of
the rectangular BiTeI sample indicated by $\rS_{i}$, $i=1,2,3,4$.

In order to study the properties of the interlayer exchange interaction of topological BiTeI we assume that the surface states present on the surfaces $\rS_{i}$ are helical and satisfy~\cite{ss2,sstates,ss3,ss4,Bernevig,Rev1,Rev2,sstates}
\begin{equation}\label{HelicalH}
  H_{\rS_{i}}(\bppar)=\hbar v_{F}(\bsigma\times\bppar)_{\hn_{i}}\;,
\end{equation}
where $\hn_{i}$ is the normal to the surface $\rS_{i}$, and $v_{F}$ is
the Fermi velocity of the surface states. Adapting
Eq.~\eqref{internonsim} to describe the 2D metallic states that
mediate the exchange interaction between two magnetic chains, Fig.~\ref{fig8}(b), one arrives at
\begin{eqnarray}\label{internonsim2d}
I^{i}(z)=-\sum_{a,b=x,y,z}\frac{J^{2}_{0}S^{\Top}_{a}S^{\Bot}_{b}c}{2(2\pi)^{2}A_{0}}\int_{-{\pi}/{c}}^{{\pi}/{c}}dq_{z}e^{iq_z z}\nonumber\\ \times \int_{{\rm1DBZ}}d{ q_{j}}\chi^{i}_{ab}(q_{j},q_{z}) \sum_{j\in F_{{\rm T}}}e^{i q_{j}j}\;,
\end{eqnarray}
here, $i=(1,2,3,4)$ is the surface index [Fig.~\ref{fig8}(a)],
$S^{\Top}_{a}$ ($S^{\Bot}_{b}$) is the spin of the top (bottom)
magnetic layer, $j=x$ or $y$ depending on the surface $\rS_{i}$,
$\chi^{i}_{ab}(q_{j},q_{z})$ is the $ab$ component of the
spin-susceptibility for a given surface $\rS_{i}$, and $A_{0}$ is the
area of the 2D BZ enclosing the surface states. It is convenient to
analyze the interlayer exchange coupling in Eq.~\eqref{internonsim2d} in the
set of local coordinates of each surface that transforms the
Hamiltonian in Eq.~\eqref{HelicalH} to $H(\bppar)=\hbar
v_{F}(\bsigma\times\bppar)_{\zh}$ for all surfaces. The advantage of
this transformation is that the susceptibility components
$\chi^{i}_{ab}=\chi^{j}_{ab}$ for $i\ne j$. This transformation is
achieved by making the local $z$-axis normal to each surface and the
local $y$-axis pointing to the global $z$-direction. This
also requires the transformation of the magnetic
layers' spins $S^{(\Top,\Bot)}_{a,b}$ in the global coordinates to
$\mathcal{S}^{(\Top,\Bot)}_{c,d}$ in the local coordinates as
indicated in Table~\ref{table} and the corresponding transformation of
the susceptibility components $\chi^{i}_{ab}\rightarrow\chi_{cd}$.
In the local coordinate frame, the interlayer exchange coupling associated with each surface is given by
\begin{eqnarray}\label{internonsim2dlocal}
I^{i}(y)=&&-\sum_{a,b=x,y,z}{\rm sgn}(\mathcal{S}^{i,\Top}_{c}\mathcal{S}^{i,\Bot}_{d})\frac{J^{2}_{0}S^{\Top}_{a}S^{\Bot}_{b}c}{2(2\pi)^{2}A_{0}}\int_{-{\pi}/{c}}^{{\pi}/{c}}dq_{y}e^{iq_y y}\nonumber\\&& \times \int_{{\rm1DBZ}}d{ q_{x}}\chi_{cd}(q_{x},q_{y}) \sum_{x\in F_{{\rm T}}}e^{i q_{x}x}\;,
\end{eqnarray}
here, $\mathcal{S}^{i,(\Top,\Bot)}_{c,d}$ are the locally transformed spins of the top and bottom magnetic layers, corresponding to $S^{(\Top,\Bot)}_{a,b}$ in the global coordinates, $\chi_{cd}(q_{y})$ is the susceptibility component in local coordinates corresponding to $\chi^{i}_{ab}$ in global coordinates. The transformation of the spins and the spin susceptibility indices between global and local coordinates is given in Table~\ref{table}.
\begin{table}[h!] 
  \centering
\begin{tabular}{||c|c |c ||}
  \hline \hline
    Global Spins &Surface& Local Spins\\ [0.5ex]
  \hline\hline
 $(S_{x},S_{y},S_{z})$ & 1 & $(-\mathcal{S}^{1}_{x},\mathcal{S}^{1}_{z},\mathcal{S}^{1}_{y})$ \\
   $(S_{x},S_{y},S_{z})$  & 2 &$(\mathcal{S}^{2}_{z},\mathcal{S}^{2}_{x},\mathcal{S}^{2}_{y})$ \\
   $(S_{x},S_{y},S_{z})$  & 3 & $(\mathcal{S}^{3}_{x},-\mathcal{S}^{3}_{z},\mathcal{S}^{3}_{y})$ \\
   $(S_{x},S_{y},S_{z})$  & 4 &$(-\mathcal{S}^{4}_{z},-\mathcal{S}^{4}_{x},\mathcal{S}^{4}_{y})$\\
  \hline\hline
\end{tabular}
  \caption{Transformation of the global spins into the local coordinate system of the surfaces ${\rm S}_{(1,2,3,4)}$ in Fig.~\ref{fig8}. The indices of the global susceptibility transform as the indices of the global spins. }
\label{table}
\end{table}
The local $x$-dimensions of the ferromagnetic layers satisfy periodic
boundary conditions  since they are assumed to be large compared to the interlayer
distance. The last sum in Eq.~\eqref{internonsim2dlocal} is then
nonzero only for $q_{x}=0$. Recalling that the length of the projected
1D BZ of BiTeI is $(2\pi)^2c/(2\pi A_{0})$ [where $(2\pi)^2/A_{0}$ is the area of the 2D BZ containing the surface states], the interlayer exchange coupling can be written as
\begin{eqnarray}\label{internonsim2d2local}
I^{i}_{ab}(y)=-&&\frac{{\rm sgn}(\mathcal{S}^{i,\Top}_{c}\mathcal{S}^{i,\Bot}_{d})}{2}\left(\frac{J_{0}}{A_{0}}\right)^2\frac{S^{\Top}_{a}S^{\Bot}_{b}c^{2}}{2\pi }\int_{-\pi/c}^{\pi/c}{dq_{y}}\nonumber\\&& \times e^{iq_{y}y}\chi_{cd}(q_{x}=0,q_{y}).
\end{eqnarray}
The components of spin susceptibility, $\chi_{cd}$, for a helical system described by $H=\hbar v_{F}({\bm \sigma}\times {\bppar})_{\hat{z}}$
consist of helicity-preserving and helicity-mixing contributions. For simplicity,
hereafter we omit the $q_{x}=0$ argument in $\chi_{cd}$ and denote
$\chi_{cd}(q_{y}) \equiv \chi_{cd}(q_{x}=0,q_{y})$. Since the
surface states are characterized by their helicity, the spin
susceptibility can be written as
$\chi_{cd}(q_{y})=\sum_{\mu,\nu=\pm}\chi^{\mu\nu}_{cd}(q_{y})$ with
\begin{eqnarray}\label{chispinsurf}
\chi^{\mu\nu}_{cd}(q_{y}) &=& \frac{-\mu^{2}_{B}}{(2\pi)^{2}}\int_{-\pi/c}^{\pi/c}dk_{y} \int_{{\rm1DBZ}}{{dk_{x} }}\nonumber\\&&\times\frac{f(E_{{\bkpar},\mu})-f(E_{{\bkpar+\bq_{y}},\nu})}{E_{{\bkpar},\mu}-E_{{\bkpar+\bq_{y}},\nu}+i\delta}\mathcal{F}^{\mu\nu}_{cd}(\bkpar,{\bkpar+\bq_{y}}),\nonumber\\
\end{eqnarray}
where $\mu,\nu=\pm$ are the helicities of the surface states,
$f(E_{{\bkpar},\mu})$ is the Fermi function,
$\mathcal{F}^{\mu\nu}_{cd}({\bkpar},{\bkpar+\bqpar})$ is the form
factor defined in Eq.~\eqref{chispin} with the state $|{\bkpar},\mu\rangle=(-i, \mu e^{-i\phi_{k}})^{\dag}e^{i{\bkpar}\cdot {\bm
    r}}/\sqrt{2}$. By considering that the largest critical spanning
vector $2k_{F}=2E_{c}/(\hbar v_{F})$, where $E_{c}\approx0.2$eV is the
cutoff energy of the low-energy theory of the surface
states~\cite{Cutsurface}, one obtains  a period of oscillation that is larger
than $2c$, thus, the limits of the integral in Eq.~\eqref{chispinsurf} will
be restricted only by the Fermi functions, \textit{i.e.},
\begin{eqnarray}\label{chispinsurf1}
\chi^{\mu\nu}_{cd}(q_{y}) &=& \frac{-\mu^{2}_{B}}{(2\pi)^{2}}\int dk_{y} \int{{dk_{x} }}\nonumber\\&&\times\frac{f(E_{{\bkpar},\mu})-f(E_{{\bkpar+\bq_{y}},\nu})}{E_{{\bkpar},\mu}-E_{{\bkpar+\bq_{y}},\nu}+i\delta}\mathcal{F}^{\mu\nu}_{cd}({\bkpar},{\bkpar+\bq_{y}}).\nonumber\\
\end{eqnarray}

In general, for a  2D electron system described by the Hamiltonian $H=\hbar
v_{F}({\bsigma}\times {\bppar})_{\zh}$ we evaluated the spin susceptibility tensor components (see Appendix~\ref{appa})
\begin{eqnarray}\label{chispinsurf2}
\chi_{cd}({\bqpar}) &=& \frac{-\mu^{2}_{B}}{(2\pi)^{2}}\sum_{\mu,\nu=\pm} \int{{d^{2}\bkpar }}\nonumber\\&&\times\frac{f(E_{{\bkpar},\mu})-f(E_{{\bkpar}+{\bqpar},\nu})}{E_{{\bkpar},\mu}-E_{{\bkpar}+{\bqpar},\nu}+i\delta}\mathcal{F}^{\mu\nu}_{cd}({\bkpar},{\bkpar}+{\bqpar}),\nonumber\\
\end{eqnarray}
and find that~\cite{sufacesusc3}
\begin{equation}\label{susctensor}
\chi({\bqpar}) =\left(
                 \begin{array}{ccc}
                   g_{1}(x)\cos^{2}(\phi_{q}) & \frac{g_{1}(x)}{2}\sin(2\phi_{q}) & g_{2}(x)\cos(\phi_{q})  \\
                   \frac{g_{1}(x)}{2}\sin(2\phi_{q})  &  g_{1}(x)\sin^{2}(\phi_{q}) & g_{2}(x)\sin(\phi_{q}) \\
                   g^{*}_{2}(x)\cos(\phi_{q}) & g^{*}_{2}(x)\sin(\phi_{q}) & g_{3}(x) \\
                 \end{array}
               \right)
\end{equation}
where $\phi_{q}=\tan^{-1}(q_{x}/q_{y})$, $x=2k_{F}/\qpar$, $\qpar=\sqrt{q^{2}_{x}+q^2_{y}}$, $k_{F}$ is the Fermi momentum,  and
\begin{eqnarray}
&&g_{1}(x) =\frac{-\mu^{2}_{B}}{4\pi\hbar v_{F}}{\rm Re}\left[\sqrt{1-x^2}+\frac{q}{2}\sin^{-1}\left(\sqrt{1-x^2}\right)\right],\nonumber  \\
&&g_{2}(x)  =\frac{-i\mu^{2}_{B}q}{8\pi\hbar v_{F}}\left[1-\frac{1}{2}{\rm Re}\left(\sqrt{1-x^2}\right)\right], \nonumber \\
&&g_{3}(x)  = \frac{-\mu^{2}_{B}}{2\pi\hbar v_{F}}\left\{k_{F}+\frac{q}{2}{\rm Re}\left[\sin^{-1}\left(\sqrt{1-x^2}\right)\right]\right\}.
\end{eqnarray}
In the literature there has been a disagreement on the forms of the
functions $g_{1,2,3}(x)$~\cite{sufacesusc1,sufacesusc2}.  In order to
find these functions we derived the renormalized susceptibility, which
required the substraction of intrinsic susceptibility (the
susceptibility at $E_{F}=0$) at $\bqpar=0$,  $\chi^{{\rm
    intrinsic}}_{cd}(0)$,  from the total susceptibility
$\chi_{cd}(\bqpar)$. This method is consistent with
Refs.~\cite{Ando} and \cite{Hwang} and leads to identical
results for the polarization function therein.
Taking $q_{x}=0$, $\phi_{q}=\pm \pi/2$ [$+$ ($-$) for positive (negative) values of $q_{y}$], the susceptibility tensor reduces to
\begin{equation}\label{susctensor2}
\chi(q_{y}) =\left(
                 \begin{array}{ccc}
                  0& 0 &0 \\
                   0  &  g_{1}(x) & {\rm sgn }(q_{y})g_{2}(x) \\
                   0& {\rm sgn }(q_{y}) g^{*}_{2}(x)& g_{3}(x) \\
                 \end{array}
               \right)\;,
\end{equation}
where $x=2k_{F}/q_{y}$ in this case. Since the elements
$\chi_{cd}(q_{y})=0$ for $(c,d)=\{(x,x),(x,y),(x,z),(y,x),(z,x)\}$,
their associated interlayer exchange coupling is zero. With the remaining
non-vanishing susceptibility elements we find that the interlayer
exchange coupling for the different surfaces, when expressed in the global
coordinates of the system as indicated in Fig.~\ref{fig8}, is given by
\begin{figure}[t]
    \begin{center}
            \subfigure{
            \includegraphics[width=0.95\columnwidth]{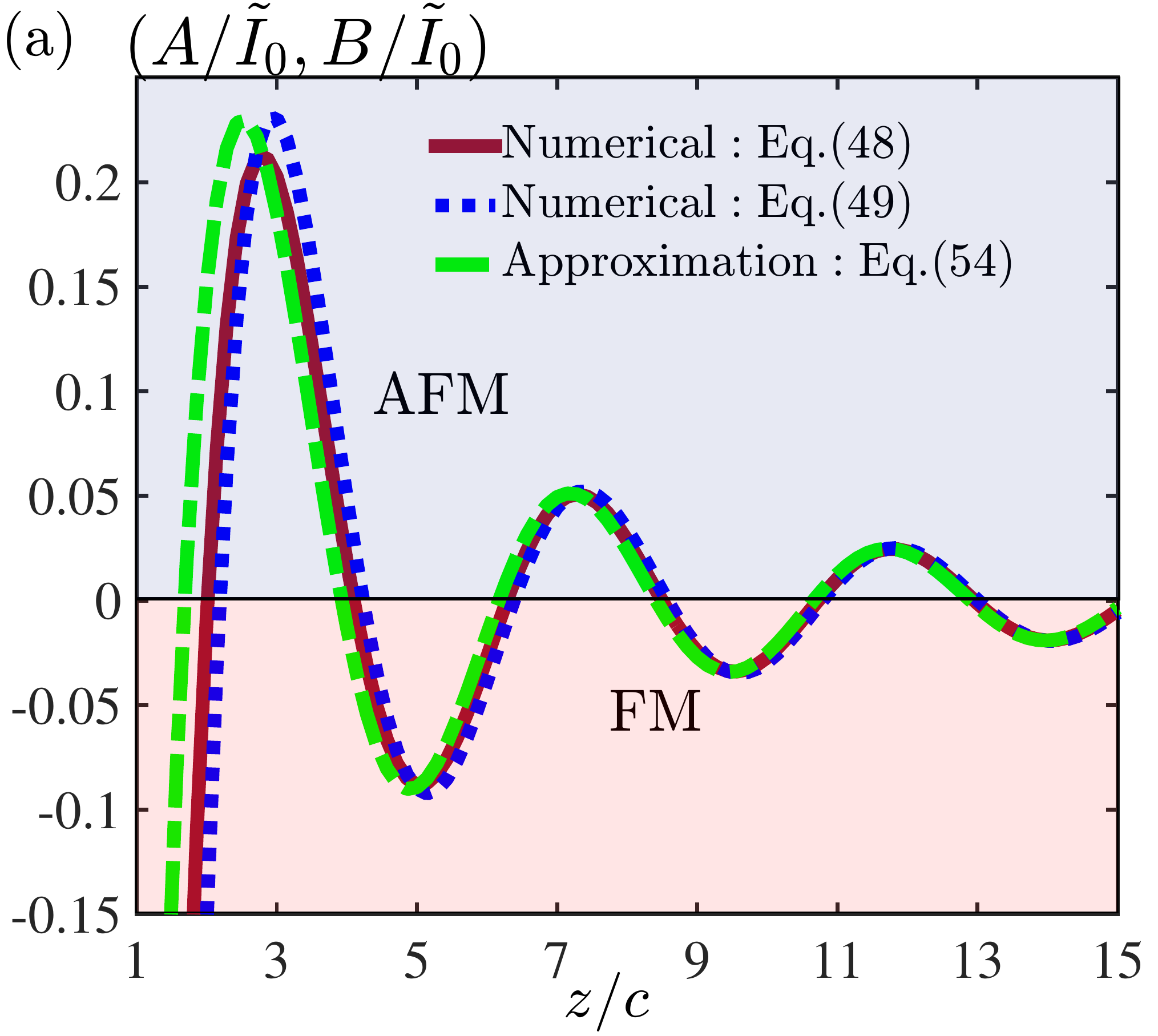}}
        \subfigure{
            \includegraphics[width=0.95\columnwidth]{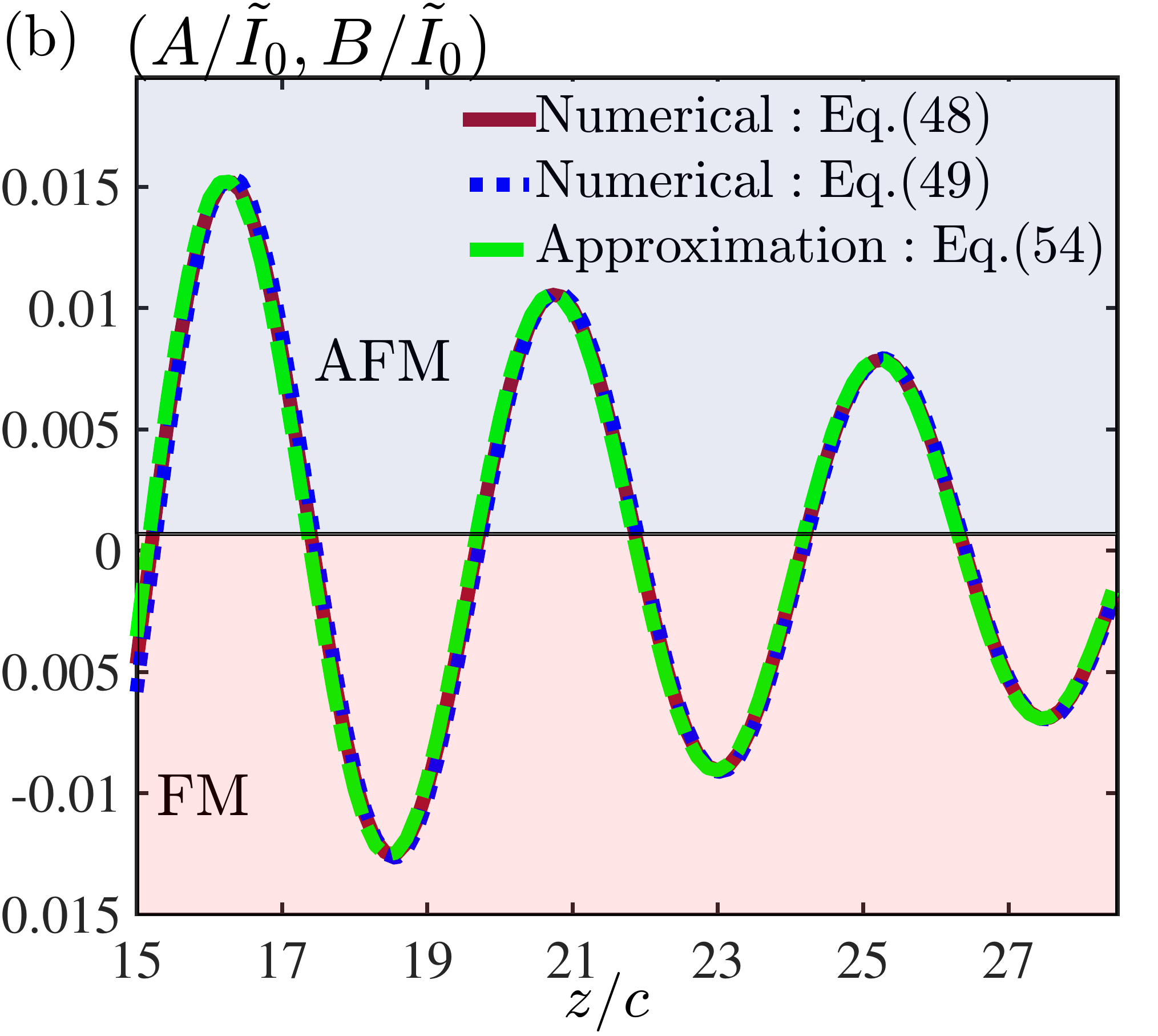}}
                \end{center}
    \caption{Thickness dependence of amplitude for collinear parts of
      the interlayer exchange coupling per surface for (a) relatively thin
      samples of BiTeI and (b) thick samples of  BiTeI. In (a) and (b)
      we take the parameter values $\hbar v_{F}=1$ eV\AA ~and
      $E_{F}=0.1$ eV. In both (a) and (b), the collinear exchange
      interaction oscillates with a period $\pi/k_{F}\approx4.5c$ and decays as $z^{-3/2}$. }
\label{fig9}
\end{figure}
\begin{figure}[t]
    \begin{center}
            \subfigure{
            \includegraphics[width=0.95\columnwidth]{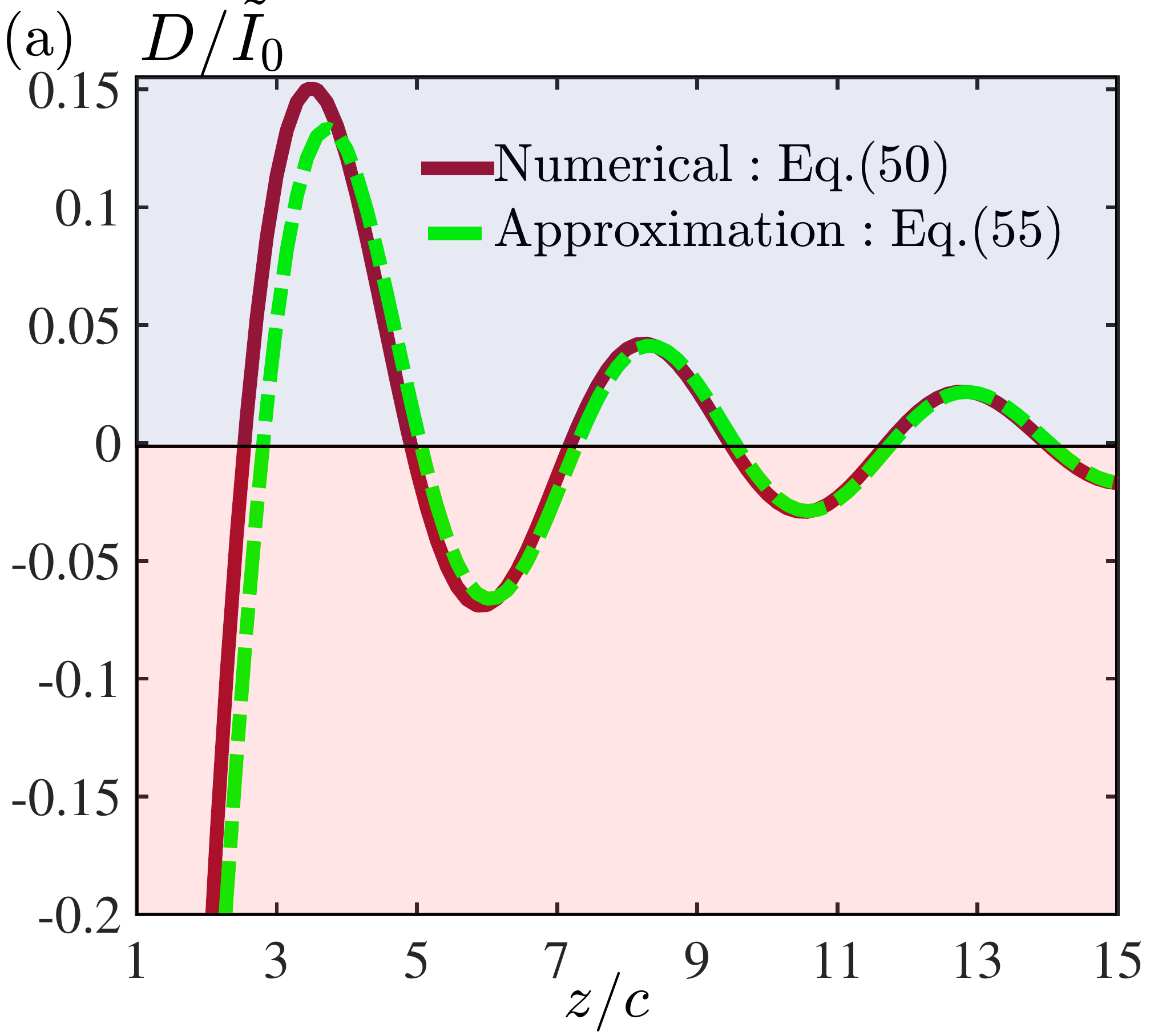}}
        \subfigure{
            \includegraphics[width=0.95\columnwidth ]{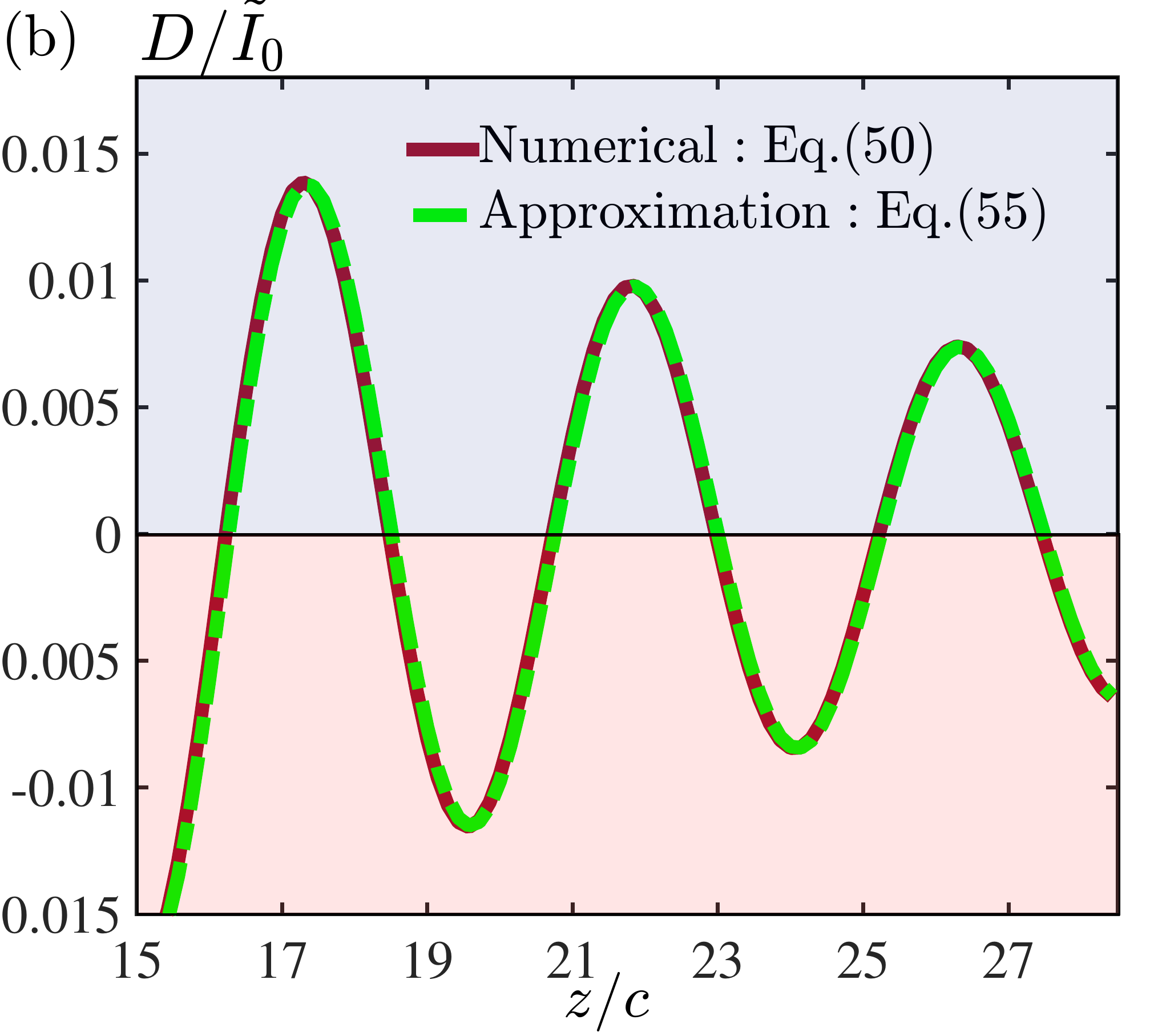}}
                \end{center}
    \caption{Thickness dependence of amplitude for DM (non-collinear)
      part of the interlayer exchange coupling per surface for (a) relatively
      thin samples of BiTeI and (b) thick samples of  BiTeI. In this
     figure we take the parameter values  $\hbar v_{F}=1$ eV\AA ~and $E_{F}=0.1$
      eV. In both (a) and (b), the DM interaction oscillates with a period $\pi/k_{F}\approx4.5c$ and decays as $z^{-3/2}$. }
\label{fig10}
\end{figure}
\begin{equation}\label{interlayersurfaces}
I^{i}(z)=AS^{\Top}_{z}S^{\Bot}_{z}+BS^{\Top}_{| \hat{n}_{i}|}S^{\Bot}_{| \hat{n}_{i}|}+D({\bm S}^{\Top}\times{\bm S}^{\Bot})_{\zh\times \hn_{i}}\;,
\end{equation}
where $\hn_i$ is the normal to the surface ${\rm S}_{i}$ in global coordinates, and
\begin{eqnarray}\label{Aint}
  A &=&-\frac{1}{2(2\pi)}\left(\frac{J_{0}c}{A_{0}}\right)^{2}\int_{-\infty}^{\infty}e^{iq_{z}z}\chi_{yy}(q_{z})dq_{z}\;, \\ \label{Bint}
  B&=& -\frac{1}{2(2\pi)}\left(\frac{J_{0}c}{A_{0}}\right)^{2}\int_{-\infty}^{\infty}e^{iq_{z}z}\chi_{zz}(q_{z})dq_{z}\;, \\ \label{Dint}
  D &=&  -\frac{1}{2(2\pi)}\left(\frac{J_{0}c}{A_{0}}\right)^{2}\int_{-\infty}^{\infty}e^{iq_{z}z}\chi_{yz}(q_{z})dq_{z}\;,
\end{eqnarray}
where
\begin{eqnarray}\label{chiab}
  &&\chi_{yy}(q_{z}) = \frac{-\mu^{2}_{B}}{4\pi\hbar v_{F}}{\rm Re}\left[\sqrt{1-x^2}+\frac{|q_z|}{2}\sin^{-1}\left(\sqrt{1-x^2}\right)\right],   \nonumber \\
   &&\chi_{zz}(q_{z}) = \frac{-\mu^{2}_{B}}{2\pi\hbar v_{F}}\left\{k_{F}+\frac{|q_{z}|}{2}{\rm Re}\left[\sin^{-1}\left(\sqrt{1-x^2}\right)\right]\right\},   \nonumber\\
  &&\chi_{yz}(q_{z})=\frac{-i\mu^{2}_{B}q_{z}}{8\pi\hbar v_{F}}\left[1-\frac{1}{2}{\rm Re}\left(\sqrt{1-x^2}\right)\right],
\end{eqnarray}
and $x=2k_{F}/q_{z}$. The integrals in Eqs.~\eqref{Aint}-\eqref{Dint} are written  in the global coordinates, where this is achieved by replacing $(y,q_{y})\rightarrow (z,q_{z})$ in Eq.~\eqref{susctensor2}. Notice that the integration over $q_z$ in Eqs.~\eqref{Aint}-\eqref{Dint} can be
extended to $\pm \infty$, since all critical spanning vectors are
much smaller than $\pi/c$ within the range of Fermi energy
$E_{F} < 0.2$eV considered in the low-energy effective theory
for BiTeI surface states.

Unlike non-topological BiTeI, the magnetic interlayer exchange coupling, Eq.~\eqref{interlayersurfaces}, mediated by the surface states of topological BiTeI allows for the coupling of collinear and non-collinear spins. The coupling between non-collinear spins is mediated by the DM interaction~\cite{DM1,DM2} which in itself arises due to the spin-momentum coupling of the surface states. The appearance of the DM-mediated interlayer coupling is intertwined with the appearance of the topological phase of BiTeI, hence, the experimental measurements of such an interaction can be used as an indicator of a topological phase transition in this material.

Having found the general expression of the interlayer exchange coupling in the topological phase of BiTeI, Eq.~\eqref{interlayersurfaces}, we proceed to determine its dependence on the sample thickness in the next section.

\section{Spatial Dependence of Exchange Interaction in the
Topological Phase}\label{sec7}
In the trivial phase of BiTeI we have found that the interlayer
exchange interaction for the system in Fig.~\ref{fig2}(a) decays with the
thickness of BiTeI as $z^{-2}$ for perpendicularly magnetized layers
and $z^{-1}$ for layers with magnetization parallel to the
interface. In this section we investigate the change in this $z$ dependence as BiTeI becomes a strong topological insulator.

The thickness dependence of the interlayer exchange coupling of topological
BiTeI in Eq.~\eqref{interlayersurfaces} is numerically determined by the integrals
in Eqs.~\eqref{Aint}-\eqref{Dint}.
Due to the presence of a Kohn anomaly at $|q_{z}|=2k_{F}$ the long-range behaviour of the interlayer exchange interaction between the magnetic chains, Fig.~\ref{fig8}(b), can be obtained by integrating near the Kohn anomaly  $q_{z}\in [2k_{F}-\epsilon,2k_{F}+\epsilon]$ where $\epsilon\ll 2k_{F}$. Taking into account that $\chi_{zz,yy}(q_{z})=\chi_{zz,yy}(-q_{z})$ are even in $q_{z}$ and $\chi_{yz}(q_{z})=-\chi_{yz}(q_{z})$ is odd in $q_{z}$, the integrals in Eqs.~\eqref{Aint}-\eqref{Dint} close to the Kohn anomaly become
\begin{eqnarray}\label{integralsSK}
  &&A \approx \gamma\sqrt{k_{F}}\int_{2k_{F}-\epsilon}^{2k_{F}+\epsilon}{\rm Re}\left(\sqrt{q_{z}-2k_{F}}\right)\cos(q_{z}z)dq_{z},\nonumber \\
 &&B\approx\gamma \int_{2k_{F}-\epsilon}^{2k_{F}+\epsilon}\left[k_{F}+\sqrt{k_{F}}{\rm Re}\left(\sqrt{q_{z}-2k_{F}}\right)\right]\cos(q_{z}z)dq_{z},\nonumber\\
&& D\approx\gamma \int_{2k_{F}-\epsilon}^{2k_{F}+\epsilon}\left[\sqrt{k_{F}}{\rm Re}\left(\sqrt{q_{z}-2k_{F}}\right)-2k_{F}\right]\sin(q_{z}z)dq_{z},\nonumber\\
\end{eqnarray}
where $\gamma=[(J_{0}\mu_{B}c)/(2\pi A_{0}\sqrt{\hbar v_{F}})]^{2}$. Recalling that the domain of the function ${\rm Re}(\sqrt{q_{z}-2k_{F}})$ is $q_{z}\ge 2k_{F}$, integrating Eq.~\eqref{integralsSK} by parts once and then changing to the variable $q'=\sqrt{q_{z}-2k_{F}}$, one obtains
\begin{eqnarray}
  A &=& B\approx-\frac{\gamma \sqrt{k_{F}}}{z}\int_{0}^{\sqrt{\epsilon}}\sin(q'^{2}z+2k_{F}z)dq'\;, \nonumber \\
  D&\approx& \frac{\gamma \sqrt{k_{F}}}{z}\int_{0}^{\sqrt{\epsilon}}\cos(q'^{2}z+2k_{F}z)dq'\;.
\end{eqnarray}
Since the dominant contribution of the previous integral comes from $q'=0$, we can extend its upper limit to $+\infty$ arriving at
\begin{eqnarray}\label{longrangesurface}
  A &=& B\approx-\tilde{I}_{0}\left(\frac{c}{z}\right)^{3/2}\cos\left(2k_{F}z-\frac{\pi}{4}\right)\;, \\
  D&\approx&-\tilde{I}_{0}\left(\frac{c}{z}\right)^{3/2}\sin\left(2k_{F}z-\frac{\pi}{4}\right)\;,
\end{eqnarray}
where
\begin{equation}
\tilde{I}_{0}=\frac{\sqrt{k_{F}c\pi}}{2\hbar v_{F}}\left(\frac{J_{0}\mu_{B}}{2\pi A_{0}}\right)^{2}\;.
\end{equation}
Then the interlayer exchange coupling between the magnetic chains for each
surface is
\begin{eqnarray}\label{FinalinterlayerS}
 && I^{i}(z) =-\tilde{I}_{0}\left(\frac{c}{z}\right)^{3/2}\bigg[\left(S^{\Top}_{z}S^{\Bot}_{z}+S^{\Top}_{| \nh_{i}|}S^{B}_{| \hn_{i}|}\right)  \nonumber\\ && \times \cos\left(2k_{F}z-\frac{\pi}{4}\right) + ({\bm S}^{\Top}\times{\bm S}^{B})_{\zh \times \hn_{i}}\sin\left(2k_{F}z-\frac{\pi}{4}\right) \bigg]. \nonumber \\
\end{eqnarray}

A comparison between the numerical calculation of the integrals
Eqs.~\eqref{Aint}-\eqref{Dint} with the analytical formulas in
Eq.~\eqref{longrangesurface} in Figs.~\ref{fig9}(a) and \ref{fig10}(a) reveals a close agreement  for thicknesses $z/c<5$. As
the sample thickness exceed $z/c>5$, the numerical and analytical
results become essentially equal and overlap with each other as shown in Figs.~\ref{fig9}(b) and \ref{fig10}(b). The interlayer exchange interaction mediated by the different surface states is characterized by a single period of oscillation, $\pi/k_{F}$, and the envelope of these oscillations decays with the thickness of the sample as $z^{-3/2}$.

We have seen that the magnetic exchange of the system in Fig.~\ref{fig2}(a) for topological BiTeI reduces to the exchange between the magnetic chains at the edges of the sample and it is mediated by the surface states, Fig.~\ref{fig8}(b). Unlike the trivial phase of BiTeI, the magnetic interlayer exchange interaction per surface in topological BiTeI couples both collinear and non-collinear spins due to the appearance of the DM interaction, and it decays with the thickness of BiTeI as $z^{-3/2}$.
\begin{figure}[t]
            \includegraphics[width=0.85\columnwidth ]{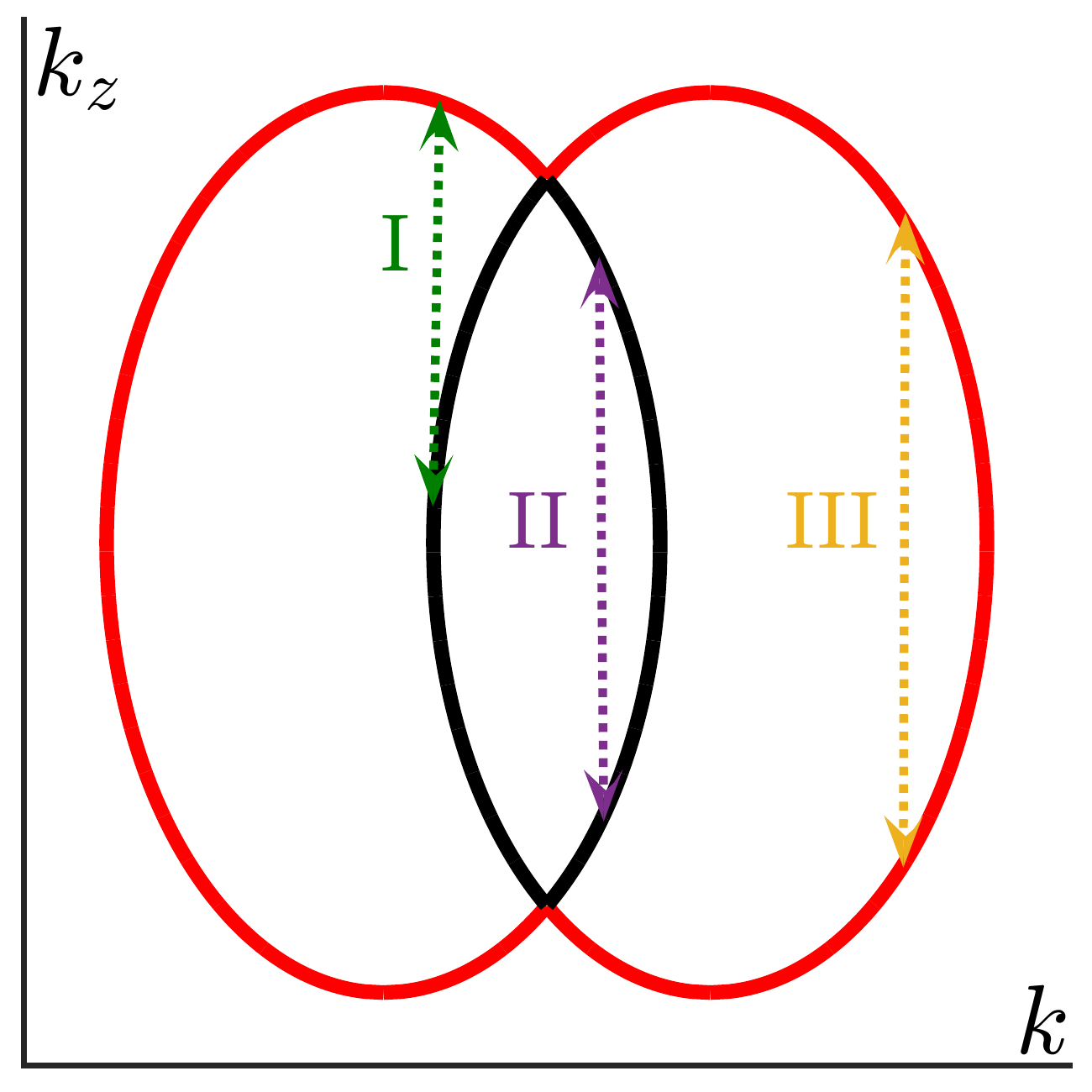}
    \caption{Schematic representation of the types of
      electronic transitions along $k_z$ through the Fermi surface:
      interband transitions between states with different helicities
      (I) and intraband transitions between states with the same
      helicities (II and III). Red (light) color indicates states with
      negative       helicity and black (dark) color indicates states
      with positive helicity.}
\label{vtrans}
\end{figure}

\section{Discussion}\label{discuss}
In this section, we start by first highlighting the physical differences of the
interlayer exchange interaction mediated by BiTeI in its trivial and
topological phases. In the trivial phase of BiTeI, we have found that the interlayer RKKY interaction
  is collinear and anisotropic for spin alignment between the out-of-plane and
  in-plane directions. This can be understood from a clear picture of the underlying
  physical processes contributing to the interlayer RKKY interaction,
  which are $\bkpar$-preserving interband and intraband transitions along $k_z$  through the Fermi
surface (Fig.~\ref{vtrans}).
\begin{table}[t] 
  \centering
\begin{tabular}{||c|c |c ||}
  \hline \hline
  \shortstack{ Matrix Element 
 $(x,y,z)$
  } &\shortstack{Interband\\($\nu=-\mu$)}&\shortstack{ Intraband \\ ($\nu=\mu$)}\\ [0.5ex]
  \hline\hline
 $\langle \mu,k'_{z},\bkpar|\sigma_{x}|\bkpar,k_{z},\nu\rangle$ &$i\mu e^{iq_{z}z}\cos(\phi_{k})$&$\mu e^{iq_{z}z}\sin(\phi_{k})$ \\
   $\langle \mu,k'_{z},\bkpar|\sigma_{y}|\bkpar,k_{z},\nu\rangle$&$i\mu e^{iq_{z}z}\sin(\phi_{k})$& $-\mu e^{iq_{z}z}\cos(\phi_{k})$ \\
 $\langle \mu,k'_{z},\bkpar|\sigma_{z}|\bkpar,k_{z},\nu\rangle$&$\mu e^{iq_{z}z}$& 0 \\
  \hline\hline
\end{tabular}
  \caption{Matrix elements of the $\bkpar$-preserving interband and
    intraband transitions along $k_z$ due to interaction with a
    ferromagnetic layer with magnetization in the $x$, $y$ or $z$-directions. Here, $q_{z}=k_{z}-k'_{z}$ and $|{\bm k},\mu\rangle$ are given by Eq.~\eqref{eigenstates}.}
\label{table2}
\end{table}
For an electron undergoing an interband
transition between bands with different helicities, the requirement of
in-plane momentum conservation means that its spin must
flip (shown as process I in Fig.~\ref{vtrans}).
This spin flip can be achieved through interacting with a
ferromagnetic layer magnetized either in the out-of plane or in-plane
directions, since the corresponding matrix elements are all nonvanishing, as shown in Table~\ref{table2}.
On the other hand, an
electron's spin is preserved for intraband transitions  (shown as
processes II and III in Fig.~\ref{vtrans}). This is only
possible via interacting with a ferromagnetic layer with an in-plane
magnetization, since the matrix element corresponding to the interaction with a $z$-polarized ferromagnet is zero, as shown in Table~\ref{table2}.
It follows from Table~\ref{table2} that
the interlayer exchange interaction is collinear with vanishing
interaction between magnetizations along different directions, because the product of any two
different matrix elements within the same type of transitions averages
out to zero over all directions \cite{note2}.
Futhermore, since the interlayer exchange coupling between layers with
out-of-plane magnetizations $I_{zz}$ is contributed only by interband
transitions, it is different from that between layers with in-plane
magnetizations $I_{xx,yy}$ that is contributed by both interband and intraband
transitions. The asymptotic behavior of the two cases are
distinguished by their distinct power laws going as $z^{-2}$ and
$z^{-1}$ respectively at large
thicknesses.

In the topological phase of BiTeI, we have shown that the interlayer
exchange interaction becomes limited to the magnetic chains residing at the edges
of the sample and it is mediated via the spin-momentum-locked surface
states. Unlike the trivial phase of BiTeI,
surface helical electrons in the topological phase mediate the RKKY interaction and due to their SOC they allow for magnetic
exchange between non-collinear spins arising from the DM
interaction. In general, the collinear and DM contributions of the
exchange coupling on each surface oscillate with a single period
determined by the Fermi wavelength and decay with the thickness of
BiTeI as $z^{-3/2}$. The DM term is present on each surface but the
sign of its coupling is dependent on the surface orientation
(\textit{e.g.}, it has opposite signs on opposite surfaces)
while the collinear terms carry the same sign for all surfaces.

We now discuss how the main features of the interlayer exchange
interaction in the trivial and topological phases may allow for
the experimental detection of the topological phase transition via the
measurement of this interaction.
Two setups can be devised for the detection of the topological phase
transition in BiTeI. First, using ferromagnets with orthogonal
magnetizations, the collinear exchange terms will vanish  in the
topological phase and the exchange is contributed only by
the non-collinear DM terms from each surface. Since the non-collinear
terms are absent in the trivial phase, the interlayer exchange coupling for
this setup will be zero when the pressure is below the critical
pressure for topological phase transition. Beyond this pressure, the
interlayer exchange coupling will be dominated by the surface DM terms. As
noted in Eq.~\eqref{FinalinterlayerS}, the DM term has opposite signs
for surfaces with opposite normal vectors $\hn_{i}$, and all the DM
contributions will cancel from pairs of opposite surfaces in a BiTeI sample
with an even number of perfectly aligned side surfaces. Therefore, experimental detection of a non-vanishing interlayer exchange coupling in the
topological phase requires a sample with an odd number of side surfaces (\textit{e.g.}, a
pentagonal prism) or irregular side surfaces with minimum cancellation
of DM coupling from opposite surfaces. Another possibility is to avoid
cancellation from opposite surfaces altogether by localizing the measurement geometry on only one side surface.

Second, using ferromagnets with parallel magnetizations, the DM part
of the exchange will vanish in the topological phase and the exchange
interaction is contributed only by collinear terms in both
phases.
Even though both phases are characterized by collinear exchange terms,
there is a distinct dependence of the exchange on the thickness,
$z^{-1}$ ($z^{-2}$) for magnetization parallel (orthogonal) to the interface in the trivial phase and $z^{-3/2}$ in the topological phase, and the change of this thickness dependence can serve as an indicator
for the topological phase transition, with the advantage that this
measurement scheme does not rely on the number of surfaces simultaneously measured.

Experimentally, the interlayer exchange coupling in heterostructures composed
of magnetic and non-magnetic materials can be determined by the
magneto-optical Kerr effect, magneto-resistance oscillations,
polarized neutron reflectrometry and ferromagnetic resonance
experiments~\cite{layered1,layered2,exp1,exp2,exp3,exp4,exp5,exp6,exp7,exp8,exp9,exp10}. In
the trivial phase of BiTeI
the observation of the phenomena described
in our work requires a quantitative determination of the interlayer
exchange coupling for both ferromagnetically and
anti-ferromagnetically coupled systems, together with  the ability to
discern the in-plane and out-of-plane components of the coupling. To
this end, we suggest that experiments relying on ferromagnetic
resonances would be a suitable platform for this
observation~\cite{exp7,exp8,exp9,exp10}. The detection of the
interlayer exchange coupling in the topological phase and the
topological phase transition requires pressure-controlled measurements
of magneto-resistance, spin susceptibility or ferromagnetic
resonances. These methods have been previously utilized to study the
pressure effects on the interlayer exchange coupling in Fr/Cr multilayers~\cite{pressexp1}, 2D ferromagnets~\cite{pressexp2} and FeCoB/Ru/FeCoB heterostructures~\cite{pressexp3}. These well-established
experimental methods coupled with the controlled growth of bulk Rashba
semiconductors~\cite{BiTeI1,BiTeI2,BiTeI3,BiTeI4,BiTeI5} should make the
observation of the unconventional interlayer exchange interaction mediated by
these exotic materials readily accessible.

\section{Conclusions}\label{conc}
We have
presented a theory for the interlayer exchange coupling between two
ferromagnets deposited on opposite surfaces of the bulk Rashba
semiconductor BiTeI, in its non-topological and topological
phases. Our work highlights
the unconventional dependence of the exchange interaction on the BiTeI spacer's topological phase and thickness, as
well as the ferromagnets' spin orientations.

In the non-topological phase of BiTeI, our calculation of the
interlayer exchange coupling revealed that the latter only couples collinear
spins and is strongly dependent on the magnetization direction of the
ferromagnets. If the ferromagnets are deposited on opposite surfaces along
the stacking direction of BiTeI and have an out-of-plane magnetization direction,
then the interlayer exchange coupling
behaves in a qualitatively similar way to that in
a metallic spacer with an ellipsoidal Fermi surface.
The interlayer exchange coupling shows a single period of oscillation and decays
with the thickness of the spacer $z$ as $z^{-2}$, and the only
effect introduced by the Rashba SOC is the renormalization of the
amplitude of the exchange interaction. However,
if the ferromagnets have an in-plane magnetization direction,
the interlayer exchange interaction exhibits significant qualitative differences
compared to a spin-degenerate metal.
The exchange interaction displays two periods of spatial
oscillations and decays as $z^{-1}$ with an amplitude that is proportional to
$\alpha^{2}$ (where $\alpha$ is the strength of the Rashba SOC) due to the interplay between the Rashba SOC and interfacing
spins.

In the topological phase of BiTeI, only the surface states can mediate
the interlayer exchange interaction resulting in a coupling between
the one-dimensional spin chains at the edges of the two ferromagnets.
In addition to collinear spins coupling, coupling of non-collinear
spins emerges due to the presence of DM interaction, and both are
characterized by a single oscillation period given by the Fermi
wavelength and decay with the thickness of BiTeI as $z^{-3/2}$.
The qualitative differences in the interlayer exchange coupling of
BiTeI between the trivial and topological phases can be used as a
signature to detect the topological phase transition in this exotic
material. The theory and findings we obtained for the topological
phase of BiTeI are also applicable to other strong topological insulators such as
Bi$_{2}$Se$_{3}$ and Bi$_{2}$Te$_{3}$.

\acknowledgments
We thank Tim Mewes for fruitful discussions. This work was supported by the U.S. Department of Energy, Office of Science, Basic Energy Sciences under Early Career Award $\#$DE-SC0019326.

\begin{figure*}[t]
    \begin{center}
            \subfigure{
            \includegraphics[width=0.9\columnwidth]{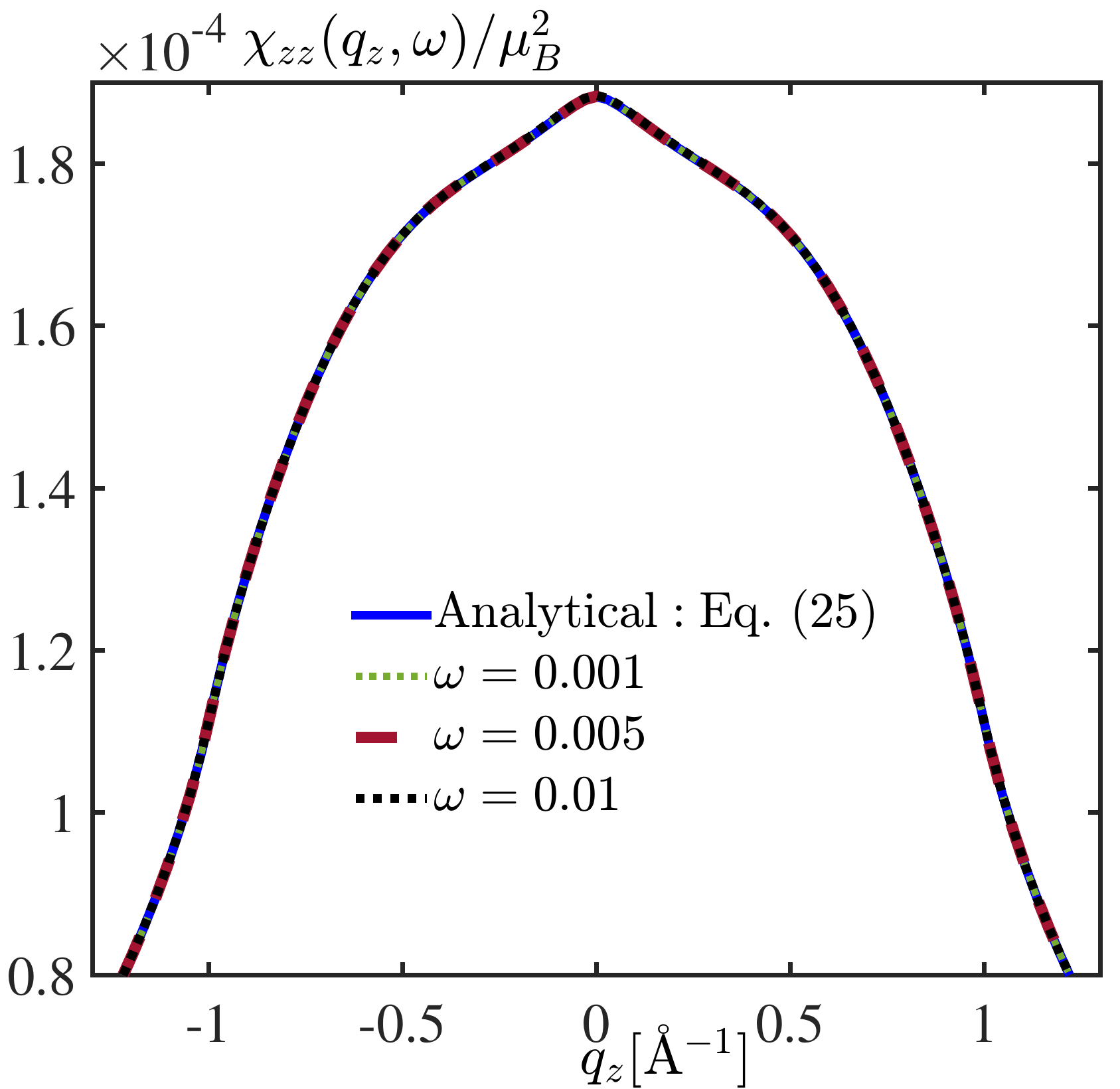}}%
        \subfigure{
            \includegraphics[width=0.9\columnwidth]{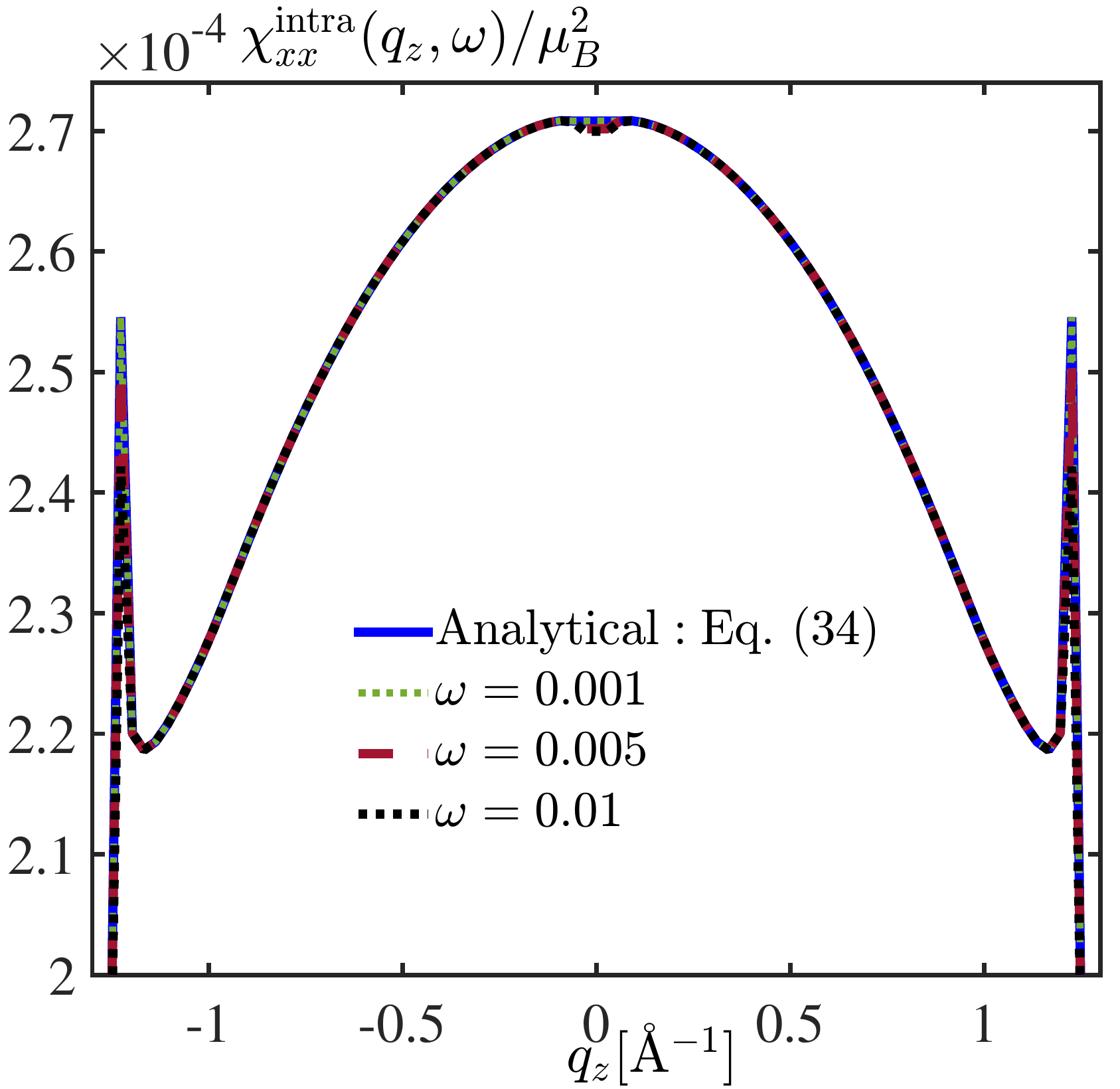}}
                \end{center}
    \caption{Comparison of $\lim_{\omega\rightarrow0}\chi^{\mu\nu}_{ss'}(\bm{q},\omega)$ to the analytical expressions in Eq.~\eqref{chizzfinal} (left panel) and \eqref{chispinxx} (right panel). In both panels we can notice the equivalence of $\lim_{\omega\rightarrow0}\chi^{\mu\nu}_{ss'}(\bm{q},\omega)$ and the analytical forms.}
\label{fig7}
\end{figure*}
\appendix
\section{Spin Susceptibility of Spin-Orbit-Coupled Systems}\label{appa}
In this section we provide a derivation of the non-interacting spin susceptibility of a
spin-orbit-coupled electron gas from the Matsubara Green's function
formalism.  The spin susceptibility can be written as
\begin{eqnarray}
\chi_{ab}(\bq,iq_{n})=-\mu^{2}_{B}\frac{1}{\beta}\sum_{ik_{n}}\sum_{\bk}{\rm Tr}\left\{G_{\bk}(ik_{n})\sigma_{a}\right.\nonumber\\\left. \times G_{\bk+\bq}(ik_{n}+q_{n})\sigma_{b}\right\}\;,\label{greensusc}
\end{eqnarray}
where $\beta=1/(k_{B}T)$, $k_{B}$ is the Boltzmann constant and $T$ is
the temperature, $k_{n}$ and $q_{n}$ are the fermionic Matsubara
frequencies, and Tr denotes a trace. The Matsubara Green's function $G_{\bk}(ik_{n})$ is
\begin{equation}\label{fmatsubara}
  G_{\bk}(ik_{n})=\sum_{\mu=\pm}\frac{|\bk,\mu\rangle\langle\mu,\bk|}{ik_{n}-E_{\bk,\mu}}\;,
\end{equation}
where $\mu = \pm$ denotes the band index and $|\bk,\mu\rangle$ the
spinor wavefunction of the $\mu^{\mathrm{th}}$-band energy eigenstate. Substituting
Eq.~\eqref{fmatsubara} into Eq.~\eqref{greensusc} yields
\begin{eqnarray}
\chi_{ab}(\bq,iq_{n}) =-\mu^{2}_{B}\frac{1}{\beta}\sum_{ik_{n}}\sum_{\bk}\sum_{\mu,\nu}{\rm Tr}\left\{\frac{|\bk,\mu\rangle\langle\mu,\bk|}{ik_{n}-E_{\bk,\mu}}\sigma_{a} \right.\nonumber \\
\left.\times\frac{|\bk+\bq,\nu\rangle\langle\nu,\bk+\bq|}{ik_{n}+iq_{n}-E_{\bk+\bq,\nu}}\sigma_{b}\right\}.\nonumber\\
\end{eqnarray}
Invariance of the trace under cyclic permutations leads to
\begin{eqnarray}
&&\chi_{ab}(\bq,iq_{n})
    =-\mu^{2}_{B}\frac{1}{\beta}\sum_{ik_{n}}\sum_{\bk}\sum_{\mu,\nu}\nonumber\\
   &&\left[\frac{\langle
 \mu,\bk |\sigma_{a}|\bk+\bq, \nu\rangle \langle \nu,\bk+\bq|\sigma_{b}|\bk, \mu\rangle }{(ik_{n}-E_{\bk,\mu})(ik_{n}+iq_{n}-E_{\bk+\bq,\nu})}\right]\;.
\end{eqnarray}
Performing the Matsubara sum, the above reduces to 
\begin{eqnarray}\label{sucmats2}
\chi_{ab}(\bq,iq_{n})
  &=&-\mu^{2}_{B}\sum_{\bk}\sum_{\mu,\nu}\frac{f(E_{\bk,\mu})-f(E_{\bk+\bq,\nu})}{iq_{n}+E_{\bk,\mu}-E_{\bk+\bq,\nu}}\nonumber\\
  &&\times \mathcal{F}^{\mu\nu}_{ab}(\bk,\bk+\bq)\;.
\end{eqnarray}
where $f(E_{\bk,\mu})$ is the Fermi function, and $\mathcal{F}^{\mu\nu}_{ab}(\bk,\bk+\bq)=\langle
\mu,\bk |\sigma_{a}|\bk+\bq, \nu\rangle \langle
\nu,\bk+\bq|\sigma_{b}|\bk, \mu\rangle$ is a form factor.
Analytically continuing to the real frequency
$iq_{n}\rightarrow\omega+i\delta$ ($\delta=0^{+}$), one obtains the
final expression of the retarded spin susceptibility
\begin{eqnarray}\label{spinsucesfeqfinal}
  \chi_{ab}(\bq,\omega)&=&-\mu^{2}_{B}\sum_{\bk}\sum_{\mu,\nu}\frac{f(E_{\bk,\mu})-f(E_{\bk+\bq,\nu})}{E_{\bk,\mu}-E_{\bk+\bq,\nu}+\omega+i\delta}\nonumber\\
  &&\times \mathcal{F}^{\mu\nu}_{ab}(\bk,\bk+\bq).
\end{eqnarray}
Eq.~\eqref{spinsucesfeqfinal} in its static limit,
$\omega\rightarrow0$, reduces to Eq.~\eqref{chispin} for three
dimensions and $\bqpar=0$, and for two dimensions it reduces to Eq.~\eqref{chispinsurf2} in the main text.

\section{Static Spin Susceptibility of Bulk Rashba Semiconductors }\label{app}

In this appendix we discuss the static limit of the spin susceptibility in a bulk Rashba semiconductor. Here we start with the dynamic spin susceptibility, \textit{i.e.},
\begin{eqnarray}\label{chispinomega}
&&\chi^{\mu\nu}_{ab}(\bm{q},\omega) = -\mu^{2}_{B}\int{\frac{d^{3}\bk }{(2\pi)^{3}}}\nonumber\\&&\times\frac{f(E_{{\bm k},\mu})-f(E_{{\bm k}+{\bm q},\nu})}{E_{{\bm k},\mu}-E_{{\bm k}+{\bm q},\nu}+\omega+i\delta}\mathcal{F}^{\mu\nu}_{ab}({\bm k},{\bm k}+{\bm q}).
\end{eqnarray}
The real part of the spin susceptibility is
\begin{eqnarray}\label{chispinomega1}
&&\chi^{\mu\nu}_{ab}(\bm{q},\omega) = -\mu^{2}_{B}\mathcal{P}\int{\frac{d^{3}\bk }{(2\pi)^{3}}}\nonumber\\&&\times\frac{f(E_{{\bm k},\mu})-f(E_{{\bm k}+{\bm q},\nu})}{E_{{\bm k},\mu}-E_{{\bm k}+{\bm q},\nu}+\omega}\mathcal{F}^{\mu\nu}_{ab}({\bm k},{\bm k}+{\bm q}),
\end{eqnarray}
where $\mathcal{P}$ denotes the principal value of the integral. The latter integral, in its most general form, has contribution from complex and real poles, and in order to correctly account for these contributions in the static limit, one needs to consider $\lim_{\omega\rightarrow0}\chi^{\mu\nu}_{ss'}(\bm{q},\omega)$. However, we have two distinct cases that can be treated differently. First: the case in which the complex poles have vanishing contributions as $\omega\rightarrow0$. In this case, we can directly take $\omega\rightarrow0$ in the integral in Eq.~\eqref{chispinomega1}, such that
\begin{eqnarray}\label{chispinomega2}
&&\chi^{\mu\nu}_{ab}(\bm{q},0) = \mu^{2}_{B}\mathcal{P}\int{\frac{d^{3}\bk }{(2\pi)^{3}}}\nonumber\\&&\times\frac{f(E_{{\bm k},\mu})-f(E_{{\bm k}+{\bm q},\nu})}{E_{{\bm k}+{\bm q},\nu}-E_{{\bm k},\mu}}\mathcal{F}^{\mu\nu}_{ab}({\bm k},{\bm k}+{\bm q}),
\end{eqnarray}
 An example of such a case is the 3D electron gas~\cite{RKKY1}. Second: the case in which the complex poles of the integral in Eq.~\eqref{chispinomega1} have non-vanishing contributions as $\omega\rightarrow0$. In this case taking $\omega\rightarrow0$ in Eq.~\eqref{chispinomega1} leads to unphysical results. Instead, one needs to find $\lim_{\omega\rightarrow0}\chi^{\mu\nu}_{ss'}(\bm{q},\omega)$ to correctly account for the contribution from all poles. This scenario is displayed in the Rashba SOC metal in 2D~\cite{static}.

For a bulk Rashba semiconductor, even though it may seem similar to
the 2D Rashba SOC metal, we find that for $(q_{x},q_{y})=(0,0)$ the
contributions of the complex poles vanish as $\omega\rightarrow0$ because of the additional $k_{z}$ dependence, and
that the static spin susceptibility is given by
Eq.~\eqref{chispin2}. We have numerically verified the equivalence of
our analytical forms of the static spin susceptibility,
Eqs.~\eqref{chizzfinal} and \eqref{chispinxx}, and
$\lim_{\omega\rightarrow0}\chi^{\mu\nu}_{ss'}(\bm{q},\omega)$, as
shown in Fig.~\ref{fig7}.

\bibliographystyle{apsrev4-2}
\bibliography{refsv2_3}

\begin{thebibliography}{89}%
\makeatletter
\providecommand \@ifxundefined [1]{%
 \@ifx{#1\undefined}
}%
\providecommand \@ifnum [1]{%
 \ifnum #1\expandafter \@firstoftwo
 \else \expandafter \@secondoftwo
 \fi
}%
\providecommand \@ifx [1]{%
 \ifx #1\expandafter \@firstoftwo
 \else \expandafter \@secondoftwo
 \fi
}%
\providecommand \natexlab [1]{#1}%
\providecommand \enquote  [1]{``#1''}%
\providecommand \bibnamefont  [1]{#1}%
\providecommand \bibfnamefont [1]{#1}%
\providecommand \citenamefont [1]{#1}%
\providecommand \href@noop [0]{\@secondoftwo}%
\providecommand \href [0]{\begingroup \@sanitize@url \@href}%
\providecommand \@href[1]{\@@startlink{#1}\@@href}%
\providecommand \@@href[1]{\endgroup#1\@@endlink}%
\providecommand \@sanitize@url [0]{\catcode `\\12\catcode `\$12\catcode
  `\&12\catcode `\#12\catcode `\^12\catcode `\_12\catcode `\%12\relax}%
\providecommand \@@startlink[1]{}%
\providecommand \@@endlink[0]{}%
\providecommand \url  [0]{\begingroup\@sanitize@url \@url }%
\providecommand \@url [1]{\endgroup\@href {#1}{\urlprefix }}%
\providecommand \urlprefix  [0]{URL }%
\providecommand \Eprint [0]{\href }%
\providecommand \doibase [0]{https://doi.org/}%
\providecommand \selectlanguage [0]{\@gobble}%
\providecommand \bibinfo  [0]{\@secondoftwo}%
\providecommand \bibfield  [0]{\@secondoftwo}%
\providecommand \translation [1]{[#1]}%
\providecommand \BibitemOpen [0]{}%
\providecommand \bibitemStop [0]{}%
\providecommand \bibitemNoStop [0]{.\EOS\space}%
\providecommand \EOS [0]{\spacefactor3000\relax}%
\providecommand \BibitemShut  [1]{\csname bibitem#1\endcsname}%
\let\auto@bib@innerbib\@empty
\bibitem [{\citenamefont {Nitta}\ \emph {et~al.}(1997)\citenamefont {Nitta},
  \citenamefont {Akazaki}, \citenamefont {Takayanagi},\ and\ \citenamefont
  {Enoki}}]{Rashba1}%
  \BibitemOpen
  \bibfield  {author} {\bibinfo {author} {\bibfnamefont {J.}~\bibnamefont
  {Nitta}}, \bibinfo {author} {\bibfnamefont {T.}~\bibnamefont {Akazaki}},
  \bibinfo {author} {\bibfnamefont {H.}~\bibnamefont {Takayanagi}},\ and\
  \bibinfo {author} {\bibfnamefont {T.}~\bibnamefont {Enoki}},\ }\href
  {https://doi.org/10.1103/PhysRevLett.78.1335} {\bibfield  {journal} {\bibinfo
   {journal} {Phys. Rev. Lett.}\ }\textbf {\bibinfo {volume} {78}},\ \bibinfo
  {pages} {1335} (\bibinfo {year} {1997})}\BibitemShut {NoStop}%
\bibitem [{\citenamefont {Ast}\ \emph {et~al.}(2007)\citenamefont {Ast},
  \citenamefont {Henk}, \citenamefont {Ernst}, \citenamefont {Moreschini},
  \citenamefont {Falub}, \citenamefont {Pacil\'e}, \citenamefont {Bruno},
  \citenamefont {Kern},\ and\ \citenamefont {Grioni}}]{Rashba2}%
  \BibitemOpen
  \bibfield  {author} {\bibinfo {author} {\bibfnamefont {C.~R.}\ \bibnamefont
  {Ast}}, \bibinfo {author} {\bibfnamefont {J.}~\bibnamefont {Henk}}, \bibinfo
  {author} {\bibfnamefont {A.}~\bibnamefont {Ernst}}, \bibinfo {author}
  {\bibfnamefont {L.}~\bibnamefont {Moreschini}}, \bibinfo {author}
  {\bibfnamefont {M.~C.}\ \bibnamefont {Falub}}, \bibinfo {author}
  {\bibfnamefont {D.}~\bibnamefont {Pacil\'e}}, \bibinfo {author}
  {\bibfnamefont {P.}~\bibnamefont {Bruno}}, \bibinfo {author} {\bibfnamefont
  {K.}~\bibnamefont {Kern}},\ and\ \bibinfo {author} {\bibfnamefont
  {M.}~\bibnamefont {Grioni}},\ }\href
  {https://doi.org/10.1103/PhysRevLett.98.186807} {\bibfield  {journal}
  {\bibinfo  {journal} {Phys. Rev. Lett.}\ }\textbf {\bibinfo {volume} {98}},\
  \bibinfo {pages} {186807} (\bibinfo {year} {2007})}\BibitemShut {NoStop}%
\bibitem [{\citenamefont {LaShell}\ \emph {et~al.}(1996)\citenamefont
  {LaShell}, \citenamefont {McDougall},\ and\ \citenamefont
  {Jensen}}]{Rashba3}%
  \BibitemOpen
  \bibfield  {author} {\bibinfo {author} {\bibfnamefont {S.}~\bibnamefont
  {LaShell}}, \bibinfo {author} {\bibfnamefont {B.~A.}\ \bibnamefont
  {McDougall}},\ and\ \bibinfo {author} {\bibfnamefont {E.}~\bibnamefont
  {Jensen}},\ }\href {https://doi.org/10.1103/PhysRevLett.77.3419} {\bibfield
  {journal} {\bibinfo  {journal} {Phys. Rev. Lett.}\ }\textbf {\bibinfo
  {volume} {77}},\ \bibinfo {pages} {3419} (\bibinfo {year}
  {1996})}\BibitemShut {NoStop}%
\bibitem [{\citenamefont {Gierz}\ \emph {et~al.}(2009)\citenamefont {Gierz},
  \citenamefont {Suzuki}, \citenamefont {Frantzeskakis}, \citenamefont {Pons},
  \citenamefont {Ostanin}, \citenamefont {Ernst}, \citenamefont {Henk},
  \citenamefont {Grioni}, \citenamefont {Kern},\ and\ \citenamefont
  {Ast}}]{Rashba4}%
  \BibitemOpen
  \bibfield  {author} {\bibinfo {author} {\bibfnamefont {I.}~\bibnamefont
  {Gierz}}, \bibinfo {author} {\bibfnamefont {T.}~\bibnamefont {Suzuki}},
  \bibinfo {author} {\bibfnamefont {E.}~\bibnamefont {Frantzeskakis}}, \bibinfo
  {author} {\bibfnamefont {S.}~\bibnamefont {Pons}}, \bibinfo {author}
  {\bibfnamefont {S.}~\bibnamefont {Ostanin}}, \bibinfo {author} {\bibfnamefont
  {A.}~\bibnamefont {Ernst}}, \bibinfo {author} {\bibfnamefont
  {J.}~\bibnamefont {Henk}}, \bibinfo {author} {\bibfnamefont {M.}~\bibnamefont
  {Grioni}}, \bibinfo {author} {\bibfnamefont {K.}~\bibnamefont {Kern}},\ and\
  \bibinfo {author} {\bibfnamefont {C.~R.}\ \bibnamefont {Ast}},\ }\href
  {https://doi.org/10.1103/PhysRevLett.103.046803} {\bibfield  {journal}
  {\bibinfo  {journal} {Phys. Rev. Lett.}\ }\textbf {\bibinfo {volume} {103}},\
  \bibinfo {pages} {046803} (\bibinfo {year} {2009})}\BibitemShut {NoStop}%
\bibitem [{\citenamefont {Koroteev}\ \emph {et~al.}(2004)\citenamefont
  {Koroteev}, \citenamefont {Bihlmayer}, \citenamefont {Gayone}, \citenamefont
  {Chulkov}, \citenamefont {Bl\"ugel}, \citenamefont {Echenique},\ and\
  \citenamefont {Hofmann}}]{Rashba5}%
  \BibitemOpen
  \bibfield  {author} {\bibinfo {author} {\bibfnamefont {Y.~M.}\ \bibnamefont
  {Koroteev}}, \bibinfo {author} {\bibfnamefont {G.}~\bibnamefont {Bihlmayer}},
  \bibinfo {author} {\bibfnamefont {J.~E.}\ \bibnamefont {Gayone}}, \bibinfo
  {author} {\bibfnamefont {E.~V.}\ \bibnamefont {Chulkov}}, \bibinfo {author}
  {\bibfnamefont {S.}~\bibnamefont {Bl\"ugel}}, \bibinfo {author}
  {\bibfnamefont {P.~M.}\ \bibnamefont {Echenique}},\ and\ \bibinfo {author}
  {\bibfnamefont {P.}~\bibnamefont {Hofmann}},\ }\href
  {https://doi.org/10.1103/PhysRevLett.93.046403} {\bibfield  {journal}
  {\bibinfo  {journal} {Phys. Rev. Lett.}\ }\textbf {\bibinfo {volume} {93}},\
  \bibinfo {pages} {046403} (\bibinfo {year} {2004})}\BibitemShut {NoStop}%
\bibitem [{\citenamefont {Santander-Syro}\ \emph {et~al.}(2014)\citenamefont
  {Santander-Syro}, \citenamefont {Fortuna}, \citenamefont {Bareille},
  \citenamefont {R{\"{o}}del}, \citenamefont {Landolt}, \citenamefont {Plumb},
  \citenamefont {Dil},\ and\ \citenamefont {Radovic}}]{Rashba6}%
  \BibitemOpen
  \bibfield  {author} {\bibinfo {author} {\bibfnamefont {A.~F.}\ \bibnamefont
  {Santander-Syro}}, \bibinfo {author} {\bibfnamefont {F.}~\bibnamefont
  {Fortuna}}, \bibinfo {author} {\bibfnamefont {C.}~\bibnamefont {Bareille}},
  \bibinfo {author} {\bibfnamefont {T.~C.}\ \bibnamefont {R{\"{o}}del}},
  \bibinfo {author} {\bibfnamefont {G.}~\bibnamefont {Landolt}}, \bibinfo
  {author} {\bibfnamefont {N.~C.}\ \bibnamefont {Plumb}}, \bibinfo {author}
  {\bibfnamefont {J.~H.}\ \bibnamefont {Dil}},\ and\ \bibinfo {author}
  {\bibfnamefont {M.}~\bibnamefont {Radovic}},\ }\href
  {https://doi.org/10.1038/nmat4107} {\bibfield  {journal} {\bibinfo  {journal}
  {Nature Materials}\ }\textbf {\bibinfo {volume} {13}},\ \bibinfo {pages}
  {1085} (\bibinfo {year} {2014})}\BibitemShut {NoStop}%
\bibitem [{\citenamefont {Bychkov}\ and\ \citenamefont
  {Rashba}(1984)}]{Rashba}%
  \BibitemOpen
  \bibfield  {author} {\bibinfo {author} {\bibfnamefont {Y.~A.}\ \bibnamefont
  {Bychkov}}\ and\ \bibinfo {author} {\bibfnamefont {E.~I.}\ \bibnamefont
  {Rashba}},\ }\href {http://www.jetpletters.ac.ru/ps/1264/article_19121.shtml}
  {\bibfield  {journal} {\bibinfo  {journal} {JETP Lett.}\ }\textbf {\bibinfo
  {volume} {39}},\ \bibinfo {pages} {78} (\bibinfo {year} {1984})}\BibitemShut
  {NoStop}%
\bibitem [{\citenamefont {Bahramy}\ and\ \citenamefont {Ogawa}(2017)}]{BiTeI1}%
  \BibitemOpen
  \bibfield  {author} {\bibinfo {author} {\bibfnamefont {M.~S.}\ \bibnamefont
  {Bahramy}}\ and\ \bibinfo {author} {\bibfnamefont {N.}~\bibnamefont
  {Ogawa}},\ }\bibfield  {journal} {\bibinfo  {journal} {Advanced Materials}\
  }\textbf {\bibinfo {volume} {29}},\ \href
  {https://doi.org/10.1002/adma.201605911} {10.1002/adma.201605911} (\bibinfo
  {year} {2017})\BibitemShut {NoStop}%
\bibitem [{\citenamefont {Ishizaka}\ \emph {et~al.}(2011)\citenamefont
  {Ishizaka}, \citenamefont {Bahramy}, \citenamefont {Murakawa}, \citenamefont
  {Sakano}, \citenamefont {Shimojima}, \citenamefont {Sonobe}, \citenamefont
  {Koizumi}, \citenamefont {Shin}, \citenamefont {Miyahara}, \citenamefont
  {Kimura}, \citenamefont {Miyamoto}, \citenamefont {Okuda}, \citenamefont
  {Namatame}, \citenamefont {Taniguchi}, \citenamefont {Arita}, \citenamefont
  {Nagaosa}, \citenamefont {Kobayashi}, \citenamefont {Murakami}, \citenamefont
  {Kumai}, \citenamefont {Kaneko},\ and\ \citenamefont {Onose}}]{BiTeI2}%
  \BibitemOpen
  \bibfield  {author} {\bibinfo {author} {\bibfnamefont {K.}~\bibnamefont
  {Ishizaka}}, \bibinfo {author} {\bibfnamefont {M.~S.}\ \bibnamefont
  {Bahramy}}, \bibinfo {author} {\bibfnamefont {H.}~\bibnamefont {Murakawa}},
  \bibinfo {author} {\bibfnamefont {M.}~\bibnamefont {Sakano}}, \bibinfo
  {author} {\bibfnamefont {T.}~\bibnamefont {Shimojima}}, \bibinfo {author}
  {\bibfnamefont {T.}~\bibnamefont {Sonobe}}, \bibinfo {author} {\bibfnamefont
  {K.}~\bibnamefont {Koizumi}}, \bibinfo {author} {\bibfnamefont
  {S.}~\bibnamefont {Shin}}, \bibinfo {author} {\bibfnamefont {H.}~\bibnamefont
  {Miyahara}}, \bibinfo {author} {\bibfnamefont {A.}~\bibnamefont {Kimura}},
  \bibinfo {author} {\bibfnamefont {K.}~\bibnamefont {Miyamoto}}, \bibinfo
  {author} {\bibfnamefont {T.}~\bibnamefont {Okuda}}, \bibinfo {author}
  {\bibfnamefont {H.}~\bibnamefont {Namatame}}, \bibinfo {author}
  {\bibfnamefont {M.}~\bibnamefont {Taniguchi}}, \bibinfo {author}
  {\bibfnamefont {R.}~\bibnamefont {Arita}}, \bibinfo {author} {\bibfnamefont
  {N.}~\bibnamefont {Nagaosa}}, \bibinfo {author} {\bibfnamefont
  {K.}~\bibnamefont {Kobayashi}}, \bibinfo {author} {\bibfnamefont
  {Y.}~\bibnamefont {Murakami}}, \bibinfo {author} {\bibfnamefont
  {R.}~\bibnamefont {Kumai}}, \bibinfo {author} {\bibfnamefont
  {Y.}~\bibnamefont {Kaneko}},\ and\ \bibinfo {author} {\bibfnamefont
  {Y.}~\bibnamefont {Onose}},\ }\href {https://doi.org/10.1038/nmat3051}
  {\bibfield  {journal} {\bibinfo  {journal} {Nature Materials}\ }\textbf
  {\bibinfo {volume} {10}},\ \bibinfo {pages} {521} (\bibinfo {year}
  {2011})}\BibitemShut {NoStop}%
\bibitem [{\citenamefont {Murakawa}\ \emph {et~al.}(2013)\citenamefont
  {Murakawa}, \citenamefont {Bahramy}, \citenamefont {Tokunaga}, \citenamefont
  {Kohama}, \citenamefont {Bell}, \citenamefont {Kaneko}, \citenamefont
  {Nagaosa}, \citenamefont {Hwang},\ and\ \citenamefont {Tokura}}]{BiTeI3}%
  \BibitemOpen
  \bibfield  {author} {\bibinfo {author} {\bibfnamefont {H.}~\bibnamefont
  {Murakawa}}, \bibinfo {author} {\bibfnamefont {M.~S.}\ \bibnamefont
  {Bahramy}}, \bibinfo {author} {\bibfnamefont {M.}~\bibnamefont {Tokunaga}},
  \bibinfo {author} {\bibfnamefont {Y.}~\bibnamefont {Kohama}}, \bibinfo
  {author} {\bibfnamefont {C.}~\bibnamefont {Bell}}, \bibinfo {author}
  {\bibfnamefont {Y.}~\bibnamefont {Kaneko}}, \bibinfo {author} {\bibfnamefont
  {N.}~\bibnamefont {Nagaosa}}, \bibinfo {author} {\bibfnamefont {H.~Y.}\
  \bibnamefont {Hwang}},\ and\ \bibinfo {author} {\bibfnamefont
  {Y.}~\bibnamefont {Tokura}},\ }\href
  {https://doi.org/10.1126/science.1242247} {\bibfield  {journal} {\bibinfo
  {journal} {Science}\ }\textbf {\bibinfo {volume} {342}},\ \bibinfo {pages}
  {1490} (\bibinfo {year} {2013})}\BibitemShut {NoStop}%
\bibitem [{\citenamefont {Kanou}\ and\ \citenamefont
  {Sasagawa}(2013)}]{BiTeI4}%
  \BibitemOpen
  \bibfield  {author} {\bibinfo {author} {\bibfnamefont {M.}~\bibnamefont
  {Kanou}}\ and\ \bibinfo {author} {\bibfnamefont {T.}~\bibnamefont
  {Sasagawa}},\ }\href {http://stacks.iop.org/0953-8984/25/i=13/a=135801}
  {\bibfield  {journal} {\bibinfo  {journal} {Journal of Physics: Condensed
  Matter}\ }\textbf {\bibinfo {volume} {25}},\ \bibinfo {pages} {135801}
  (\bibinfo {year} {2013})}\BibitemShut {NoStop}%
\bibitem [{\citenamefont {Fülöp}\ \emph {et~al.}(2018)\citenamefont
  {Fülöp}, \citenamefont {Tajkov}, \citenamefont {Pető}, \citenamefont
  {Kun}, \citenamefont {Koltai}, \citenamefont {Oroszlány}, \citenamefont
  {Tóvári}, \citenamefont {Murakawa}, \citenamefont {Tokura}, \citenamefont
  {Bordács}, \citenamefont {Tapasztó},\ and\ \citenamefont
  {Csonka}}]{BiTeI5}%
  \BibitemOpen
  \bibfield  {author} {\bibinfo {author} {\bibfnamefont {B.}~\bibnamefont
  {Fülöp}}, \bibinfo {author} {\bibfnamefont {Z.}~\bibnamefont {Tajkov}},
  \bibinfo {author} {\bibfnamefont {J.}~\bibnamefont {Pető}}, \bibinfo
  {author} {\bibfnamefont {P.}~\bibnamefont {Kun}}, \bibinfo {author}
  {\bibfnamefont {J.}~\bibnamefont {Koltai}}, \bibinfo {author} {\bibfnamefont
  {L.}~\bibnamefont {Oroszlány}}, \bibinfo {author} {\bibfnamefont
  {E.}~\bibnamefont {Tóvári}}, \bibinfo {author} {\bibfnamefont
  {H.}~\bibnamefont {Murakawa}}, \bibinfo {author} {\bibfnamefont
  {Y.}~\bibnamefont {Tokura}}, \bibinfo {author} {\bibfnamefont
  {S.}~\bibnamefont {Bordács}}, \bibinfo {author} {\bibfnamefont
  {L.}~\bibnamefont {Tapasztó}},\ and\ \bibinfo {author} {\bibfnamefont
  {S.}~\bibnamefont {Csonka}},\ }\href
  {http://stacks.iop.org/2053-1583/5/i=3/a=031013} {\bibfield  {journal}
  {\bibinfo  {journal} {2D Materials}\ }\textbf {\bibinfo {volume} {5}},\
  \bibinfo {pages} {031013} (\bibinfo {year} {2018})}\BibitemShut {NoStop}%
\bibitem [{\citenamefont {Bahramy}\ \emph {et~al.}(2011)\citenamefont
  {Bahramy}, \citenamefont {Arita},\ and\ \citenamefont {Nagaosa}}]{BiTeI6}%
  \BibitemOpen
  \bibfield  {author} {\bibinfo {author} {\bibfnamefont {M.~S.}\ \bibnamefont
  {Bahramy}}, \bibinfo {author} {\bibfnamefont {R.}~\bibnamefont {Arita}},\
  and\ \bibinfo {author} {\bibfnamefont {N.}~\bibnamefont {Nagaosa}},\ }\href
  {https://doi.org/10.1103/PhysRevB.84.041202} {\bibfield  {journal} {\bibinfo
  {journal} {Phys. Rev. B}\ }\textbf {\bibinfo {volume} {84}},\ \bibinfo
  {pages} {041202} (\bibinfo {year} {2011})}\BibitemShut {NoStop}%
\bibitem [{\citenamefont {Bahramy}\ \emph {et~al.}(2012)\citenamefont
  {Bahramy}, \citenamefont {Yang}, \citenamefont {Arita},\ and\ \citenamefont
  {Nagaosa}}]{BiTeITop}%
  \BibitemOpen
  \bibfield  {author} {\bibinfo {author} {\bibfnamefont {M.~S.}\ \bibnamefont
  {Bahramy}}, \bibinfo {author} {\bibfnamefont {B.}~\bibnamefont {Yang}},
  \bibinfo {author} {\bibfnamefont {R.}~\bibnamefont {Arita}},\ and\ \bibinfo
  {author} {\bibfnamefont {N.}~\bibnamefont {Nagaosa}},\ }\href
  {https://doi.org/10.1038/ncomms1679} {\bibfield  {journal} {\bibinfo
  {journal} {Nature Communications}\ }\textbf {\bibinfo {volume} {3}},\
  \bibinfo {pages} {677} (\bibinfo {year} {2012})}\BibitemShut {NoStop}%
\bibitem [{\citenamefont {Qi}\ \emph {et~al.}(2017)\citenamefont {Qi},
  \citenamefont {Shi}, \citenamefont {Naumov}, \citenamefont {Kumar},
  \citenamefont {Sankar}, \citenamefont {Schnelle}, \citenamefont {Shekhar},
  \citenamefont {Chou}, \citenamefont {Felser}, \citenamefont {Yan},\ and\
  \citenamefont {Medvedev}}]{BiTeITop1}%
  \BibitemOpen
  \bibfield  {author} {\bibinfo {author} {\bibfnamefont {Y.}~\bibnamefont
  {Qi}}, \bibinfo {author} {\bibfnamefont {W.}~\bibnamefont {Shi}}, \bibinfo
  {author} {\bibfnamefont {P.~G.}\ \bibnamefont {Naumov}}, \bibinfo {author}
  {\bibfnamefont {N.}~\bibnamefont {Kumar}}, \bibinfo {author} {\bibfnamefont
  {R.}~\bibnamefont {Sankar}}, \bibinfo {author} {\bibfnamefont
  {W.}~\bibnamefont {Schnelle}}, \bibinfo {author} {\bibfnamefont
  {C.}~\bibnamefont {Shekhar}}, \bibinfo {author} {\bibfnamefont {F.-C.}\
  \bibnamefont {Chou}}, \bibinfo {author} {\bibfnamefont {C.}~\bibnamefont
  {Felser}}, \bibinfo {author} {\bibfnamefont {B.}~\bibnamefont {Yan}},\ and\
  \bibinfo {author} {\bibfnamefont {S.~A.}\ \bibnamefont {Medvedev}},\ }\href
  {https://doi.org/10.1002/adma.201605965} {\bibfield  {journal} {\bibinfo
  {journal} {Advanced Materials}\ }\textbf {\bibinfo {volume} {29}},\ \bibinfo
  {pages} {1605965} (\bibinfo {year} {2017})}\BibitemShut {NoStop}%
\bibitem [{\citenamefont {Xi}\ \emph {et~al.}(2013)\citenamefont {Xi},
  \citenamefont {Ma}, \citenamefont {Liu}, \citenamefont {Chen}, \citenamefont
  {Ku}, \citenamefont {Berger}, \citenamefont {Martin}, \citenamefont
  {Tanner},\ and\ \citenamefont {Carr}}]{BiTeITopExp}%
  \BibitemOpen
  \bibfield  {author} {\bibinfo {author} {\bibfnamefont {X.}~\bibnamefont
  {Xi}}, \bibinfo {author} {\bibfnamefont {C.}~\bibnamefont {Ma}}, \bibinfo
  {author} {\bibfnamefont {Z.}~\bibnamefont {Liu}}, \bibinfo {author}
  {\bibfnamefont {Z.}~\bibnamefont {Chen}}, \bibinfo {author} {\bibfnamefont
  {W.}~\bibnamefont {Ku}}, \bibinfo {author} {\bibfnamefont {H.}~\bibnamefont
  {Berger}}, \bibinfo {author} {\bibfnamefont {C.}~\bibnamefont {Martin}},
  \bibinfo {author} {\bibfnamefont {D.~B.}\ \bibnamefont {Tanner}},\ and\
  \bibinfo {author} {\bibfnamefont {G.~L.}\ \bibnamefont {Carr}},\ }\href
  {https://doi.org/10.1103/PhysRevLett.111.155701} {\bibfield  {journal}
  {\bibinfo  {journal} {Phys. Rev. Lett.}\ }\textbf {\bibinfo {volume} {111}},\
  \bibinfo {pages} {155701} (\bibinfo {year} {2013})}\BibitemShut {NoStop}%
\bibitem [{\citenamefont {VanGennep}\ \emph {et~al.}(2017)\citenamefont
  {VanGennep}, \citenamefont {Linscheid}, \citenamefont {Jackson},
  \citenamefont {Weir}, \citenamefont {Vohra}, \citenamefont {Berger},
  \citenamefont {Stewart}, \citenamefont {Hennig}, \citenamefont {Hirschfeld},\
  and\ \citenamefont {Hamlin}}]{BiTeITopExp3}%
  \BibitemOpen
  \bibfield  {author} {\bibinfo {author} {\bibfnamefont {D.}~\bibnamefont
  {VanGennep}}, \bibinfo {author} {\bibfnamefont {A.}~\bibnamefont
  {Linscheid}}, \bibinfo {author} {\bibfnamefont {D.~E.}\ \bibnamefont
  {Jackson}}, \bibinfo {author} {\bibfnamefont {S.~T.}\ \bibnamefont {Weir}},
  \bibinfo {author} {\bibfnamefont {Y.~K.}\ \bibnamefont {Vohra}}, \bibinfo
  {author} {\bibfnamefont {H.}~\bibnamefont {Berger}}, \bibinfo {author}
  {\bibfnamefont {G.~R.}\ \bibnamefont {Stewart}}, \bibinfo {author}
  {\bibfnamefont {R.~G.}\ \bibnamefont {Hennig}}, \bibinfo {author}
  {\bibfnamefont {P.~J.}\ \bibnamefont {Hirschfeld}},\ and\ \bibinfo {author}
  {\bibfnamefont {J.~J.}\ \bibnamefont {Hamlin}},\ }\href
  {https://doi.org/10.1088/1361-648x/aa5567} {\bibfield  {journal} {\bibinfo
  {journal} {Journal of Physics: Condensed Matter}\ }\textbf {\bibinfo {volume}
  {29}},\ \bibinfo {pages} {09LT02} (\bibinfo {year} {2017})}\BibitemShut
  {NoStop}%
\bibitem [{\citenamefont {Tran}\ \emph {et~al.}(2014)\citenamefont {Tran},
  \citenamefont {Levallois}, \citenamefont {Lerch}, \citenamefont {Teyssier},
  \citenamefont {Kuzmenko}, \citenamefont {Aut\`es}, \citenamefont {Yazyev},
  \citenamefont {Ubaldini}, \citenamefont {Giannini}, \citenamefont {van~der
  Marel},\ and\ \citenamefont {Akrap}}]{BiTeITopExp2}%
  \BibitemOpen
  \bibfield  {author} {\bibinfo {author} {\bibfnamefont {M.~K.}\ \bibnamefont
  {Tran}}, \bibinfo {author} {\bibfnamefont {J.}~\bibnamefont {Levallois}},
  \bibinfo {author} {\bibfnamefont {P.}~\bibnamefont {Lerch}}, \bibinfo
  {author} {\bibfnamefont {J.}~\bibnamefont {Teyssier}}, \bibinfo {author}
  {\bibfnamefont {A.~B.}\ \bibnamefont {Kuzmenko}}, \bibinfo {author}
  {\bibfnamefont {G.}~\bibnamefont {Aut\`es}}, \bibinfo {author} {\bibfnamefont
  {O.~V.}\ \bibnamefont {Yazyev}}, \bibinfo {author} {\bibfnamefont
  {A.}~\bibnamefont {Ubaldini}}, \bibinfo {author} {\bibfnamefont
  {E.}~\bibnamefont {Giannini}}, \bibinfo {author} {\bibfnamefont
  {D.}~\bibnamefont {van~der Marel}},\ and\ \bibinfo {author} {\bibfnamefont
  {A.}~\bibnamefont {Akrap}},\ }\href
  {https://doi.org/10.1103/PhysRevLett.112.047402} {\bibfield  {journal}
  {\bibinfo  {journal} {Phys. Rev. Lett.}\ }\textbf {\bibinfo {volume} {112}},\
  \bibinfo {pages} {047402} (\bibinfo {year} {2014})}\BibitemShut {NoStop}%
\bibitem [{\citenamefont {Facio}\ \emph {et~al.}(2018)\citenamefont {Facio},
  \citenamefont {Efremov}, \citenamefont {Koepernik}, \citenamefont {You},
  \citenamefont {Sodemann},\ and\ \citenamefont {van~den
  Brink}}]{BiTeITopTheory}%
  \BibitemOpen
  \bibfield  {author} {\bibinfo {author} {\bibfnamefont {J.~I.}\ \bibnamefont
  {Facio}}, \bibinfo {author} {\bibfnamefont {D.}~\bibnamefont {Efremov}},
  \bibinfo {author} {\bibfnamefont {K.}~\bibnamefont {Koepernik}}, \bibinfo
  {author} {\bibfnamefont {J.-S.}\ \bibnamefont {You}}, \bibinfo {author}
  {\bibfnamefont {I.}~\bibnamefont {Sodemann}},\ and\ \bibinfo {author}
  {\bibfnamefont {J.}~\bibnamefont {van~den Brink}},\ }\href
  {https://doi.org/10.1103/PhysRevLett.121.246403} {\bibfield  {journal}
  {\bibinfo  {journal} {Phys. Rev. Lett.}\ }\textbf {\bibinfo {volume} {121}},\
  \bibinfo {pages} {246403} (\bibinfo {year} {2018})}\BibitemShut {NoStop}%
\bibitem [{\citenamefont {Liu}\ and\ \citenamefont
  {Vanderbilt}(2014)}]{BiTeITopTheory1}%
  \BibitemOpen
  \bibfield  {author} {\bibinfo {author} {\bibfnamefont {J.}~\bibnamefont
  {Liu}}\ and\ \bibinfo {author} {\bibfnamefont {D.}~\bibnamefont
  {Vanderbilt}},\ }\href {https://doi.org/10.1103/PhysRevB.90.155316}
  {\bibfield  {journal} {\bibinfo  {journal} {Phys. Rev. B}\ }\textbf {\bibinfo
  {volume} {90}},\ \bibinfo {pages} {155316} (\bibinfo {year}
  {2014})}\BibitemShut {NoStop}%
\bibitem [{\citenamefont {Rusinov}\ \emph {et~al.}(2016)\citenamefont
  {Rusinov}, \citenamefont {Menshchikova}, \citenamefont {Sklyadneva},
  \citenamefont {Heid}, \citenamefont {Bohnen},\ and\ \citenamefont
  {Chulkov}}]{BiTeITopTheory2}%
  \BibitemOpen
  \bibfield  {author} {\bibinfo {author} {\bibfnamefont {I.~P.}\ \bibnamefont
  {Rusinov}}, \bibinfo {author} {\bibfnamefont {T.~V.}\ \bibnamefont
  {Menshchikova}}, \bibinfo {author} {\bibfnamefont {I.~Y.}\ \bibnamefont
  {Sklyadneva}}, \bibinfo {author} {\bibfnamefont {R.}~\bibnamefont {Heid}},
  \bibinfo {author} {\bibfnamefont {K.-P.}\ \bibnamefont {Bohnen}},\ and\
  \bibinfo {author} {\bibfnamefont {E.~V.}\ \bibnamefont {Chulkov}},\ }\href
  {https://doi.org/10.1088/1367-2630/18/11/113003} {\bibfield  {journal}
  {\bibinfo  {journal} {New Journal of Physics}\ }\textbf {\bibinfo {volume}
  {18}},\ \bibinfo {pages} {113003} (\bibinfo {year} {2016})}\BibitemShut
  {NoStop}%
\bibitem [{\citenamefont {Joshi}(2016)}]{Spintronics1}%
  \BibitemOpen
  \bibfield  {author} {\bibinfo {author} {\bibfnamefont {V.~K.}\ \bibnamefont
  {Joshi}},\ }\href
  {https://doi.org/https://doi.org/10.1016/j.jestch.2016.05.002} {\bibfield
  {journal} {\bibinfo  {journal} {Engineering Science and Technology, an
  International Journal}\ }\textbf {\bibinfo {volume} {19}},\ \bibinfo {pages}
  {1503 } (\bibinfo {year} {2016})}\BibitemShut {NoStop}%
\bibitem [{\citenamefont {\ifmmode \check{Z}\else
  \v{Z}\fi{}uti\ifmmode~\acute{c}\else \'{c}\fi{}}\ \emph
  {et~al.}(2004)\citenamefont {\ifmmode \check{Z}\else
  \v{Z}\fi{}uti\ifmmode~\acute{c}\else \'{c}\fi{}}, \citenamefont {Fabian},\
  and\ \citenamefont {Das~Sarma}}]{Spintronics2}%
  \BibitemOpen
  \bibfield  {author} {\bibinfo {author} {\bibfnamefont {I.}~\bibnamefont
  {\ifmmode \check{Z}\else \v{Z}\fi{}uti\ifmmode~\acute{c}\else \'{c}\fi{}}},
  \bibinfo {author} {\bibfnamefont {J.}~\bibnamefont {Fabian}},\ and\ \bibinfo
  {author} {\bibfnamefont {S.}~\bibnamefont {Das~Sarma}},\ }\href
  {https://doi.org/10.1103/RevModPhys.76.323} {\bibfield  {journal} {\bibinfo
  {journal} {Rev. Mod. Phys.}\ }\textbf {\bibinfo {volume} {76}},\ \bibinfo
  {pages} {323} (\bibinfo {year} {2004})}\BibitemShut {NoStop}%
\bibitem [{\citenamefont {Hasan}\ and\ \citenamefont {Kane}(2010)}]{Rev1}%
  \BibitemOpen
  \bibfield  {author} {\bibinfo {author} {\bibfnamefont {M.~Z.}\ \bibnamefont
  {Hasan}}\ and\ \bibinfo {author} {\bibfnamefont {C.~L.}\ \bibnamefont
  {Kane}},\ }\href {https://doi.org/10.1103/RevModPhys.82.3045} {\bibfield
  {journal} {\bibinfo  {journal} {Rev. Mod. Phys.}\ }\textbf {\bibinfo {volume}
  {82}},\ \bibinfo {pages} {3045} (\bibinfo {year} {2010})}\BibitemShut
  {NoStop}%
\bibitem [{\citenamefont {Qi}\ and\ \citenamefont {Zhang}(2011)}]{Rev2}%
  \BibitemOpen
  \bibfield  {author} {\bibinfo {author} {\bibfnamefont {X.-L.}\ \bibnamefont
  {Qi}}\ and\ \bibinfo {author} {\bibfnamefont {S.-C.}\ \bibnamefont {Zhang}},\
  }\href {https://doi.org/10.1103/RevModPhys.83.1057} {\bibfield  {journal}
  {\bibinfo  {journal} {Rev. Mod. Phys.}\ }\textbf {\bibinfo {volume} {83}},\
  \bibinfo {pages} {1057} (\bibinfo {year} {2011})}\BibitemShut {NoStop}%
\bibitem [{\citenamefont {Bernevig}\ and\ \citenamefont
  {Hughes}(2013)}]{Bernevig}%
  \BibitemOpen
  \bibfield  {author} {\bibinfo {author} {\bibfnamefont {B.~A.}\ \bibnamefont
  {Bernevig}}\ and\ \bibinfo {author} {\bibfnamefont {T.~L.}\ \bibnamefont
  {Hughes}},\ }\href {https://press.princeton.edu/titles/10039.html} {\emph
  {\bibinfo {title} {Topological Insulators and Topological Superconductors}}}\
  (\bibinfo  {publisher} {Princeton University Press},\ \bibinfo {address} {New
  York},\ \bibinfo {year} {2013})\BibitemShut {NoStop}%
\bibitem [{\citenamefont {Gr\"unberg}\ \emph {et~al.}(1986)\citenamefont
  {Gr\"unberg}, \citenamefont {Schreiber}, \citenamefont {Pang}, \citenamefont
  {Brodsky},\ and\ \citenamefont {Sowers}}]{layered1}%
  \BibitemOpen
  \bibfield  {author} {\bibinfo {author} {\bibfnamefont {P.}~\bibnamefont
  {Gr\"unberg}}, \bibinfo {author} {\bibfnamefont {R.}~\bibnamefont
  {Schreiber}}, \bibinfo {author} {\bibfnamefont {Y.}~\bibnamefont {Pang}},
  \bibinfo {author} {\bibfnamefont {M.~B.}\ \bibnamefont {Brodsky}},\ and\
  \bibinfo {author} {\bibfnamefont {H.}~\bibnamefont {Sowers}},\ }\href
  {https://doi.org/10.1103/PhysRevLett.57.2442} {\bibfield  {journal} {\bibinfo
   {journal} {Phys. Rev. Lett.}\ }\textbf {\bibinfo {volume} {57}},\ \bibinfo
  {pages} {2442} (\bibinfo {year} {1986})}\BibitemShut {NoStop}%
\bibitem [{\citenamefont {Baibich}\ \emph {et~al.}(1988)\citenamefont
  {Baibich}, \citenamefont {Broto}, \citenamefont {Fert}, \citenamefont
  {Van~Dau}, \citenamefont {Petroff}, \citenamefont {Etienne}, \citenamefont
  {Creuzet}, \citenamefont {Friederich},\ and\ \citenamefont
  {Chazelas}}]{layered2}%
  \BibitemOpen
  \bibfield  {author} {\bibinfo {author} {\bibfnamefont {M.~N.}\ \bibnamefont
  {Baibich}}, \bibinfo {author} {\bibfnamefont {J.~M.}\ \bibnamefont {Broto}},
  \bibinfo {author} {\bibfnamefont {A.}~\bibnamefont {Fert}}, \bibinfo {author}
  {\bibfnamefont {F.~N.}\ \bibnamefont {Van~Dau}}, \bibinfo {author}
  {\bibfnamefont {F.}~\bibnamefont {Petroff}}, \bibinfo {author} {\bibfnamefont
  {P.}~\bibnamefont {Etienne}}, \bibinfo {author} {\bibfnamefont
  {G.}~\bibnamefont {Creuzet}}, \bibinfo {author} {\bibfnamefont
  {A.}~\bibnamefont {Friederich}},\ and\ \bibinfo {author} {\bibfnamefont
  {J.}~\bibnamefont {Chazelas}},\ }\href
  {https://doi.org/10.1103/PhysRevLett.61.2472} {\bibfield  {journal} {\bibinfo
   {journal} {Phys. Rev. Lett.}\ }\textbf {\bibinfo {volume} {61}},\ \bibinfo
  {pages} {2472} (\bibinfo {year} {1988})}\BibitemShut {NoStop}%
\bibitem [{\citenamefont {Ruderman}\ and\ \citenamefont
  {Kittel}(1954)}]{RKKY1}%
  \BibitemOpen
  \bibfield  {author} {\bibinfo {author} {\bibfnamefont {M.~A.}\ \bibnamefont
  {Ruderman}}\ and\ \bibinfo {author} {\bibfnamefont {C.}~\bibnamefont
  {Kittel}},\ }\href {https://doi.org/10.1103/PhysRev.96.99} {\bibfield
  {journal} {\bibinfo  {journal} {Phys. Rev.}\ }\textbf {\bibinfo {volume}
  {96}},\ \bibinfo {pages} {99} (\bibinfo {year} {1954})}\BibitemShut {NoStop}%
\bibitem [{\citenamefont {Kasuya}(1956)}]{RKKY2}%
  \BibitemOpen
  \bibfield  {author} {\bibinfo {author} {\bibfnamefont {T.}~\bibnamefont
  {Kasuya}},\ }\href {https://doi.org/10.1143/PTP.16.45} {\bibfield  {journal}
  {\bibinfo  {journal} {Progress of Theoretical Physics}\ }\textbf {\bibinfo
  {volume} {16}},\ \bibinfo {pages} {45} (\bibinfo {year} {1956})}\BibitemShut
  {NoStop}%
\bibitem [{\citenamefont {Yosida}(1957)}]{RKKY3}%
  \BibitemOpen
  \bibfield  {author} {\bibinfo {author} {\bibfnamefont {K.}~\bibnamefont
  {Yosida}},\ }\href {https://doi.org/10.1103/PhysRev.106.893} {\bibfield
  {journal} {\bibinfo  {journal} {Phys. Rev.}\ }\textbf {\bibinfo {volume}
  {106}},\ \bibinfo {pages} {893} (\bibinfo {year} {1957})}\BibitemShut
  {NoStop}%
\bibitem [{\citenamefont {Black-Schaffer}(2010)}]{Graphenerkky}%
  \BibitemOpen
  \bibfield  {author} {\bibinfo {author} {\bibfnamefont {A.~M.}\ \bibnamefont
  {Black-Schaffer}},\ }\href {https://doi.org/10.1103/PhysRevB.81.205416}
  {\bibfield  {journal} {\bibinfo  {journal} {Phys. Rev. B}\ }\textbf {\bibinfo
  {volume} {81}},\ \bibinfo {pages} {205416} (\bibinfo {year}
  {2010})}\BibitemShut {NoStop}%
\bibitem [{\citenamefont {Hatami}\ \emph {et~al.}(2014)\citenamefont {Hatami},
  \citenamefont {Kernreiter},\ and\ \citenamefont {Z\"ulicke}}]{MOS2RKKY}%
  \BibitemOpen
  \bibfield  {author} {\bibinfo {author} {\bibfnamefont {H.}~\bibnamefont
  {Hatami}}, \bibinfo {author} {\bibfnamefont {T.}~\bibnamefont {Kernreiter}},\
  and\ \bibinfo {author} {\bibfnamefont {U.}~\bibnamefont {Z\"ulicke}},\ }\href
  {https://doi.org/10.1103/PhysRevB.90.045412} {\bibfield  {journal} {\bibinfo
  {journal} {Phys. Rev. B}\ }\textbf {\bibinfo {volume} {90}},\ \bibinfo
  {pages} {045412} (\bibinfo {year} {2014})}\BibitemShut {NoStop}%
\bibitem [{\citenamefont {Mastrogiuseppe}\ \emph {et~al.}(2014)\citenamefont
  {Mastrogiuseppe}, \citenamefont {Sandler},\ and\ \citenamefont
  {Ulloa}}]{MOS2RKKY2}%
  \BibitemOpen
  \bibfield  {author} {\bibinfo {author} {\bibfnamefont {D.}~\bibnamefont
  {Mastrogiuseppe}}, \bibinfo {author} {\bibfnamefont {N.}~\bibnamefont
  {Sandler}},\ and\ \bibinfo {author} {\bibfnamefont {S.~E.}\ \bibnamefont
  {Ulloa}},\ }\href {https://doi.org/10.1103/PhysRevB.90.161403} {\bibfield
  {journal} {\bibinfo  {journal} {Phys. Rev. B}\ }\textbf {\bibinfo {volume}
  {90}},\ \bibinfo {pages} {161403} (\bibinfo {year} {2014})}\BibitemShut
  {NoStop}%
\bibitem [{\citenamefont {Mastrogiuseppe}\ \emph {et~al.}(2016)\citenamefont
  {Mastrogiuseppe}, \citenamefont {Sandler},\ and\ \citenamefont
  {Ulloa}}]{DiracSemimetaRKKY}%
  \BibitemOpen
  \bibfield  {author} {\bibinfo {author} {\bibfnamefont {D.}~\bibnamefont
  {Mastrogiuseppe}}, \bibinfo {author} {\bibfnamefont {N.}~\bibnamefont
  {Sandler}},\ and\ \bibinfo {author} {\bibfnamefont {S.~E.}\ \bibnamefont
  {Ulloa}},\ }\href {https://doi.org/10.1103/PhysRevB.93.094433} {\bibfield
  {journal} {\bibinfo  {journal} {Phys. Rev. B}\ }\textbf {\bibinfo {volume}
  {93}},\ \bibinfo {pages} {094433} (\bibinfo {year} {2016})}\BibitemShut
  {NoStop}%
\bibitem [{\citenamefont {Chang}\ \emph {et~al.}(2015)\citenamefont {Chang},
  \citenamefont {Zhou}, \citenamefont {Wang}, \citenamefont {Shan},\ and\
  \citenamefont {Xiao}}]{DiracSemimetaRKKY2}%
  \BibitemOpen
  \bibfield  {author} {\bibinfo {author} {\bibfnamefont {H.-R.}\ \bibnamefont
  {Chang}}, \bibinfo {author} {\bibfnamefont {J.}~\bibnamefont {Zhou}},
  \bibinfo {author} {\bibfnamefont {S.-X.}\ \bibnamefont {Wang}}, \bibinfo
  {author} {\bibfnamefont {W.-Y.}\ \bibnamefont {Shan}},\ and\ \bibinfo
  {author} {\bibfnamefont {D.}~\bibnamefont {Xiao}},\ }\href
  {https://doi.org/10.1103/PhysRevB.92.241103} {\bibfield  {journal} {\bibinfo
  {journal} {Phys. Rev. B}\ }\textbf {\bibinfo {volume} {92}},\ \bibinfo
  {pages} {241103} (\bibinfo {year} {2015})}\BibitemShut {NoStop}%
\bibitem [{\citenamefont {Bruno}\ and\ \citenamefont
  {Chappert}(1991)}]{LayersRKKY1}%
  \BibitemOpen
  \bibfield  {author} {\bibinfo {author} {\bibfnamefont {P.}~\bibnamefont
  {Bruno}}\ and\ \bibinfo {author} {\bibfnamefont {C.}~\bibnamefont
  {Chappert}},\ }\href {https://doi.org/10.1103/PhysRevLett.67.1602} {\bibfield
   {journal} {\bibinfo  {journal} {Phys. Rev. Lett.}\ }\textbf {\bibinfo
  {volume} {67}},\ \bibinfo {pages} {1602} (\bibinfo {year}
  {1991})}\BibitemShut {NoStop}%
\bibitem [{\citenamefont {Bruno}\ and\ \citenamefont
  {Chappert}(1992)}]{LayersRKKY2}%
  \BibitemOpen
  \bibfield  {author} {\bibinfo {author} {\bibfnamefont {P.}~\bibnamefont
  {Bruno}}\ and\ \bibinfo {author} {\bibfnamefont {C.}~\bibnamefont
  {Chappert}},\ }\href {https://doi.org/10.1103/PhysRevB.46.261} {\bibfield
  {journal} {\bibinfo  {journal} {Phys. Rev. B}\ }\textbf {\bibinfo {volume}
  {46}},\ \bibinfo {pages} {261} (\bibinfo {year} {1992})}\BibitemShut
  {NoStop}%
\bibitem [{\citenamefont {Stiles}(2005)}]{Stiles2005}%
  \BibitemOpen
  \bibfield  {author} {\bibinfo {author} {\bibfnamefont {M.}~\bibnamefont
  {Stiles}},\ }\bibinfo {title} {Interlayer exchange coupling},\ in\ \href
  {https://doi.org/10.1007/3-540-27163-5_4} {\emph {\bibinfo {booktitle}
  {Ultrathin Magnetic Structures III: Fundamentals of Nanomagnetism}}},\
  \bibinfo {editor} {edited by\ \bibinfo {editor} {\bibfnamefont {J.~A.~C.}\
  \bibnamefont {Bland}}\ and\ \bibinfo {editor} {\bibfnamefont
  {B.}~\bibnamefont {Heinrich}}}\ (\bibinfo  {publisher} {Springer Berlin
  Heidelberg},\ \bibinfo {address} {Berlin, Heidelberg},\ \bibinfo {year}
  {2005})\ pp.\ \bibinfo {pages} {99--142}\BibitemShut {NoStop}%
\bibitem [{\citenamefont {Mourik}\ \emph {et~al.}(2012)\citenamefont {Mourik},
  \citenamefont {Zuo}, \citenamefont {Frolov}, \citenamefont {Plissard},
  \citenamefont {Bakkers},\ and\ \citenamefont {Kouwenhoven}}]{Majoranawire}%
  \BibitemOpen
  \bibfield  {author} {\bibinfo {author} {\bibfnamefont {V.}~\bibnamefont
  {Mourik}}, \bibinfo {author} {\bibfnamefont {K.}~\bibnamefont {Zuo}},
  \bibinfo {author} {\bibfnamefont {S.~M.}\ \bibnamefont {Frolov}}, \bibinfo
  {author} {\bibfnamefont {S.~R.}\ \bibnamefont {Plissard}}, \bibinfo {author}
  {\bibfnamefont {E.~P. A.~M.}\ \bibnamefont {Bakkers}},\ and\ \bibinfo
  {author} {\bibfnamefont {L.~P.}\ \bibnamefont {Kouwenhoven}},\ }\href
  {https://doi.org/10.1126/science.1222360} {\bibfield  {journal} {\bibinfo
  {journal} {Science}\ }\textbf {\bibinfo {volume} {336}},\ \bibinfo {pages}
  {1003} (\bibinfo {year} {2012})}\BibitemShut {NoStop}%
\bibitem [{\citenamefont {Je}\ \emph {et~al.}(2013)\citenamefont {Je},
  \citenamefont {Kim}, \citenamefont {Yoo}, \citenamefont {Min}, \citenamefont
  {Lee},\ and\ \citenamefont {Choe}}]{domainwall}%
  \BibitemOpen
  \bibfield  {author} {\bibinfo {author} {\bibfnamefont {S.-G.}\ \bibnamefont
  {Je}}, \bibinfo {author} {\bibfnamefont {D.-H.}\ \bibnamefont {Kim}},
  \bibinfo {author} {\bibfnamefont {S.-C.}\ \bibnamefont {Yoo}}, \bibinfo
  {author} {\bibfnamefont {B.-C.}\ \bibnamefont {Min}}, \bibinfo {author}
  {\bibfnamefont {K.-J.}\ \bibnamefont {Lee}},\ and\ \bibinfo {author}
  {\bibfnamefont {S.-B.}\ \bibnamefont {Choe}},\ }\href
  {https://doi.org/10.1103/PhysRevB.88.214401} {\bibfield  {journal} {\bibinfo
  {journal} {Phys. Rev. B}\ }\textbf {\bibinfo {volume} {88}},\ \bibinfo
  {pages} {214401} (\bibinfo {year} {2013})}\BibitemShut {NoStop}%
\bibitem [{\citenamefont {Finocchio}\ \emph {et~al.}(2016)\citenamefont
  {Finocchio}, \citenamefont {Büttner}, \citenamefont {Tomasello},
  \citenamefont {Carpentieri},\ and\ \citenamefont {Kläui}}]{skyrmions}%
  \BibitemOpen
  \bibfield  {author} {\bibinfo {author} {\bibfnamefont {G.}~\bibnamefont
  {Finocchio}}, \bibinfo {author} {\bibfnamefont {F.}~\bibnamefont {Büttner}},
  \bibinfo {author} {\bibfnamefont {R.}~\bibnamefont {Tomasello}}, \bibinfo
  {author} {\bibfnamefont {M.}~\bibnamefont {Carpentieri}},\ and\ \bibinfo
  {author} {\bibfnamefont {M.}~\bibnamefont {Kläui}},\ }\href
  {http://stacks.iop.org/0022-3727/49/i=42/a=423001} {\bibfield  {journal}
  {\bibinfo  {journal} {Journal of Physics D: Applied Physics}\ }\textbf
  {\bibinfo {volume} {49}},\ \bibinfo {pages} {423001} (\bibinfo {year}
  {2016})}\BibitemShut {NoStop}%
\bibitem [{\citenamefont {Ye}\ \emph {et~al.}(2015)\citenamefont {Ye},
  \citenamefont {Checkelsky}, \citenamefont {Kagawa},\ and\ \citenamefont
  {Tokura}}]{Transport2BiTeI}%
  \BibitemOpen
  \bibfield  {author} {\bibinfo {author} {\bibfnamefont {L.}~\bibnamefont
  {Ye}}, \bibinfo {author} {\bibfnamefont {J.~G.}\ \bibnamefont {Checkelsky}},
  \bibinfo {author} {\bibfnamefont {F.}~\bibnamefont {Kagawa}},\ and\ \bibinfo
  {author} {\bibfnamefont {Y.}~\bibnamefont {Tokura}},\ }\href
  {https://doi.org/10.1103/PhysRevB.91.201104} {\bibfield  {journal} {\bibinfo
  {journal} {Phys. Rev. B}\ }\textbf {\bibinfo {volume} {91}},\ \bibinfo
  {pages} {201104} (\bibinfo {year} {2015})}\BibitemShut {NoStop}%
\bibitem [{\citenamefont {Lee}\ \emph {et~al.}(2011)\citenamefont {Lee},
  \citenamefont {Schober}, \citenamefont {Bahramy}, \citenamefont {Murakawa},
  \citenamefont {Onose}, \citenamefont {Arita}, \citenamefont {Nagaosa},\ and\
  \citenamefont {Tokura}}]{OpticalBiTeI}%
  \BibitemOpen
  \bibfield  {author} {\bibinfo {author} {\bibfnamefont {J.~S.}\ \bibnamefont
  {Lee}}, \bibinfo {author} {\bibfnamefont {G.~A.~H.}\ \bibnamefont {Schober}},
  \bibinfo {author} {\bibfnamefont {M.~S.}\ \bibnamefont {Bahramy}}, \bibinfo
  {author} {\bibfnamefont {H.}~\bibnamefont {Murakawa}}, \bibinfo {author}
  {\bibfnamefont {Y.}~\bibnamefont {Onose}}, \bibinfo {author} {\bibfnamefont
  {R.}~\bibnamefont {Arita}}, \bibinfo {author} {\bibfnamefont
  {N.}~\bibnamefont {Nagaosa}},\ and\ \bibinfo {author} {\bibfnamefont
  {Y.}~\bibnamefont {Tokura}},\ }\href
  {https://doi.org/10.1103/PhysRevLett.107.117401} {\bibfield  {journal}
  {\bibinfo  {journal} {Phys. Rev. Lett.}\ }\textbf {\bibinfo {volume} {107}},\
  \bibinfo {pages} {117401} (\bibinfo {year} {2011})}\BibitemShut {NoStop}%
\bibitem [{\citenamefont {Bell}\ \emph {et~al.}(2013)\citenamefont {Bell},
  \citenamefont {Bahramy}, \citenamefont {Murakawa}, \citenamefont
  {Checkelsky}, \citenamefont {Arita}, \citenamefont {Kaneko}, \citenamefont
  {Onose}, \citenamefont {Tokunaga}, \citenamefont {Kohama}, \citenamefont
  {Nagaosa}, \citenamefont {Tokura},\ and\ \citenamefont
  {Hwang}}]{TransportBiTeI}%
  \BibitemOpen
  \bibfield  {author} {\bibinfo {author} {\bibfnamefont {C.}~\bibnamefont
  {Bell}}, \bibinfo {author} {\bibfnamefont {M.~S.}\ \bibnamefont {Bahramy}},
  \bibinfo {author} {\bibfnamefont {H.}~\bibnamefont {Murakawa}}, \bibinfo
  {author} {\bibfnamefont {J.~G.}\ \bibnamefont {Checkelsky}}, \bibinfo
  {author} {\bibfnamefont {R.}~\bibnamefont {Arita}}, \bibinfo {author}
  {\bibfnamefont {Y.}~\bibnamefont {Kaneko}}, \bibinfo {author} {\bibfnamefont
  {Y.}~\bibnamefont {Onose}}, \bibinfo {author} {\bibfnamefont
  {M.}~\bibnamefont {Tokunaga}}, \bibinfo {author} {\bibfnamefont
  {Y.}~\bibnamefont {Kohama}}, \bibinfo {author} {\bibfnamefont
  {N.}~\bibnamefont {Nagaosa}}, \bibinfo {author} {\bibfnamefont
  {Y.}~\bibnamefont {Tokura}},\ and\ \bibinfo {author} {\bibfnamefont {H.~Y.}\
  \bibnamefont {Hwang}},\ }\href {https://doi.org/10.1103/PhysRevB.87.081109}
  {\bibfield  {journal} {\bibinfo  {journal} {Phys. Rev. B}\ }\textbf {\bibinfo
  {volume} {87}},\ \bibinfo {pages} {081109} (\bibinfo {year}
  {2013})}\BibitemShut {NoStop}%
\bibitem [{\citenamefont {Kohn}(1959)}]{kohnanomaly}%
  \BibitemOpen
  \bibfield  {author} {\bibinfo {author} {\bibfnamefont {W.}~\bibnamefont
  {Kohn}},\ }\href {https://doi.org/10.1103/PhysRevLett.2.393} {\bibfield
  {journal} {\bibinfo  {journal} {Phys. Rev. Lett.}\ }\textbf {\bibinfo
  {volume} {2}},\ \bibinfo {pages} {393} (\bibinfo {year} {1959})}\BibitemShut
  {NoStop}%
\bibitem [{\citenamefont {Unguris}\ \emph {et~al.}(1991)\citenamefont
  {Unguris}, \citenamefont {Celotta},\ and\ \citenamefont {Pierce}}]{exp1}%
  \BibitemOpen
  \bibfield  {author} {\bibinfo {author} {\bibfnamefont {J.}~\bibnamefont
  {Unguris}}, \bibinfo {author} {\bibfnamefont {R.~J.}\ \bibnamefont
  {Celotta}},\ and\ \bibinfo {author} {\bibfnamefont {D.~T.}\ \bibnamefont
  {Pierce}},\ }\href {https://doi.org/10.1103/PhysRevLett.67.140} {\bibfield
  {journal} {\bibinfo  {journal} {Phys. Rev. Lett.}\ }\textbf {\bibinfo
  {volume} {67}},\ \bibinfo {pages} {140} (\bibinfo {year} {1991})}\BibitemShut
  {NoStop}%
\bibitem [{\citenamefont {Purcell}\ \emph {et~al.}(1991)\citenamefont
  {Purcell}, \citenamefont {Folkerts}, \citenamefont {Johnson}, \citenamefont
  {McGee}, \citenamefont {Jager}, \citenamefont {aan~de Stegge}, \citenamefont
  {Zeper}, \citenamefont {Hoving},\ and\ \citenamefont {Gr\"unberg}}]{exp2}%
  \BibitemOpen
  \bibfield  {author} {\bibinfo {author} {\bibfnamefont {S.~T.}\ \bibnamefont
  {Purcell}}, \bibinfo {author} {\bibfnamefont {W.}~\bibnamefont {Folkerts}},
  \bibinfo {author} {\bibfnamefont {M.~T.}\ \bibnamefont {Johnson}}, \bibinfo
  {author} {\bibfnamefont {N.~W.~E.}\ \bibnamefont {McGee}}, \bibinfo {author}
  {\bibfnamefont {K.}~\bibnamefont {Jager}}, \bibinfo {author} {\bibfnamefont
  {J.}~\bibnamefont {aan~de Stegge}}, \bibinfo {author} {\bibfnamefont {W.~B.}\
  \bibnamefont {Zeper}}, \bibinfo {author} {\bibfnamefont {W.}~\bibnamefont
  {Hoving}},\ and\ \bibinfo {author} {\bibfnamefont {P.}~\bibnamefont
  {Gr\"unberg}},\ }\href {https://doi.org/10.1103/PhysRevLett.67.903}
  {\bibfield  {journal} {\bibinfo  {journal} {Phys. Rev. Lett.}\ }\textbf
  {\bibinfo {volume} {67}},\ \bibinfo {pages} {903} (\bibinfo {year}
  {1991})}\BibitemShut {NoStop}%
\bibitem [{\citenamefont {Yafet}(1987)}]{Yafet}%
  \BibitemOpen
  \bibfield  {author} {\bibinfo {author} {\bibfnamefont {Y.}~\bibnamefont
  {Yafet}},\ }\href {https://doi.org/10.1103/PhysRevB.36.3948} {\bibfield
  {journal} {\bibinfo  {journal} {Phys. Rev. B}\ }\textbf {\bibinfo {volume}
  {36}},\ \bibinfo {pages} {3948} (\bibinfo {year} {1987})}\BibitemShut
  {NoStop}%
\bibitem [{\citenamefont {Van~Vleck}(1927{\natexlab{a}})}]{vanvleck1}%
  \BibitemOpen
  \bibfield  {author} {\bibinfo {author} {\bibfnamefont {J.~H.}\ \bibnamefont
  {Van~Vleck}},\ }\href {https://doi.org/10.1103/PhysRev.29.727} {\bibfield
  {journal} {\bibinfo  {journal} {Phys. Rev.}\ }\textbf {\bibinfo {volume}
  {29}},\ \bibinfo {pages} {727} (\bibinfo {year}
  {1927}{\natexlab{a}})}\BibitemShut {NoStop}%
\bibitem [{\citenamefont {Van~Vleck}(1927{\natexlab{b}})}]{vanvleck2}%
  \BibitemOpen
  \bibfield  {author} {\bibinfo {author} {\bibfnamefont {J.~H.}\ \bibnamefont
  {Van~Vleck}},\ }\href {https://doi.org/10.1103/PhysRev.30.31} {\bibfield
  {journal} {\bibinfo  {journal} {Phys. Rev.}\ }\textbf {\bibinfo {volume}
  {30}},\ \bibinfo {pages} {31} (\bibinfo {year}
  {1927}{\natexlab{b}})}\BibitemShut {NoStop}%
\bibitem [{\citenamefont {Van~Vleck}(1928)}]{vanvleck3}%
  \BibitemOpen
  \bibfield  {author} {\bibinfo {author} {\bibfnamefont {J.~H.}\ \bibnamefont
  {Van~Vleck}},\ }\href {https://doi.org/10.1103/PhysRev.31.587} {\bibfield
  {journal} {\bibinfo  {journal} {Phys. Rev.}\ }\textbf {\bibinfo {volume}
  {31}},\ \bibinfo {pages} {587} (\bibinfo {year} {1928})}\BibitemShut
  {NoStop}%
\bibitem [{\citenamefont {Zhou}\ \emph {et~al.}(2015)\citenamefont {Zhou},
  \citenamefont {Shan},\ and\ \citenamefont {Xiao}}]{spinresp}%
  \BibitemOpen
  \bibfield  {author} {\bibinfo {author} {\bibfnamefont {J.}~\bibnamefont
  {Zhou}}, \bibinfo {author} {\bibfnamefont {W.-Y.}\ \bibnamefont {Shan}},\
  and\ \bibinfo {author} {\bibfnamefont {D.}~\bibnamefont {Xiao}},\ }\href
  {https://doi.org/10.1103/PhysRevB.91.241302} {\bibfield  {journal} {\bibinfo
  {journal} {Phys. Rev. B}\ }\textbf {\bibinfo {volume} {91}},\ \bibinfo
  {pages} {241302} (\bibinfo {year} {2015})}\BibitemShut {NoStop}%
\bibitem [{\citenamefont {Yip}(2002)}]{vvnew}%
  \BibitemOpen
  \bibfield  {author} {\bibinfo {author} {\bibfnamefont {S.~K.}\ \bibnamefont
  {Yip}},\ }\href {https://doi.org/10.1103/PhysRevB.65.144508} {\bibfield
  {journal} {\bibinfo  {journal} {Phys. Rev. B}\ }\textbf {\bibinfo {volume}
  {65}},\ \bibinfo {pages} {144508} (\bibinfo {year} {2002})}\BibitemShut
  {NoStop}%
\bibitem [{\citenamefont {Kittel}(1969)}]{kittelbook}%
  \BibitemOpen
  \bibfield  {author} {\bibinfo {author} {\bibfnamefont {C.}~\bibnamefont
  {Kittel}},\ }\href
  {https://doi.org/https://doi.org/10.1016/S0081-1947(08)60030-2} {\emph
  {\bibinfo {title} {Indirect Exchange Interactions in Metals}}},\ edited by\
  \bibinfo {editor} {\bibfnamefont {F.}~\bibnamefont {Seitz}}, \bibinfo
  {editor} {\bibfnamefont {D.}~\bibnamefont {Turnbull}},\ and\ \bibinfo
  {editor} {\bibfnamefont {H.}~\bibnamefont {Ehrenreich}},\ \bibinfo {series}
  {Solid State Physics}, Vol.~\bibinfo {volume} {22}\ (\bibinfo  {publisher}
  {Academic Press},\ \bibinfo {address} {New York},\ \bibinfo {year} {1969})\
  pp.\ \bibinfo {pages} {1 -- 26}\BibitemShut {NoStop}%
\bibitem [{not({\natexlab{a}})}]{note}%
  \BibitemOpen
  \href@noop {} {\emph {\bibinfo {title} {{\rm The exchange between two
  magnetic chains at the edge of a 2D conventional metal is found, in the
  continuum limit, by integrating the magnetic exchange between two single
  impurities ($J({\bm r})\propto \sin(2k_{F}r)/r^{2}$ where $r$ is the distance
  between the impurities~\cite{kittelbook}) over the length of the these
  chains. This leads to an interlayer exchange that decays with the distance
  between the chains $z$ as $1/z$.}}}}\BibitemShut {Stop}%
\bibitem [{\citenamefont {Liu}\ \emph {et~al.}(2016)\citenamefont {Liu},
  \citenamefont {Roy},\ and\ \citenamefont {Sau}}]{sufacesusc1}%
  \BibitemOpen
  \bibfield  {author} {\bibinfo {author} {\bibfnamefont {C.-X.}\ \bibnamefont
  {Liu}}, \bibinfo {author} {\bibfnamefont {B.}~\bibnamefont {Roy}},\ and\
  \bibinfo {author} {\bibfnamefont {J.~D.}\ \bibnamefont {Sau}},\ }\href
  {https://doi.org/10.1103/PhysRevB.94.235421} {\bibfield  {journal} {\bibinfo
  {journal} {Phys. Rev. B}\ }\textbf {\bibinfo {volume} {94}},\ \bibinfo
  {pages} {235421} (\bibinfo {year} {2016})}\BibitemShut {NoStop}%
\bibitem [{\citenamefont {Garate}\ and\ \citenamefont
  {Franz}(2010)}]{sufacesusc2}%
  \BibitemOpen
  \bibfield  {author} {\bibinfo {author} {\bibfnamefont {I.}~\bibnamefont
  {Garate}}\ and\ \bibinfo {author} {\bibfnamefont {M.}~\bibnamefont {Franz}},\
  }\href {https://doi.org/10.1103/PhysRevB.81.172408} {\bibfield  {journal}
  {\bibinfo  {journal} {Phys. Rev. B}\ }\textbf {\bibinfo {volume} {81}},\
  \bibinfo {pages} {172408} (\bibinfo {year} {2010})}\BibitemShut {NoStop}%
\bibitem [{\citenamefont {Zhu}\ \emph {et~al.}(2011)\citenamefont {Zhu},
  \citenamefont {Yao}, \citenamefont {Zhang},\ and\ \citenamefont
  {Chang}}]{rkkysurf1}%
  \BibitemOpen
  \bibfield  {author} {\bibinfo {author} {\bibfnamefont {J.-J.}\ \bibnamefont
  {Zhu}}, \bibinfo {author} {\bibfnamefont {D.-X.}\ \bibnamefont {Yao}},
  \bibinfo {author} {\bibfnamefont {S.-C.}\ \bibnamefont {Zhang}},\ and\
  \bibinfo {author} {\bibfnamefont {K.}~\bibnamefont {Chang}},\ }\href
  {https://doi.org/10.1103/PhysRevLett.106.097201} {\bibfield  {journal}
  {\bibinfo  {journal} {Phys. Rev. Lett.}\ }\textbf {\bibinfo {volume} {106}},\
  \bibinfo {pages} {097201} (\bibinfo {year} {2011})}\BibitemShut {NoStop}%
\bibitem [{\citenamefont {Abanin}\ and\ \citenamefont
  {Pesin}(2011)}]{rkkysurf2}%
  \BibitemOpen
  \bibfield  {author} {\bibinfo {author} {\bibfnamefont {D.~A.}\ \bibnamefont
  {Abanin}}\ and\ \bibinfo {author} {\bibfnamefont {D.~A.}\ \bibnamefont
  {Pesin}},\ }\href {https://doi.org/10.1103/PhysRevLett.106.136802} {\bibfield
   {journal} {\bibinfo  {journal} {Phys. Rev. Lett.}\ }\textbf {\bibinfo
  {volume} {106}},\ \bibinfo {pages} {136802} (\bibinfo {year}
  {2011})}\BibitemShut {NoStop}%
\bibitem [{\citenamefont {Efimkin}\ and\ \citenamefont
  {Galitski}(2014)}]{rkkysurf3}%
  \BibitemOpen
  \bibfield  {author} {\bibinfo {author} {\bibfnamefont {D.~K.}\ \bibnamefont
  {Efimkin}}\ and\ \bibinfo {author} {\bibfnamefont {V.}~\bibnamefont
  {Galitski}},\ }\href {https://doi.org/10.1103/PhysRevB.89.115431} {\bibfield
  {journal} {\bibinfo  {journal} {Phys. Rev. B}\ }\textbf {\bibinfo {volume}
  {89}},\ \bibinfo {pages} {115431} (\bibinfo {year} {2014})}\BibitemShut
  {NoStop}%
\bibitem [{\citenamefont {Culcer}(2011)}]{Ref21}%
  \BibitemOpen
  \bibfield  {author} {\bibinfo {author} {\bibfnamefont {D.}~\bibnamefont
  {Culcer}},\ }\href {https://doi.org/10.1103/PhysRevB.84.235411} {\bibfield
  {journal} {\bibinfo  {journal} {Phys. Rev. B}\ }\textbf {\bibinfo {volume}
  {84}},\ \bibinfo {pages} {235411} (\bibinfo {year} {2011})}\BibitemShut
  {NoStop}%
\bibitem [{\citenamefont {Ho}\ and\ \citenamefont {Jalil}(2017)}]{Ref22}%
  \BibitemOpen
  \bibfield  {author} {\bibinfo {author} {\bibfnamefont {C.~S.}\ \bibnamefont
  {Ho}}\ and\ \bibinfo {author} {\bibfnamefont {M.~B.~A.}\ \bibnamefont
  {Jalil}},\ }\href {https://doi.org/10.1063/1.4977072} {\bibfield  {journal}
  {\bibinfo  {journal} {AIP Advances}\ }\textbf {\bibinfo {volume} {7}},\
  \bibinfo {pages} {055926} (\bibinfo {year} {2017})}\BibitemShut {NoStop}%
\bibitem [{\citenamefont {Zyuzin}\ and\ \citenamefont {Loss}(2014)}]{Ref23}%
  \BibitemOpen
  \bibfield  {author} {\bibinfo {author} {\bibfnamefont {A.~A.}\ \bibnamefont
  {Zyuzin}}\ and\ \bibinfo {author} {\bibfnamefont {D.}~\bibnamefont {Loss}},\
  }\href {https://doi.org/10.1103/PhysRevB.90.125443} {\bibfield  {journal}
  {\bibinfo  {journal} {Phys. Rev. B}\ }\textbf {\bibinfo {volume} {90}},\
  \bibinfo {pages} {125443} (\bibinfo {year} {2014})}\BibitemShut {NoStop}%
\bibitem [{\citenamefont {Shiranzaei}\ \emph {et~al.}(2017)\citenamefont
  {Shiranzaei}, \citenamefont {Cheraghchi},\ and\ \citenamefont
  {Parhizgar}}]{ref11}%
  \BibitemOpen
  \bibfield  {author} {\bibinfo {author} {\bibfnamefont {M.}~\bibnamefont
  {Shiranzaei}}, \bibinfo {author} {\bibfnamefont {H.}~\bibnamefont
  {Cheraghchi}},\ and\ \bibinfo {author} {\bibfnamefont {F.}~\bibnamefont
  {Parhizgar}},\ }\href {https://doi.org/10.1103/PhysRevB.96.024413} {\bibfield
   {journal} {\bibinfo  {journal} {Phys. Rev. B}\ }\textbf {\bibinfo {volume}
  {96}},\ \bibinfo {pages} {024413} (\bibinfo {year} {2017})}\BibitemShut
  {NoStop}%
\bibitem [{\citenamefont {Zhang}\ \emph {et~al.}(2012)\citenamefont {Zhang},
  \citenamefont {Kane},\ and\ \citenamefont {Mele}}]{ss2}%
  \BibitemOpen
  \bibfield  {author} {\bibinfo {author} {\bibfnamefont {F.}~\bibnamefont
  {Zhang}}, \bibinfo {author} {\bibfnamefont {C.~L.}\ \bibnamefont {Kane}},\
  and\ \bibinfo {author} {\bibfnamefont {E.~J.}\ \bibnamefont {Mele}},\ }\href
  {https://doi.org/10.1103/PhysRevB.86.081303} {\bibfield  {journal} {\bibinfo
  {journal} {Phys. Rev. B}\ }\textbf {\bibinfo {volume} {86}},\ \bibinfo
  {pages} {081303} (\bibinfo {year} {2012})}\BibitemShut {NoStop}%
\bibitem [{\citenamefont {Chu}\ \emph {et~al.}(2011)\citenamefont {Chu},
  \citenamefont {Shi},\ and\ \citenamefont {Shen}}]{sstates}%
  \BibitemOpen
  \bibfield  {author} {\bibinfo {author} {\bibfnamefont {R.-L.}\ \bibnamefont
  {Chu}}, \bibinfo {author} {\bibfnamefont {J.}~\bibnamefont {Shi}},\ and\
  \bibinfo {author} {\bibfnamefont {S.-Q.}\ \bibnamefont {Shen}},\ }\href
  {https://doi.org/10.1103/PhysRevB.84.085312} {\bibfield  {journal} {\bibinfo
  {journal} {Phys. Rev. B}\ }\textbf {\bibinfo {volume} {84}},\ \bibinfo
  {pages} {085312} (\bibinfo {year} {2011})}\BibitemShut {NoStop}%
\bibitem [{\citenamefont {Zhang}\ \emph {et~al.}(2009)\citenamefont {Zhang},
  \citenamefont {Ran},\ and\ \citenamefont {Vishwanath}}]{ss3}%
  \BibitemOpen
  \bibfield  {author} {\bibinfo {author} {\bibfnamefont {Y.}~\bibnamefont
  {Zhang}}, \bibinfo {author} {\bibfnamefont {Y.}~\bibnamefont {Ran}},\ and\
  \bibinfo {author} {\bibfnamefont {A.}~\bibnamefont {Vishwanath}},\ }\href
  {https://doi.org/10.1103/PhysRevB.79.245331} {\bibfield  {journal} {\bibinfo
  {journal} {Phys. Rev. B}\ }\textbf {\bibinfo {volume} {79}},\ \bibinfo
  {pages} {245331} (\bibinfo {year} {2009})}\BibitemShut {NoStop}%
\bibitem [{\citenamefont {Kane}\ and\ \citenamefont {Mele}(2005)}]{ss4}%
  \BibitemOpen
  \bibfield  {author} {\bibinfo {author} {\bibfnamefont {C.~L.}\ \bibnamefont
  {Kane}}\ and\ \bibinfo {author} {\bibfnamefont {E.~J.}\ \bibnamefont
  {Mele}},\ }\href {https://doi.org/10.1103/PhysRevLett.95.146802} {\bibfield
  {journal} {\bibinfo  {journal} {Phys. Rev. Lett.}\ }\textbf {\bibinfo
  {volume} {95}},\ \bibinfo {pages} {146802} (\bibinfo {year}
  {2005})}\BibitemShut {NoStop}%
\bibitem [{\citenamefont {Fu}(2009)}]{Cutsurface}%
  \BibitemOpen
  \bibfield  {author} {\bibinfo {author} {\bibfnamefont {L.}~\bibnamefont
  {Fu}},\ }\href {https://doi.org/10.1103/PhysRevLett.103.266801} {\bibfield
  {journal} {\bibinfo  {journal} {Phys. Rev. Lett.}\ }\textbf {\bibinfo
  {volume} {103}},\ \bibinfo {pages} {266801} (\bibinfo {year}
  {2009})}\BibitemShut {NoStop}%
\bibitem [{\citenamefont {Huang}\ \emph {et~al.}(2006)\citenamefont {Huang},
  \citenamefont {Chang},\ and\ \citenamefont {Lin}}]{sufacesusc3}%
  \BibitemOpen
  \bibfield  {author} {\bibinfo {author} {\bibfnamefont {W.-M.}\ \bibnamefont
  {Huang}}, \bibinfo {author} {\bibfnamefont {C.-H.}\ \bibnamefont {Chang}},\
  and\ \bibinfo {author} {\bibfnamefont {H.-H.}\ \bibnamefont {Lin}},\ }\href
  {https://doi.org/10.1103/PhysRevB.73.241307} {\bibfield  {journal} {\bibinfo
  {journal} {Phys. Rev. B}\ }\textbf {\bibinfo {volume} {73}},\ \bibinfo
  {pages} {241307} (\bibinfo {year} {2006})}\BibitemShut {NoStop}%
\bibitem [{\citenamefont {Ando}(2006)}]{Ando}%
  \BibitemOpen
  \bibfield  {author} {\bibinfo {author} {\bibfnamefont {T.}~\bibnamefont
  {Ando}},\ }\href {https://doi.org/10.1143/JPSJ.75.074716} {\bibfield
  {journal} {\bibinfo  {journal} {Journal of the Physical Society of Japan}\
  }\textbf {\bibinfo {volume} {75}},\ \bibinfo {pages} {074716} (\bibinfo
  {year} {2006})}\BibitemShut {NoStop}%
\bibitem [{\citenamefont {Hwang}\ and\ \citenamefont
  {Das~Sarma}(2007)}]{Hwang}%
  \BibitemOpen
  \bibfield  {author} {\bibinfo {author} {\bibfnamefont {E.~H.}\ \bibnamefont
  {Hwang}}\ and\ \bibinfo {author} {\bibfnamefont {S.}~\bibnamefont
  {Das~Sarma}},\ }\href {https://doi.org/10.1103/PhysRevB.75.205418} {\bibfield
   {journal} {\bibinfo  {journal} {Phys. Rev. B}\ }\textbf {\bibinfo {volume}
  {75}},\ \bibinfo {pages} {205418} (\bibinfo {year} {2007})}\BibitemShut
  {NoStop}%
\bibitem [{\citenamefont {Dzyaloshinsky}(1958)}]{DM1}%
  \BibitemOpen
  \bibfield  {author} {\bibinfo {author} {\bibfnamefont {I.}~\bibnamefont
  {Dzyaloshinsky}},\ }\href
  {https://doi.org/https://doi.org/10.1016/0022-3697(58)90076-3} {\bibfield
  {journal} {\bibinfo  {journal} {Journal of Physics and Chemistry of Solids}\
  }\textbf {\bibinfo {volume} {4}},\ \bibinfo {pages} {241 } (\bibinfo {year}
  {1958})}\BibitemShut {NoStop}%
\bibitem [{\citenamefont {Moriya}(1960)}]{DM2}%
  \BibitemOpen
  \bibfield  {author} {\bibinfo {author} {\bibfnamefont {T.}~\bibnamefont
  {Moriya}},\ }\href {https://doi.org/10.1103/PhysRev.120.91} {\bibfield
  {journal} {\bibinfo  {journal} {Phys. Rev.}\ }\textbf {\bibinfo {volume}
  {120}},\ \bibinfo {pages} {91} (\bibinfo {year} {1960})}\BibitemShut
  {NoStop}%
\bibitem [{not({\natexlab{b}})}]{note2}%
  \BibitemOpen
  \href@noop {} {\emph {\bibinfo {title} {{\rm In the case of two isolated
  magnetic impurities placed along the c-axis of bulk Rashba semiconductors,
  the DM interaction is found to vanish \cite{HK}, consistent with our
  finding.}}}}\BibitemShut {Stop}%
\bibitem [{\citenamefont {Bennett}\ \emph {et~al.}(1990)\citenamefont
  {Bennett}, \citenamefont {Schwarzacher},\ and\ \citenamefont
  {Egelhoff}}]{exp3}%
  \BibitemOpen
  \bibfield  {author} {\bibinfo {author} {\bibfnamefont {W.~R.}\ \bibnamefont
  {Bennett}}, \bibinfo {author} {\bibfnamefont {W.}~\bibnamefont
  {Schwarzacher}},\ and\ \bibinfo {author} {\bibfnamefont {W.~F.}\ \bibnamefont
  {Egelhoff}},\ }\href {https://doi.org/10.1103/PhysRevLett.65.3169} {\bibfield
   {journal} {\bibinfo  {journal} {Phys. Rev. Lett.}\ }\textbf {\bibinfo
  {volume} {65}},\ \bibinfo {pages} {3169} (\bibinfo {year}
  {1990})}\BibitemShut {NoStop}%
\bibitem [{\citenamefont {Parkin}\ \emph {et~al.}(1990)\citenamefont {Parkin},
  \citenamefont {More},\ and\ \citenamefont {Roche}}]{exp4}%
  \BibitemOpen
  \bibfield  {author} {\bibinfo {author} {\bibfnamefont {S.~S.~P.}\
  \bibnamefont {Parkin}}, \bibinfo {author} {\bibfnamefont {N.}~\bibnamefont
  {More}},\ and\ \bibinfo {author} {\bibfnamefont {K.~P.}\ \bibnamefont
  {Roche}},\ }\href {https://doi.org/10.1103/PhysRevLett.64.2304} {\bibfield
  {journal} {\bibinfo  {journal} {Phys. Rev. Lett.}\ }\textbf {\bibinfo
  {volume} {64}},\ \bibinfo {pages} {2304} (\bibinfo {year}
  {1990})}\BibitemShut {NoStop}%
\bibitem [{\citenamefont {Lang}\ \emph {et~al.}(2014)\citenamefont {Lang},
  \citenamefont {Montazeri}, \citenamefont {Onbasli}, \citenamefont {Kou},
  \citenamefont {Fan}, \citenamefont {Upadhyaya}, \citenamefont {Yao},
  \citenamefont {Liu}, \citenamefont {Jiang}, \citenamefont {Jiang},
  \citenamefont {Wong}, \citenamefont {Yu}, \citenamefont {Tang}, \citenamefont
  {Nie}, \citenamefont {He}, \citenamefont {Schwartz}, \citenamefont {Wang},
  \citenamefont {Ross},\ and\ \citenamefont {Wang}}]{exp5}%
  \BibitemOpen
  \bibfield  {author} {\bibinfo {author} {\bibfnamefont {M.}~\bibnamefont
  {Lang}}, \bibinfo {author} {\bibfnamefont {M.}~\bibnamefont {Montazeri}},
  \bibinfo {author} {\bibfnamefont {M.~C.}\ \bibnamefont {Onbasli}}, \bibinfo
  {author} {\bibfnamefont {X.}~\bibnamefont {Kou}}, \bibinfo {author}
  {\bibfnamefont {Y.}~\bibnamefont {Fan}}, \bibinfo {author} {\bibfnamefont
  {P.}~\bibnamefont {Upadhyaya}}, \bibinfo {author} {\bibfnamefont
  {K.}~\bibnamefont {Yao}}, \bibinfo {author} {\bibfnamefont {F.}~\bibnamefont
  {Liu}}, \bibinfo {author} {\bibfnamefont {Y.}~\bibnamefont {Jiang}}, \bibinfo
  {author} {\bibfnamefont {W.}~\bibnamefont {Jiang}}, \bibinfo {author}
  {\bibfnamefont {K.~L.}\ \bibnamefont {Wong}}, \bibinfo {author}
  {\bibfnamefont {G.}~\bibnamefont {Yu}}, \bibinfo {author} {\bibfnamefont
  {J.}~\bibnamefont {Tang}}, \bibinfo {author} {\bibfnamefont {T.}~\bibnamefont
  {Nie}}, \bibinfo {author} {\bibfnamefont {L.}~\bibnamefont {He}}, \bibinfo
  {author} {\bibfnamefont {R.~N.}\ \bibnamefont {Schwartz}}, \bibinfo {author}
  {\bibfnamefont {Y.}~\bibnamefont {Wang}}, \bibinfo {author} {\bibfnamefont
  {C.~A.}\ \bibnamefont {Ross}},\ and\ \bibinfo {author} {\bibfnamefont
  {K.~L.}\ \bibnamefont {Wang}},\ }\href {https://doi.org/10.1021/nl500973k}
  {\bibfield  {journal} {\bibinfo  {journal} {Nano Letters}\ }\textbf {\bibinfo
  {volume} {14}},\ \bibinfo {pages} {3459} (\bibinfo {year}
  {2014})}\BibitemShut {NoStop}%
\bibitem [{\citenamefont {Spurgeon}\ \emph {et~al.}(2014)\citenamefont
  {Spurgeon}, \citenamefont {Sloppy}, \citenamefont {Kepaptsoglou},
  \citenamefont {Balachandran}, \citenamefont {Nejati}, \citenamefont
  {Karthik}, \citenamefont {Damodaran}, \citenamefont {Johnson}, \citenamefont
  {Ambaye}, \citenamefont {Goyette}, \citenamefont {Lauter}, \citenamefont
  {Ramasse}, \citenamefont {Idrobo}, \citenamefont {Lau}, \citenamefont
  {Lofland}, \citenamefont {Rondinelli}, \citenamefont {Martin},\ and\
  \citenamefont {Taheri}}]{exp6}%
  \BibitemOpen
  \bibfield  {author} {\bibinfo {author} {\bibfnamefont {S.~R.}\ \bibnamefont
  {Spurgeon}}, \bibinfo {author} {\bibfnamefont {J.~D.}\ \bibnamefont
  {Sloppy}}, \bibinfo {author} {\bibfnamefont {D.~M.~D.}\ \bibnamefont
  {Kepaptsoglou}}, \bibinfo {author} {\bibfnamefont {P.~V.}\ \bibnamefont
  {Balachandran}}, \bibinfo {author} {\bibfnamefont {S.}~\bibnamefont
  {Nejati}}, \bibinfo {author} {\bibfnamefont {J.}~\bibnamefont {Karthik}},
  \bibinfo {author} {\bibfnamefont {A.~R.}\ \bibnamefont {Damodaran}}, \bibinfo
  {author} {\bibfnamefont {C.~L.}\ \bibnamefont {Johnson}}, \bibinfo {author}
  {\bibfnamefont {H.}~\bibnamefont {Ambaye}}, \bibinfo {author} {\bibfnamefont
  {R.}~\bibnamefont {Goyette}}, \bibinfo {author} {\bibfnamefont
  {V.}~\bibnamefont {Lauter}}, \bibinfo {author} {\bibfnamefont {Q.~M.}\
  \bibnamefont {Ramasse}}, \bibinfo {author} {\bibfnamefont {J.~C.}\
  \bibnamefont {Idrobo}}, \bibinfo {author} {\bibfnamefont {K.~K.~S.}\
  \bibnamefont {Lau}}, \bibinfo {author} {\bibfnamefont {S.~E.}\ \bibnamefont
  {Lofland}}, \bibinfo {author} {\bibfnamefont {J.~M.}\ \bibnamefont
  {Rondinelli}}, \bibinfo {author} {\bibfnamefont {L.~W.}\ \bibnamefont
  {Martin}},\ and\ \bibinfo {author} {\bibfnamefont {M.~L.}\ \bibnamefont
  {Taheri}},\ }\href {https://doi.org/10.1021/nn405636c} {\bibfield  {journal}
  {\bibinfo  {journal} {ACS Nano}\ }\textbf {\bibinfo {volume} {8}},\ \bibinfo
  {pages} {894} (\bibinfo {year} {2014})}\BibitemShut {NoStop}%
\bibitem [{\citenamefont {Khodadadi}\ \emph {et~al.}(2017)\citenamefont
  {Khodadadi}, \citenamefont {Mohammadi}, \citenamefont {Jones}, \citenamefont
  {Srivastava}, \citenamefont {Mewes}, \citenamefont {Mewes},\ and\
  \citenamefont {Kaiser}}]{exp7}%
  \BibitemOpen
  \bibfield  {author} {\bibinfo {author} {\bibfnamefont {B.}~\bibnamefont
  {Khodadadi}}, \bibinfo {author} {\bibfnamefont {J.~B.}\ \bibnamefont
  {Mohammadi}}, \bibinfo {author} {\bibfnamefont {J.~M.}\ \bibnamefont
  {Jones}}, \bibinfo {author} {\bibfnamefont {A.}~\bibnamefont {Srivastava}},
  \bibinfo {author} {\bibfnamefont {C.}~\bibnamefont {Mewes}}, \bibinfo
  {author} {\bibfnamefont {T.}~\bibnamefont {Mewes}},\ and\ \bibinfo {author}
  {\bibfnamefont {C.}~\bibnamefont {Kaiser}},\ }\href
  {https://doi.org/10.1103/PhysRevApplied.8.014024} {\bibfield  {journal}
  {\bibinfo  {journal} {Phys. Rev. Applied}\ }\textbf {\bibinfo {volume} {8}},\
  \bibinfo {pages} {014024} (\bibinfo {year} {2017})}\BibitemShut {NoStop}%
\bibitem [{\citenamefont {Cui}\ \emph {et~al.}(2013)\citenamefont {Cui},
  \citenamefont {Khodadadi}, \citenamefont {Schäfer}, \citenamefont {Mewes},
  \citenamefont {Lu},\ and\ \citenamefont {Wolf}}]{exp8}%
  \BibitemOpen
  \bibfield  {author} {\bibinfo {author} {\bibfnamefont {Y.}~\bibnamefont
  {Cui}}, \bibinfo {author} {\bibfnamefont {B.}~\bibnamefont {Khodadadi}},
  \bibinfo {author} {\bibfnamefont {S.}~\bibnamefont {Schäfer}}, \bibinfo
  {author} {\bibfnamefont {T.}~\bibnamefont {Mewes}}, \bibinfo {author}
  {\bibfnamefont {J.}~\bibnamefont {Lu}},\ and\ \bibinfo {author}
  {\bibfnamefont {S.~A.}\ \bibnamefont {Wolf}},\ }\href
  {https://doi.org/10.1063/1.4802952} {\bibfield  {journal} {\bibinfo
  {journal} {Applied Physics Letters}\ }\textbf {\bibinfo {volume} {102}},\
  \bibinfo {pages} {162403} (\bibinfo {year} {2013})}\BibitemShut {NoStop}%
\bibitem [{\citenamefont {Klingler}\ \emph {et~al.}(2015)\citenamefont
  {Klingler}, \citenamefont {Chumak}, \citenamefont {Mewes}, \citenamefont
  {Khodadadi}, \citenamefont {Mewes}, \citenamefont {Dubs}, \citenamefont
  {Surzhenko}, \citenamefont {Hillebrands},\ and\ \citenamefont
  {Conca}}]{exp9}%
  \BibitemOpen
  \bibfield  {author} {\bibinfo {author} {\bibfnamefont {S.}~\bibnamefont
  {Klingler}}, \bibinfo {author} {\bibfnamefont {A.}~\bibnamefont {Chumak}},
  \bibinfo {author} {\bibfnamefont {T.}~\bibnamefont {Mewes}}, \bibinfo
  {author} {\bibfnamefont {B.}~\bibnamefont {Khodadadi}}, \bibinfo {author}
  {\bibfnamefont {C.}~\bibnamefont {Mewes}}, \bibinfo {author} {\bibfnamefont
  {C.}~\bibnamefont {Dubs}}, \bibinfo {author} {\bibfnamefont {O.}~\bibnamefont
  {Surzhenko}}, \bibinfo {author} {\bibfnamefont {B.}~\bibnamefont
  {Hillebrands}},\ and\ \bibinfo {author} {\bibfnamefont {A.}~\bibnamefont
  {Conca}},\ }\href {http://stacks.iop.org/0022-3727/48/i=1/a=015001}
  {\bibfield  {journal} {\bibinfo  {journal} {Journal of Physics D: Applied
  Physics}\ }\textbf {\bibinfo {volume} {48}},\ \bibinfo {pages} {015001}
  (\bibinfo {year} {2015})}\BibitemShut {NoStop}%
\bibitem [{\citenamefont {Zhang}\ \emph {et~al.}(1994)\citenamefont {Zhang},
  \citenamefont {Zhou}, \citenamefont {Wigen},\ and\ \citenamefont
  {Ounadjela}}]{exp10}%
  \BibitemOpen
  \bibfield  {author} {\bibinfo {author} {\bibfnamefont {Z.}~\bibnamefont
  {Zhang}}, \bibinfo {author} {\bibfnamefont {L.}~\bibnamefont {Zhou}},
  \bibinfo {author} {\bibfnamefont {P.~E.}\ \bibnamefont {Wigen}},\ and\
  \bibinfo {author} {\bibfnamefont {K.}~\bibnamefont {Ounadjela}},\ }\href
  {https://doi.org/10.1103/PhysRevLett.73.336} {\bibfield  {journal} {\bibinfo
  {journal} {Phys. Rev. Lett.}\ }\textbf {\bibinfo {volume} {73}},\ \bibinfo
  {pages} {336} (\bibinfo {year} {1994})}\BibitemShut {NoStop}%
\bibitem [{\citenamefont {Suenaga}\ \emph {et~al.}(2007)\citenamefont
  {Suenaga}, \citenamefont {Higashihara}, \citenamefont {Ohashi}, \citenamefont
  {Oomi}, \citenamefont {Hedo}, \citenamefont {Uwatoko}, \citenamefont {Saito},
  \citenamefont {Mitani},\ and\ \citenamefont {Takanashi}}]{pressexp1}%
  \BibitemOpen
  \bibfield  {author} {\bibinfo {author} {\bibfnamefont {K.}~\bibnamefont
  {Suenaga}}, \bibinfo {author} {\bibfnamefont {S.}~\bibnamefont
  {Higashihara}}, \bibinfo {author} {\bibfnamefont {M.}~\bibnamefont {Ohashi}},
  \bibinfo {author} {\bibfnamefont {G.}~\bibnamefont {Oomi}}, \bibinfo {author}
  {\bibfnamefont {M.}~\bibnamefont {Hedo}}, \bibinfo {author} {\bibfnamefont
  {Y.}~\bibnamefont {Uwatoko}}, \bibinfo {author} {\bibfnamefont
  {K.}~\bibnamefont {Saito}}, \bibinfo {author} {\bibfnamefont
  {S.}~\bibnamefont {Mitani}},\ and\ \bibinfo {author} {\bibfnamefont
  {K.}~\bibnamefont {Takanashi}},\ }\href
  {https://doi.org/10.1103/PhysRevLett.98.207202} {\bibfield  {journal}
  {\bibinfo  {journal} {Phys. Rev. Lett.}\ }\textbf {\bibinfo {volume} {98}},\
  \bibinfo {pages} {207202} (\bibinfo {year} {2007})}\BibitemShut {NoStop}%
\bibitem [{\citenamefont {Manaka}\ \emph {et~al.}(2001)\citenamefont {Manaka},
  \citenamefont {Yamada}, \citenamefont {Nishi},\ and\ \citenamefont
  {Goto}}]{pressexp2}%
  \BibitemOpen
  \bibfield  {author} {\bibinfo {author} {\bibfnamefont {H.}~\bibnamefont
  {Manaka}}, \bibinfo {author} {\bibfnamefont {I.}~\bibnamefont {Yamada}},
  \bibinfo {author} {\bibfnamefont {M.}~\bibnamefont {Nishi}},\ and\ \bibinfo
  {author} {\bibfnamefont {T.}~\bibnamefont {Goto}},\ }\href
  {https://doi.org/10.1143/JPSJ.70.1390} {\bibfield  {journal} {\bibinfo
  {journal} {Journal of the Physical Society of Japan}\ }\textbf {\bibinfo
  {volume} {70}},\ \bibinfo {pages} {1390} (\bibinfo {year}
  {2001})}\BibitemShut {NoStop}%
\bibitem [{\citenamefont {Li}\ \emph {et~al.}(2018)\citenamefont {Li},
  \citenamefont {Miao}, \citenamefont {Cao}, \citenamefont {Li}, \citenamefont
  {Xu}, \citenamefont {Wen}, \citenamefont {Dai}, \citenamefont {Yan},\ and\
  \citenamefont {Lü}}]{pressexp3}%
  \BibitemOpen
  \bibfield  {author} {\bibinfo {author} {\bibfnamefont {S.}~\bibnamefont
  {Li}}, \bibinfo {author} {\bibfnamefont {G.-X.}\ \bibnamefont {Miao}},
  \bibinfo {author} {\bibfnamefont {D.}~\bibnamefont {Cao}}, \bibinfo {author}
  {\bibfnamefont {Q.}~\bibnamefont {Li}}, \bibinfo {author} {\bibfnamefont
  {J.}~\bibnamefont {Xu}}, \bibinfo {author} {\bibfnamefont {Z.}~\bibnamefont
  {Wen}}, \bibinfo {author} {\bibfnamefont {Y.}~\bibnamefont {Dai}}, \bibinfo
  {author} {\bibfnamefont {S.}~\bibnamefont {Yan}},\ and\ \bibinfo {author}
  {\bibfnamefont {Y.}~\bibnamefont {Lü}},\ }\href
  {https://doi.org/10.1021/acsami.7b19684} {\bibfield  {journal} {\bibinfo
  {journal} {ACS Applied Materials \& Interfaces}\ }\textbf {\bibinfo {volume}
  {10}},\ \bibinfo {pages} {8853} (\bibinfo {year} {2018})}\BibitemShut
  {NoStop}%
\bibitem [{\citenamefont {Pletyukhov}\ and\ \citenamefont
  {Gritsev}(2006)}]{static}%
  \BibitemOpen
  \bibfield  {author} {\bibinfo {author} {\bibfnamefont {M.}~\bibnamefont
  {Pletyukhov}}\ and\ \bibinfo {author} {\bibfnamefont {V.}~\bibnamefont
  {Gritsev}},\ }\href {https://doi.org/10.1103/PhysRevB.74.045307} {\bibfield
  {journal} {\bibinfo  {journal} {Phys. Rev. B}\ }\textbf {\bibinfo {volume}
  {74}},\ \bibinfo {pages} {045307} (\bibinfo {year} {2006})}\BibitemShut
  {NoStop}%
\bibitem [{\citenamefont {Wang}\ \emph {et~al.}(2017)\citenamefont {Wang},
  \citenamefont {Chang},\ and\ \citenamefont {Zhou}}]{HK}%
  \BibitemOpen
  \bibfield  {author} {\bibinfo {author} {\bibfnamefont {S.-X.}\ \bibnamefont
  {Wang}}, \bibinfo {author} {\bibfnamefont {H.-R.}\ \bibnamefont {Chang}},\
  and\ \bibinfo {author} {\bibfnamefont {J.}~\bibnamefont {Zhou}},\ }\href
  {https://doi.org/10.1103/PhysRevB.96.115204} {\bibfield  {journal} {\bibinfo
  {journal} {Phys. Rev. B}\ }\textbf {\bibinfo {volume} {96}},\ \bibinfo
  {pages} {115204} (\bibinfo {year} {2017})}\BibitemShut {NoStop}%
\end{thebibliography}%

\end{document}